\documentclass[11pt]{article}
\usepackage{url}

\usepackage{amssymb}
\usepackage{amsthm}
\usepackage{mathtools}
\usepackage{bm}
\usepackage{amsmath}
\newcommand*{\myproofname}{Proof}
\newenvironment{myproof}[1][\myproofname]{\begin{proof}[#1]}{\end{proof}}

\usepackage{float}

\usepackage{lipsum}
\usepackage{footnote}

\usepackage{graphicx}

\usepackage{fourier}
\usepackage{array}
\usepackage{makecell}

\usepackage{longtable}
\usepackage{lipsum}

\usepackage{lscape}

\usepackage{natbib}

\usepackage[title,toc,page]{appendix}

\usepackage{subfloat}
\usepackage{subfig}

\usepackage{booktabs}

\usepackage[mathlines,pagewise]{lineno}

\newcommand{\gap}{\vspace{0.1in}}

\newtheorem{prop}{Proposition}
\newtheorem{lemma}{Lemma}

\usepackage{amssymb}

\usepackage[figuresright]{rotating}

\makeatletter
\def\ps@pprintTitle{%
	\let\@oddhead\@empty
	\let\@evenhead\@empty
	\def\@oddfoot{}%
	\let\@evenfoot\@oddfoot}
\makeatother

\title{Commuting Service Platform: Concept and Analysis}

\author{Rong Fan\textsuperscript{a} ,
	Xuegang (Jeff) Ban\textsuperscript{*}   \\
   {\footnotesize Department of Civil and Environmental Engineering, University of Washington,
   Seattle, WA 98195 U.S.A.} \\
{\footnotesize \textsuperscript{a}rongfan6@uw.edu, \textsuperscript{*}banx@uw.edu} }
\begin{document}

\maketitle
\begin{abstract}
We propose and investigate the concept of commuting service platforms (CSP) that leverage emerging mobility services to provide commuting services and connect directly commuters (employees) and their worksites (employers). By applying the two-sided market analysis framework, we show under what conditions a CSP may present the two-sidedness. Both the monopoly and duopoly CSPs are then analyzed. We show how the price allocation, i.e., the prices charged to commuters and worksites, can impact the participation and profit of the CSPs. We also add demand constraints to the duopoly model so that the participation rates of worksites and employees are (almost) the same. With demand constraints, the competition between the two CSPs becomes less intense in general. Discussions are presented on how the results and findings in this paper may help build CSP in practice and how to develop new, CSP-based travel demand management strategies.
\end{abstract}

% \linenumbers

%\newpage
%\singlespacing
\section{Introduction} \label{intro}
Commuting is ``recurring travels between home and work or study'' \citep{wiki:commuting}. Accounting for a substantial portion of the total daily trips (e.g., commuting trips are about 1/3 of the total daily trips in the US), commuting trips are important to businesses (employers), local economy, and people's daily life, which also experience the most congestion and related problems, especially in fast growing urban areas. Among all trips, they are probably the easiest to describe: Businesses (employers) lease/purchase space for their activities (often in urban areas), and abide by city ordinance. Employees work at the businesses, meeting the work and arrival times set by employers. Normally, employers make decisions that largely determine employees' work schedule. In response to such schedule, employees are commuters who create the commuting traffic (i.e., travel demand) as they decide on mode/vehicle of travel, time of travel, route, and the like. At the same time, public agencies manage/maintain the transportation infrastructure, providing proper capacity (and policies) to serve the demand. Congestion happens when demand exceeds capacity, often at specific periods of time, e.g., the peak periods that are in most cases the commuting periods. In this paper, we focus on commuting to work, while the above can also apply to ``commuting to school.''

\gap

Reducing commuting congestion and related issues have been a long-lasting challenge in transportation. In addition to infrastructure expansion (rare nowadays) and efficient traffic control schemes (such as traffic signal control, routing, etc.), adequate travel demand management (TDM) methods are crucial. TDM focuses on developing relatively longer term planning and coordination strategies to help manage people's time and modes of travel \citep{TDM2012}, with the purposes of eliminating certain trips, or switching them to more efficient modes (such as transit), or changing trip starting times (e.g., for peak spreading). TDM has been studied very extensively in the last several decades \citep{ferguson1990transportation} especially for commuting traffic. There are TDM programs in some cities in the US and around the world that developed TDM strategies to reduce congestion by promoting non single-occupancy-vehicle (SOV) travel, among others. One example is the commute trip reduction (CTR) program of the State of Washington \citep{WSDOTctr2009}. As discussed above, commuters, their employers, and transportation management agencies are the major players for commuting related decisions. Effective TDM methods should recognize and take advantage of the behavior and interactions of these major players.

\gap

In the existing urban transportation system, commuting related decisions by the major players (i.e., commuters, employers, and transportation agencies) are only loosely connected and largely isolated. Businesses (employers) are the major attractor of commuting trips, but have no or little responsibility of managing traffic or congestion; commuters (employees) form the commuting traffic in the transportation system, and to a great extent, have to follow the work schedule established by their employers and thus do not have much flexibility in their commuting schedule \citep{holguin2011impacts}; transportation agencies, who provide transportation infrastructure and system capacity, do not have any direct control on travel schedule, demands, etc. Some TDM strategies did recognize this issue and developed policies and programs to keep employers in the loop (such as employer-based transit passes, vanpooling, telecommuting, parking management, etc.) to manage commuting demands by encouraging their employees to switch from SOV travels to more efficient modes or avoid travel at all \citep{FHAW1994,Ernst1992, WSDOTctr2009}. Furthermore, many technology companies and government agencies have been implementing flexible work hours or even telecommuting policies so that their employees do not need to follow strict work schedule or go to work every day. However, such strategies are for individual employers/employees and mostly on a voluntary basis, lacking coordinations among different employers and even employees within the same employer. This results in much less significant impact to reduce commuting problems than what they could have achieved, which can be shown clearly by the steady increase in commuting related congestion, e.g., the number of hours wasted in traffic by an average US commuter in urban areas has grown for about 40\% in 6 years, from 36 hours in 2009 to 50 hours in 2015 \citep{Inrix2015}. Therefore, to make TDM strategies more effective, we need mechanisms that can more closely connect/coordinate employers and employees (and also agencies) so that the effect of commuting trips can be directly reflected in their decision makings.

\gap

At the same time, recent technology advances have produced novel mobility modes that have transformed (and will continue to transform) urban transportation. For example, mobile-app based new mobility services have led to the paradigm of mobility as a service (MaaS), the ``integration of various forms of transport services into a single mobility service accessible on demand'' \citep{MaaS2018}. MaaS connects transportation service providers and travelers directly, which includes various forms \citep{SharedMobility2016} such as ridesourcing (e.g., Uber/Lyft), ridesharing, carpooling, carsharing, bikesharing, on-demand shuttle services, among others. By focusing on all types of travels (commuting, entertainment, shopping, etc.), current MaaS only connects travelers with service providers, thus largely excluding key players of important trips (e.g., employers in commuting trips). As a result, the current form of MaaS may not be effective in solving commuting problems. After all, we have been trying to ``nudge'' travelers by technologies, incentives, etc. to the point that we probably need other innovative ways as well to collectively solve commuting problems. Meanwhile, there are also employer-sponsored transportation programs (ESTP), in which employers are directly involved with providing commuting services to their employees \citep{Employer-Sponsored2012}. This is mostly in the form of providing carpool or shuttle services to employees from home to work and vice versa, by either operating the shuttle services directly (e.g., Amazon) or by outsourcing the operations to a third party (e.g., Microsoft). For example, Amazon piloted a shuttle project to bring workers from suburb areas to its Seattle campus in 2016 \citep{Amazon2016levy}. Microsoft started the Connector program to shuttle employees from adjacent areas to its headquarter in Redmond and offices in Bellevue in 2007. Google and Apple have also implemented similar shuttle services to improve the commuting condition of their employees \citep{Google2007, Apple2015Dormehl}. More recently, industry innovators are tapping into ESTP by helping design TDM strategies \citep{Luum2019} and provide carpool services to co-workers \citep{Scoop2019}.
%Operated by MV Transportation, Microsoft's shuttle service was reported to reduce 5\% of SOV travels from 2007 to 2009 \citep{Microsoft2017Crockford, Microsoft_MVtransportation2009}.

\gap

While both MaaS and ESTP are rapidly evolving, we see a growing trend of the integration of the two. We believe that integrating MaaS and ESTP to focus on commuting trips, facilitated by proper TDM policies, may provide the needed mechanism to better connect employers and commuters, and as a result providing new ways to solve commuting challenges. In particular, we envision that such integration may produce the so-called employer-based \emph{commuting service platform} (CSP) to serve future urban commuting. CSP can help match an employer with its potential employees in the long term (called the \emph{planning} level) and provide commuting services on a daily basis (called the \emph{operational} level). We focus on the planning level analysis of CSP in this paper. Employers are becoming more motivated to join CSP because (i) they are increasingly aware of the commuting issues and have started to help directly or indirectly their employees' daily commuting (e.g., the above-discussed ESTPs), which has become one of the important strategies for them to recruit/retain the needed talents \citep{commuteSeattle2016,commuteSeattle2017,harrington2019commute}; (ii) more companies are supporting sustainability and are becoming more socially responsible (including how to deal with congestion and related issues caused by commuting); (iii) there are pressures from local communities, cities, and even states for companies to take more actions to help resolve commuting issues, e.g., the CTR program in the State of Washington. For employees, it is always in their best interest to find the optimal commuting options that can better balance work and family. Therefore, when selecting employers, possible commuting options and the commuting packages an employer can provide will have an important impact on their decisions. Therefore, CSP can be considered as a platform to connect employers and their potential employees when commuting is concerned, similar to how Amazon connects sellers and potential buyers via its online platform. Currently, a real-world, full-scale CSP does not exist yet. However, there are early prototypes of CSP. For example, for one of its business models, Scoop charges employers to provide carpool services to their employees. We expect that such early platforms will rapidly grow and evolve to the future CSP to provide a wider spectrum of commuting services.

\gap

With a CSP, an employer needs to subscribe for the platform (by paying an annual fee or per ``transaction'' fee; here a transaction means a commuting service for one of its employees) so that its employees can use the service. Employees will also be charged each time the CSP service is used. For the platform (i.e., CSP), the cost of service will still be the cost of labor and vehicle depreciation, fuel, maybe also the cost of negotiation with employers, etc. However, the source of revenue has new components. Instead of charging every commuter (employee) a fare as traditionally done, CSP has potential revenue sources from both the commuter side and the employer side. That is, CSP can closely connect employers and employees, with proper policies / management strategies from the agencies, which is similar to a two-sided market. Two-sided market is characterized by two distinct sides (e.g., employers and employees on the CSP) who get ultimate benefit by interacting through a common platform (e.g., the CSP in our study) \citep{rochet2003platform}. A platform is said to be two-sided if the price allocation but not only the aggregated price of the two sides affects the profit (or participation) \citep{rochet2006two}. Notice here that two-sided market methods have been applied to analyze MaaS where the two sides are service providers and travelers \citep{zha2016economic,djavadian2017agent}. In a CSP, however, the two sides are employers and employees (commuters), which is markedly different from the analysis for MaaS.

\gap

In this paper, we aim to conduct the economics analysis of CSP in the planning level. We are interested in understanding the interactions of employers and employees on CSP, under what conditions a CSP will be a two-sided market, and if so how to apply the two-sided market analysis framework to study the basic interactions of employers and employees on the CSP in a systematic manner. Such analyses can help understand the interactions of key players and how different policies (such as prices charged by CSP for employers and employees) by the platform may lead to different behavior of employers/employees and the resulting system effects, based on which to develop operational level methods of CSP and related TDM strategies. To begin with, we focus on a particular TDM strategy called proximate commute \citep{ProximateCommute2018} in this study. Proximate commute allows employees who work for a multi-worksites company/employer (e.g., Starbucks, Key Bank, etc.) to be assigned to worksites closer to their homes, which is beneficial to employees, employers, community, and the environment. Allowing qualified employees voluntarily swapping worksites is one of the ways that a multi-worksites employer could reduce commuting distance for its employees \citep{Mullins1995}. Here we assume that one or two CSPs are providing the commuting services to all the employees of the employer. We will apply the two-sided market analysis method to understand the interactions of worksites and employees when proximate commute is implemented, and the effect and implications of such interactions/behavior.

\gap

For the analysis, we will start with a monopoly (single) CSP that provides one type of commuting services. We will then analyze a duopoly model with two CSPs providing two types of commuting services. For the duopoly model, we are interested in the impact of worksite flexibility on commuting trips, and assume two commuting services provided by the CSPs: non-work-flex (NWF) services for which an employee needs to arrive at the worksite punctually at a particular work starting time (say 9 am in the morning), and work-flex (WF) services for which an employee has more flexibility to arrive at his/her worksite (say from 8 am to 10 am). Different players (commuters, worksites, and the CSPs) may view the two services differently: commuters may like WF due to the flexibility it provides, worksites may prefer NWF since it is easier to manage, while a CSP may prefer WF so that it does not need to send all employees to their worksites simultaneously. Therefore, understanding how the price allocation of the CSPs may influence the choices of the commuters and worksites (i.e., the two sides) regarding the two services (CSPs) will be of paramount importance to devise sensible policies and TDM strategies to encourage the use of one CSP over the other, from the perspective of managing commuting demands and related issues. Besides understanding the economic behaviors of commuters, worksites and CSPs, another concern is how the employees will be matched to the worksites in a two-sided market. For this, we add demand constraints for the duopoly model to ensure that the participation rates of employees and worksites are (almost) the same.

\gap

The proximate commute scenarios studied here is a very simplified version of a general CSP. However, we believe that the proposed CSP concept, and the two-sided market based modeling framework and analysis method developed in this paper are the first critical step and a crucial building block to establish and analyze more general CSPs and mobility service platforms for other types of trips in future urban mobility systems. Ultimately, we hope that such analyses can help develop the CSPs and next-generation TDM strategies that can better leverage the emerging systems and technologies.

\subsection{Literature review }
\subsubsection{TDM and Proximate Commute}
Originated from 1970s and 1980s, TDM promotes collaborative efforts from employers, commuters, governments to avoid the costly expansion of the transportation system \citep{ferguson1990transportation}. Since then TDM has been guiding the design of transportation and physical infrastructure and encouraging the use of transit, ridesharing, walking, biking, and telework \citep{MobilityLab2009}. There are mainly three groups of TDM strategies to : 1) improve mobility options, such as carsharing services, HOV priority lanes, walking and cycling improvement, public transportation improvement, telecommuting, flexible working hour, etc; 2) apply economic measures such as congestion price, parking regulations, etc; 3) enhance smart growth and land use policies, including transit-oriented development, location-efficient development, etc \citep{TDMcategories2009}.

\gap

In the early 1990s, Southern California launched an employer-based TDM program to improve air quality by trip reduction, which required employers with over 100 employees during peak hours to conduct trip reduction plans \citep{FHAW1994}. The cost-effectiveness of the trip reduction plans was studied, among which the commissioned program in an accounting firm resulted in a decrease of 8.4 \% in daily vehicle trips \citep{Ernst1992}. Commute trip reduction (CTR) is a specific TDM program for commuting trips, aiming to encourage travelers to drive alone less, reduce carbon emissions and keep the busiest commute routes flowing. For example, since the Washington State Legislature passed the CTR Law in 1991 \citep{kadesh1997commute}, over 1000 worksites and over 530,000 commuters have joined the state CTR program by 2009. The widely adoption of CTR program has resulted in a 9\% traffic delay reduction in the Central Puget Sound Region from 2006 to 2009 \citep{WSDOTctr2009}.

\gap

Proximate commute is a CTR strategy that allows employees of a multi-worksite organization to be assigned to the worksites close to their homes \citep{Mullins1995}. It aims to reduce commute distances by swapping workers in different worksites or taking commute distances/times into consideration when building a new worksite or recruiting new workers. Key Bank of Washington conducted a demonstration project of proximate commute in 1995. The project lasted for 15 month, during which nearly 500 employees at 30 Key Bank branches in Washington were given the opportunity to voluntarily switch to the branches closer to their homes. 17\% of eligible employees enrolled in this program, for whom commute miles reduced by 65\%. There was a 33\% reduction in the longest commute per Key Bank worksite. The results showed that proximate commute is a low-cost method for reducing employee commute time, distance, expense, which can help increase work force productivity. Employers were also willing to implement proximate commute because their investment can likely be recouped within a year through reduced absenteeism, higher morale and productivity, and other improvements.

\gap

Existing TDM strategies and CTR programs have correctly recognized the importance of involving employers.  Evaluations have also been done on the impact of involving employers in TDM to reduce traffic congestion and related issues \citep{yushimito2014two,yushimito2015correcting}. However, those programs and evaluations were often done for individual employers/employees, and lacked coordinations among different employers and employees. More critically, they have not taken the full advantage of emerging mobility options (such as ridesourcing) into consideration. In this paper, we attempts to investigate a specific TDM strategy, i.e., the next-generation proximate commute strategies that are implemented via the CSP.

\subsubsection{Two-sided market}
Two-sided markets are defined as markets where one or several platforms enable interactions between the the two sides and get the two sides ``on board" by appropriately charging each side \citep{rochet2006two}. For example, the Uber platform matches drivers and riders, and charges riders while pays wages to drivers ( wages can be considered as negative prices charged to drivers). The two sides choose to join the platform (i.e., consume the services provided by the platform) that makes them better-off. The platform bears the cost of services and charges the two sides to obtain profits. Because of the same-side negative network effects and the cross-side positive network effects, the price allocation but not only the total price of services will affect the participation rates of both sides and the profit of the platform \citep{rochet2006two}. Same-side effects capture the consumer behavior that an agent will usually be worse-off if more agents from the same side join the same platform, whereas cross-side effects exist if an agent from one side benefits from the increasing participation from the other side on a common platform. Depending on the number of platforms and the relations among them, a two-sided market may consist of a single monopoly platform or multiple competitive platforms. When there are competitive platforms, competition among platforms affects the participation and profits on each platform \citep{armstrong2007two}. Users from either side could choose to join a single platform, which is referred to as ``single-home'', or choose to use multiple platforms, which is called ``multi-home''.

\gap

\cite{rochet2003platform} provided the first comprehensive investigation of the theory of the two-sided market based on a single platform. With an analytical solution of the price allocation for different governance structures, their study unveiled how a platform makes profit by courting the two-sides. While illustrated in the context of credit cards, their study provided a benchmark model that is applicable to a wide range of applications of two-sided markets. \cite{armstrong2006competition}, \cite{armstrong2007two} analyzed two-sided markets under different degrees of product differentiation on each side of the market. They analyzed the conditions of strong product differentiation on both sides, in which case agents from both sides single-home. The conditions when sellers view the platforms as homogeneous while buyers view them as heterogeneous are also discussed. And in this case buyers still single-home, but sellers choose to multi-home. In the latter case, the platforms compete indirectly for the multi-homing sellers by attracting buyers to join, which is defined as a ``competitive bottleneck'' equilibria. ``Competitive bottleneck'' explains the observation that many platforms charge little or nothing to buyers when sellers multi-home.

\gap

Two-sided market provides a method to analyze how price structure affects profits and economic efficiency.
Take credit cards for an example. Different credit card issuers are platforms; buyers choose to own one (single-home) or multiple (multi-home) types of credit cards for purchase; sellers choose to accept one (single-home) or multiple (multi-home) types of credit cards. A transaction happens on a platform if a buyer purchases from a seller using the credit card issued by the platform. To optimize its profit, the platform needs to decide which side to bear the price burden. This usually leads the platform to make less money on one side, or even subsidize this side, and recoup its cost from the other side. The platform loses profit when subsidizing one side, and this side is regarded as a ``loss leader''. In the credit card example, buyers are usually the loss leaders and the many promotion programs by credit card issuers (such as points or rebates) are the subsidies.

\gap

There are many examples of real-world markets involving two groups of agents interacting via common platforms, which may be characterized as two-sided markets. Examples include: 1) academic publishing; 2) advertising media market; 3) payment systems, such as credit cards; 4) Internet service providers. There are a handful applications of the two-sided market theory in transportation. One example is the matching of drivers and customers in taxi or ridesourcing. By treating the ridesourcing platform as a two-sided market with customers and drivers as the two sides, \cite{zha2016economic} and \cite{wang2016pricing} studied the matching process with negative same-side externality and the positive cross-side externality.  \cite{djavadian2017agent} evaluated an agent-based stochastic day-to-day adjustment process in a two-sided market. A collection of publications in these market is summarized in Table \ref{tab:ModelExpansion}.

%\begin{landscape}
\begin{table}[H]
	\caption{Applications of Two-sided Market} \label{tab:ModelExpansion}
	\small
	\begin{tabular}{| p{.08\textwidth} | p{.1\textwidth} | p{0.1\textwidth} | p{0.5\textwidth} | p{.09\textwidth} | }
		\hline
		Field  & Platform(s) & Two sides & Findings  & Authors\\ \hline
		\textbf{Academic Journals} & Academic journals & authors ; readers  & Open access policy makes publications free to readers and charges high publication fees to authors. This policy is good when considering maximizing social welfare, but may harm readers utility, the impact or profit of the journal. &\cite{jeon2010pricing}  \\ \hline
		\textbf{Payment card} & credit card, debit card & merchants ;customers  &  Benchmark model shows that HAC rule not only benefits the multi-card platform but also raises social welfare. However, in the extended model HAC rule may no longer raise social welfare under all parameter settings. &\cite{rochet2008tying}  \\ \hline
		\textbf{Magazine} & Magazine companies & readers; advertisers & Higher demand on the reader side increases advertising rates. Higher demand on the advertiser side reduces the price of magazine to readers. & \cite{kaiser2006price}  \\ \hline
		\textbf{Internet} & Internet Service Provider (ISP) & content providers; broadband users & Network neutrality regulation increases the total surplus under certain parameter ranges for both monopoly and duopoly platforms. & \cite{economides2012network} \\
		\hline
		\textbf{Flexible mobility services} & The built environment & operators ; travelers & The differences between one-sided and two-sided market. The threshold when the network externalities lead to two-sidedness. & \cite{djavadian2017agent} \\
		\hline
		\textbf{Ride-sourcing} & Ride-sourcing services & drivers; passengers  & The matching condition/ regulation policy when the first / second best solution holds in monopoly case is found. The study of competing platforms suggests merging of platforms as competition won't lower the price level or improve social welfare & \cite{zha2016economic} \\
		\hline
	\end{tabular}
\end{table}
%\end{landscape}
Despite the above efforts of two-sided market applications, no study so far has attempted to connect employers and employees by a common platform (e.g., CSP, as proposed in this paper) when commuting trips are considered. Consequently, no study has applied the two-sided market framework to analyze the behavior and interactions of commuters and employers on CSP.

 \section{Preliminaries}
\subsection{Problem Statement} \label{sec:ProState}

%As discussed in section \ref{intro}, the proximate commute strategy has been proven to raise the satisfaction of employees and reduce commuting miles. In light of those existing programs/strategies, the envisioned CSP is practical and is promising in filling the gap that the attractors of the commuting trips (i.e., worksites), the commuters, and the mobility service companies are less coordinated. The CSP charges both worksites and employees. A worksite need to subscribe for the CSP so that its employees can be served by the CSP. Commuters still need to pay to commute with CSP, but the amount they need to pay is likely to be small. By definition, the market of the CSP is two-sided, like credit card or academic journals. Two-sidedness of CSPs is nontrivial, detailed discussions are presented in section \ref{sec:Mono_Bench}.

%\gap

\begin{figure}[H]
	\center
	\includegraphics[page=2,clip, trim=6cm 8cm 2cm 1cm, width=0.7\textwidth]{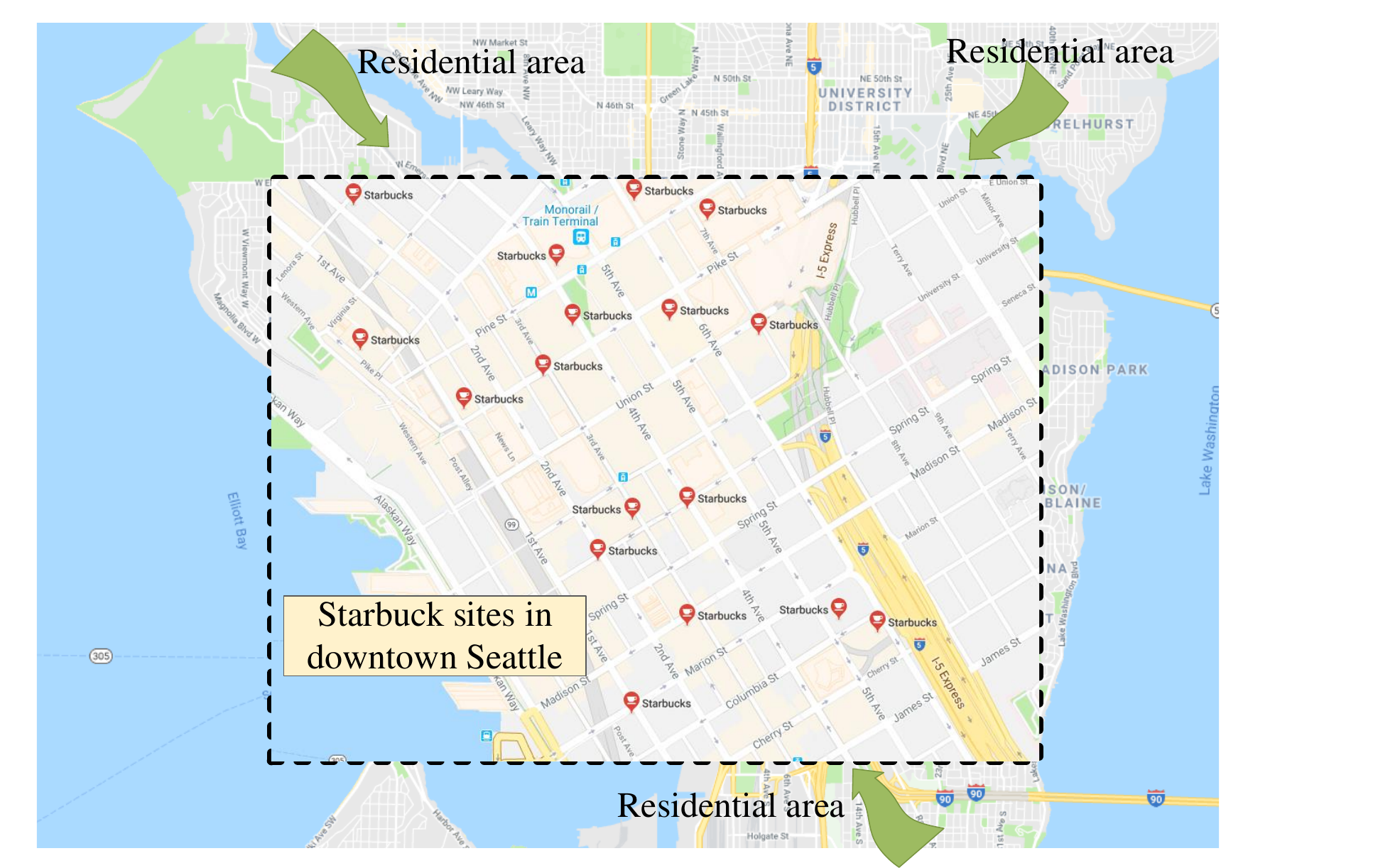}
	\caption{The distribution of Starbucks in downtown Seattle and the commuting trends}
	\label{fig:StarbucksDis}
\end{figure}

In this paper, we study the proximate commute problem of an employer with multiple worksites when there is a CSP to provide commuting services for its employees. There are many examples of such employers in urban areas. For example, as shown in Figure \ref{fig:StarbucksDis}, there are 17 Starbucks stores (worksites) in the Seattle downtown area. We will investigate in this paper when proximate commute with CSP will present two-sidedness, and when this happens how the pricing schemes and other mechanisms may impact the participation of the two sides (i.e., worksites and employees) and the profit of the CSP. We will analyze both the monopoly platform and the duopoly platforms in the two-sided markets. For the monopoly platform, there will be only one type of CSP service in the market. For the duopoly platforms, we focus on how work place flexibility may influence employees/employers' choices of the platforms. For this, as discussed above, we assume there are two types of CSPs to provide NWF services and WF services respectively. Through such investigations, we hope to gain deeper understanding of CSPs, how the price allocation of a CSP may impact its scale (i.e., the participation of the two sides) and profit, which can provide useful insight on how to design CSP and next-generation TDM strategies based on emerging technologies and mobility options such as MaaS.

We make several assumptions to simplify the real world proximate commute problem:

\gap

\noindent\textbf{(a)} One or multiple CSPs exist to provide commuting services. CSPs charge both worksites and employees for using the service.\\
\noindent\textbf{(b)} A commuter can choose which worksite to work for based on his/her own preference. \\
\noindent\textbf{(c)} The participation of worksites and employees is not pre-defined in the monopoly model (Section \ref{sec:monopolyModel}) or the duopoly model (Section \ref{sec:DuoSing}), which is determined by the market equilibrium. \\
\noindent\textbf{(d)} We assume that the employees are evenly split among the worksites on the same CSP. In section \ref{sec:competitive_constraints}, we relax Assumption (c) to add demand constraints to the duopoly model so that the participation of worksites and employees are almost the same, with small (and bounded) deviations. This essentially match the total number of employees with the total number of worksites. In the future, we can further relax this assumption by adding demand constraints to each worksite directly when the location of the worksite is specified in a transportation network.

\gap

Assumption (a) ensures that worksites and employees interact on the CSP, and the CSP's price strategy may impact their behavior of using the platform. Under assumption (b), employees are exchangeable among different worksites, which is the key concept of proximate commute. From the two-sided market analysis, we would be able to obtain the proportion of commuters/worksites participating in a platform and the resulting profits. We also start with assumption (c) such that the participation rates of employees and worksites are not pre-specified. We then relax this by adding a demand constraint to relate the participation rates of employees and worksites in assumption (d). In future research, we will specify the home locations of commuters and the locations of worksites; this may allow us to add demand constraints to each worksite directly.

\subsection{Model Setting}
An agent from either side (i.e., worksites or employees) pays a fee to join a CSP. By doing so, the agent gets a fixed benefit. Also, an agent gets better-off when the cross-side network effect increases, and gets worse-off when the same-side network effect increases. An agent chooses the CSP when s/he has higher utility. A CSP sets prices to the two sides to maximize its profit. Here is a list of notations. More specific definitions of those variables/parameters are given in Section \ref{sec:monopolyModel} and
\ref{sec:DuoSing}.
\gap

\noindent\textbf{Sets:}

\gap

\noindent \begin{tabular}{ll}
	$i$ & labels of platforms, $i \in \{ N,W \} $; $N,W$ denote NWF CSP, WF CSP, respectively. \\[5pt]
	$k$ & labels of different groups, $k \in \{ B,C \} $; $B,C$ denote worksites, commuters (employees),\\[3pt]
	&   respectively. \\[5pt]
\end{tabular}

\gap

\noindent\textbf{Variables:}

\gap

\noindent \begin{tabular}{ll}
	$q^k$ & (monopoly model) the fraction of group $k$ agents join the CSP. \\[5pt]
	$q_i^k$ & (duopoly model) the fraction of group $k$ agents single-homing on CSP $i$. \\[5pt]
	$Q^k$ & (duopoly model) the fraction of group $k$ agents multi-homing on both CSPs. We assume that \\[3pt]
	& commuters only single-home, business sites can choose to single-home or multi-home, thus $Q^B \in [0,1]$,  \\[3pt]
	& $Q^C=0$.\\[5pt]
	%  $\alpha_i$: variable, productivity per employee per day after choosing platform $i$, $i \in \{ N,W \} $. \\
	$p_i^k$ & subscription price of a group $k$ agent on CSP $i$, charged by the platform (static cost), $p_i^k \ge 0$.\\[5pt]
	$x^k$ & range from $[0,1]$, the location of a group $k$ agent at the unit interval in Hotelling Model. \\[5pt]
	$U_i^k$ & the utility of a group $k$ agent on CSP $i$.\\[5pt]
	$R_i$ & the profit of CSP $i$.\\[5pt]
\end{tabular}
\\
Here we use unit measure of group $k$ agents, thus $Q^k + q_N^k + q_W^k = 1$ in the duopoly model.

\gap

\noindent\textbf{Parameters:}

\gap

\noindent \begin{tabular}{ll}
	$U_0^k$ & the intrinsic benefit of a group $k$ agent when joining a CSP.\\[5pt]
	$t^k$ & the rate of inconvenience cost (i.e., same-side ``congestion'' effects) of the Hotelling model. \\[5pt]
	$\beta_i$ &  the rate of cross-side benefit of commuters on CSP $i$. A commuter obtains benefit $\beta_i q_i^B$ by joining\\[3pt]
	& CSP $i$, as she has the potential to choose from $ q_i^B$ worksites.\\[5pt]
	$\alpha_i$ & the rate of cross-side benefit of worksites on CSP $i$. A worksite obtains benefit $\alpha_i q_i^C$ by joining CSP $i$,  \\[3pt]
	& as it has the potential to choose from $ q_i^B$ commuters.\\[5pt]
	$b^k$ & the rate of cross-side benefit of group $k$ agents in monopoly model\\[5pt]
	$f_i^k$ & the cost of CSP $i$ to serve group $k$ agents in duopoly model. \\[5pt]
	$f^k$ & the cost of the CSP to serve group $k$ agents in monopoly model. \\[5pt]
\end{tabular}

\gap

\noindent\textbf{Auxiliary parameters:}

\gap

\noindent \begin{tabular}{ll}
	$\alpha^+, \alpha^-$ & the sum/difference of the cross-side benefit rate of worksites on the two CSPs, respectively, defined \\[3pt]
	& for analytical purpose, $\alpha^+ = \alpha_N + \alpha_W$, $\alpha^- = \alpha_N - \alpha_W$\\[5pt]
	$\beta^+, \beta^-$ & the sum/difference of the cross-side benefit rate of commuters on the two CSPs, respectively, \\[3pt]
	& defined for analytical purpose, $\beta^+ = \beta_N + \beta_W$, $\beta^- = \beta_N - \beta_W$\\[5pt]
	$\psi_W^k$ &  a term of equilibrium prices in Proposition \ref{prop:prices_competitive}, affected by different cross-side/same-side network \\[3pt]
	& effect on the two CSPs\\[5pt]
\end{tabular}

\section{Hotelling model and linear demand specification} \label{sec:HotelModel}
To model the inconvenience cost of worksites and employees joining the CSP, we apply the Hotelling model \citep{hotelling1990stability} that has been applied to many studies in two-sided markets. \cite{economides2012network} used Hotelling model for monopoly and duopoly two-sided markets. \cite{rochet2003platform}, \cite{armstrong2006competition} and \cite{kaiser2006price} applied Hotelling model for duopoly two-sided market models. In this study, we apply Hotelling model in both the monopoly platform and duopoly platforms. A classic Hotelling model depicts that customers are uniformly distributed on a unit length street, and two stores locate at the two ends ($x=0$ and $x=1$) of the street. Hotelling competition assumes that consumers purchase at those stores if and only if the minimum utility of consuming at the two stores is larger than some constant $\bar{U}$ \citep{fudenberg1991game}. Under this assumption, the Nash equilibrium is achieved when the utility of purchasing at the two stores are the same. Denote the Nash equilibrium as $x^*$, $x^* \in [0,1]$. Under equilibrium, consumers located to the left of $x^*$ choose the store at $x=0$, the rest of consumers choose the store at $x=1$. Therefore, $x$ also indicates the proportion of customers, i.e., the participation rate, who choose the store at $x=0$ ($1-x$) is the proportion of customers who choose the store at $x=1$). The role of Hotelling model in monopoly platform and duopoly platforms are similar. Here we use duopoly platforms as an example, and illustrate how the Hotelling model is applied to represent inconvenience cost.

\gap

\begin{figure}[H]
	\center
	\includegraphics[page=4,clip, trim=1cm 16cm 13cm 1cm, width=0.6\textwidth]{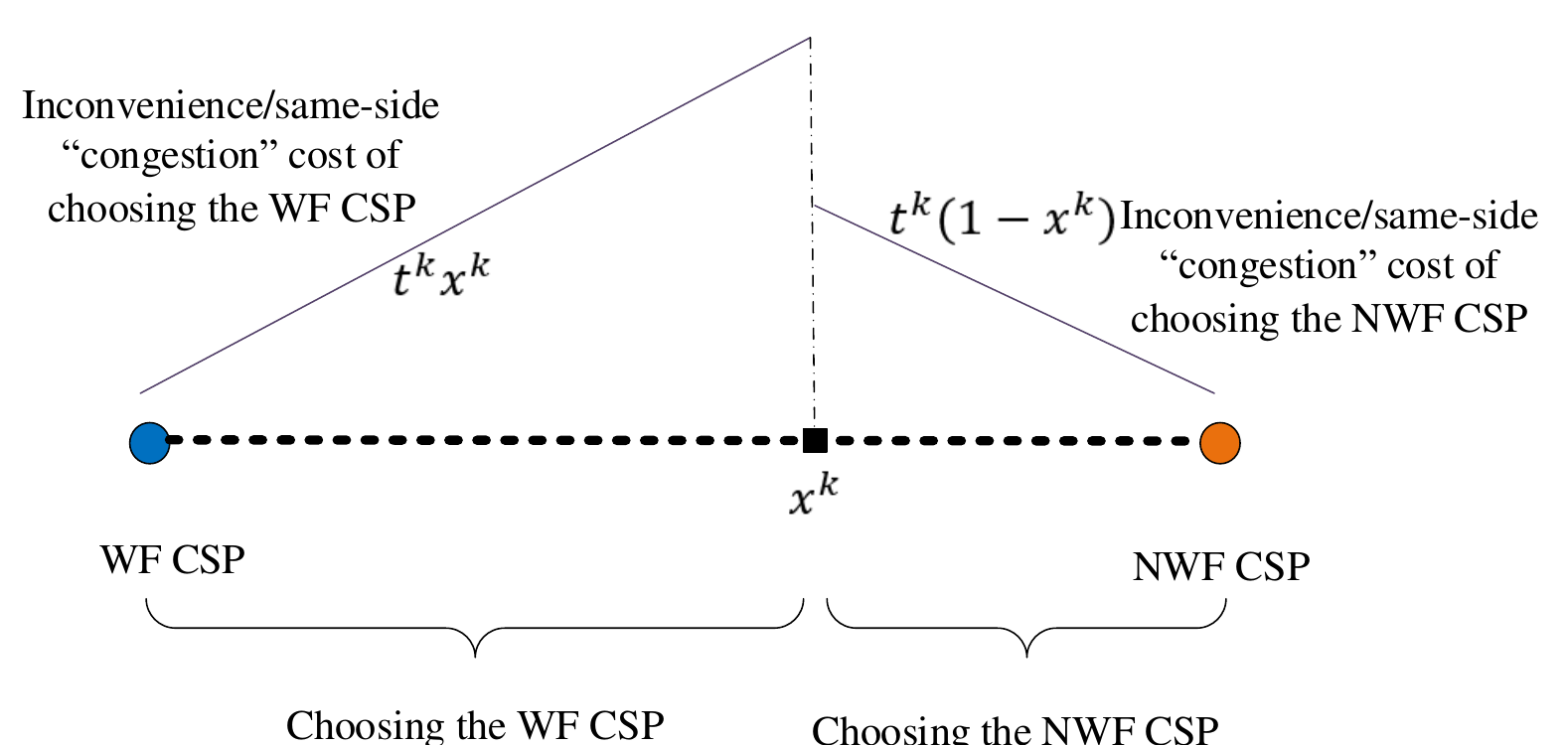}
	\caption{An illustration of Hotelling Model $(k \in \{ B,C \})$}
	\label{fig:Hotelling}
\end{figure}

In duopoly platforms, the NWF CSP ($x=1$) and the WF CSP ($x=0$) are located at the two ends of a unit interval, as shown in Figure \ref{fig:Hotelling}. The demands of worksites and commuters are both specified by Hotelling models. Take worksites as an example, worksites distribute uniformly along the unit interval. $x^B$ denotes the location of a worksite, which also indicates preference of the worksite over the WF CSP (or the NWF CSP): smaller $x^B$ implies the worksite prefers more of the WF CSP (and less of the NWF CSP) . $t^B$ denotes the rate of inconvenience cost of worksites. The worksite located at $x^B$ experiences a inconvenience cost of $t^B x^B$ when joining the WF CSP, or a inconvenience cost of $t^B(1-x^B)$ when joining the NWF CSP (shown in equation \eqref{EQutilityB} and \eqref{EQutilityB_N}). Under such setting, the cost term induced by Hotelling model ($t^B x^B$ or $t^B(1-x^B)$) reflects the same-side ``congestion'' effects. It means that a worksite will be worse-off if more other worksites join the same CSP. $t^B$ can be understood as the same-side ``congestion'' effects of the two CSPs. At the same time, $t^B$ also reflects the level of competition between the two CSP. The competition becomes stronger for worksites when $t^B$ decreases \citep{armstrong2007two}. This can be explained as follows. For the same level of $x^B$, if we decrease $t^B$, the gap of the inconvenience cost of joining the two platforms is smaller, namely $t^Bx^B - t^B(1-x^B)$ is smaller. This indicates that worksites experience less distinct inconvenience cost on the two CSPs when $t^B$ is small, which means the level of competition is high. At the equilibrium, the optimal $x^{B*}$ represents the participation rate of businesses in the WF CSP, and $1-x^{B*}$ is the participation rate of businesses in the NWF CSP.

\gap

Same analysis can be applied to the commuter side, which is omitted here. We can also apply the Hotelling model to the monopoly platform, with a similar interpretation as shown in Figure \ref{fig:Hotelling} that one end is the CSP and the other end is ``not joining the CSP''.

\section{Monopoly platform} \label{sec:monopolyModel}
In a monopoly model, agents from the two sides choose to join the CSP or be off the market. The utilities of agents are affected by the same-side and cross-side network effects. In this section, we first discuss two-sidedness of the CSP, then analyze how network effects impact the participation, price strategies, and profits of the CSP.

\subsection{Two-sidedness in commuting problems} \label{sec:Mono_Bench}
Before introducing the model, we want to answer why the envisioned CSP is a two-sided market, and why the two-side market theory is important in studying the CSP. We first present a benchmark model of the monopoly platform similar to \cite{armstrong2006competition} to illustrate the two-sidedness of a market with CSP. Note that in this benchmark model, the same-side congestion effect will not be considered, which will be added later in Section \ref{sec:Mono_hotel}. As stated above, worksites and commuters are the two sides, denoted as $k \in \{ B,C \}$. We assume unit mass for both sides, i.e., $q^B, q^C \in [0,1]$. $q^B$ and $q^C$ are then also the participation rates of the  two sides. By joining the CSP, a group $k$ agent incurs a fixed benefit $U_0^k$. In the monopoly model, there is only one CSP sending commuters to worksites. The costs for the CSP to serve the two groups are $f^B and f^C$, respectively. $f^C$ represents the per commuter transportation cost, while $f^B$ represents the negotiation cost of attracting a worksite to join the CSP. The rate of the cross-side network benefits are measured by $b^k$. By joining the CSP, each agent from group $k$ experiences a benefit of $b^k q^l$ under the assumption that s/he values the participation of the other group $l$. This is intuitively understandable. A commuter on the CSP will be better-off if more worksites join the CSP since s/he will have more choices of worksites. Similarly, a worksite on the CSP will also benefit if more commuters join the CSP since this can potentially attract more employees (commuters) to the worksite, which is the desired results when the worksite subscribes for the CSP. %On the other hand, larger participation of commuters on the CSP implies increased number of CSP commuters to a worksite, which is the desired results when a worksite subscribes the CSP.

\gap

Based on the above discussions, the utility of a group $k$ agent is determined by

\begin{linenomath}
	\begin{equation}
	U^k = U_0^k + b^k q^l - p^k \quad \forall k,l \in  \{ B,C \} \text{ and } k \ne l \label{eq:Mono_U_bench}
	\end{equation}
\end{linenomath}
Supposing that the participation of group $k$ agents on the CSP can be measured by an increasing function of utility, the participation of group $k$ agents is:
\hfill\break
\begin{linenomath}
	\begin{equation}
	q^k = \phi^k (U^k)
	\end{equation}
\end{linenomath}
From equation \eqref{eq:Mono_U_bench}, the price of group $k$ can be expressed as $p^k = U_0^k + b^k q^l - U^k$. Profit of the CSP can be written as,
\begin{linenomath}
\begin{align}
R &= ( p^B - f^B ) \phi^B (U^B) + ( p^C - f^C ) \phi^C (U^C) \nonumber \\
&=\Big[ (U_0^B + b^B \phi^C (U^C) - U^B ) - f^B \Big]\phi^B (U^B) + \Big[ (U_0^C + b^C \phi^B (U^B) - U^C) - f^C \Big]\phi^C (U^C)
\end{align}
\end{linenomath}
The equilibrium price can be obtained by maximizing the profit of the platform:
\hfill\break
\begin{linenomath}
	\begin{equation}
	p^k = f^k - U_0^k - b^l \phi^l(U^l) + \frac{\phi^k(U^k)}{[{\phi^k }(U^k)]'} \quad \forall k,l \in  \{ B,C \} \text{ and } k \ne l \label{eq:single_optimalp}
	\end{equation}
\end{linenomath}

In Proposition \ref{prop:LernerIndices}, we rewrite the equilibrium price \eqref{eq:single_optimalp} in the form of Lerner indices and elasticities \citep{lerner1934}.
\begin{prop}
\label{prop:LernerIndices}
	Write
	\hfill\break	
	\begin{linenomath}
		\begin{equation}
		\eta^B(p^B \mid q^C) = \frac{p^B [\phi^B(U^B)]'}{\phi^B(U^B)} = \frac{p^B [\phi^B(U_0^B + b^B q^C - p^B)]'}{\phi^B(U_0^B + b^B q^C - p^B)}
		\end{equation}
		\begin{equation}
		\eta^B(p^C \mid q^B) = \frac{p^C [\phi^C(U^C)]'}{\phi^C(U^C)} = \frac{p^C [\phi^C(U_0^C + b^C q^B - p^C)]'}{\phi^C(U_0^C + b^C q^B - p^C)}
		\end{equation}
	\end{linenomath}
	for a group's price elasticity of demand given the level of participation by another group. Then the optimal prices satisfy
	\hfill\break
	\begin{linenomath}
		\begin{flalign}
		\frac{p^B - (f^B - U_0^B - b^C q^C)}{p^B} = \frac{1}{\eta^B (p^B \mid q^C)}; \quad  \frac{p^C - (f^C - U_0^C - b^B q^B)}{p^C} = \frac{1}{\eta^C (p^C \mid q^B)} \label{EQ:elasticity}
		\end{flalign}
	\end{linenomath}
\end{prop}

\noindent Based on this benchmark model and Proposition \ref{prop:LernerIndices}, we illustrate two important concepts:

\gap

\textit{(i) Loss leader:} under the optimal price structure, it is possible for group $k$ agents to act as the loss leader (to be subsidized), that is, when $p^k < f^k$. From equation \eqref{EQ:elasticity}, this occurs if the group's elasticity of demand $\eta^k(p^k \mid q^l)$ is large and/or the cross-side benefit ($b^l$) enjoyed by group $l$ ($k,l \in \{ B,C \}, k \ne l$) is large. On the other hand, when $b^l$ is small and/or $\eta^k(p^k \mid q^l)$ is small, the CSP charges higher price to group $k$ agents.
%In credit card example, it usually occurs that sellers values the number of buyers on the same platform, and/or the elasticity of buyers is high, which lead the credit card platform to charge lower price to buyers. This analysis can also be applied in the commuting problem.
This implies that, if worksites value the number of commuters that join the CSP, and/or commuters' elasticity of demand is high, the CSP will subsidize the commuters to attract more commuters, thus more worksites to join the platform.

\gap

\textit{(ii) Two-sidedness:} two-sidedness is conceptually defined from either of the two aspects: 1) ``the volume of transaction on the platform depends on the allocation of price between the two sides but not only on the aggregated price level'' \citep{rochet2006two};  2) the decision of one group affects the outcomes of the other group, typically through an externality, i.e. cross-side positive network effects and same-side negative network effects \citep{rysman2009economics,caillaud2003chicken,armstrong2006competition}. Because CSP services transport employees to their worksites, no agent from one group would be willing to join the CSP unless agents from the other group also join the CSP (if no commuter joins the CSP, it's irrational for worksites to join the CSP and vice-versa). As a result, the price of one group is affected by the participation and network effects of both groups. In the benchmark model, if there is no network effects (either cross-side or same-side) and only the aggregated price level affects the transaction, the participation of the two sides will be the same, $q^B = q^C = \phi(p^B+p^C)$. The model will be reduced to $R = (p^B + p^C  - f^B - f^C)\phi(p^B+p^C)$. In the reduced model, if the CSP holds the aggregated price (i.e., $p^B + p^C$) to be constant, its profit will not change even when the price allocation differs. In this case, the model will reduce to one-sided. In the models in section \ref{sec:Mono_hotel} and \ref{sec:DuoSing}, we assume there are network effects in the market. Therefore, participation always changes with price allocation and network effects from both groups, indicating that the CSP is indeed a two-sided market. Two-sidedness is important for us to unveil how to incentivize participation on the CSP under different level of network effects.

\gap

Now that we are clear about the definition of two-sidedness, we can continue to argue that one-sided logic is not suitable for the proposed CSPs. \cite{wright2004one} listed eight fallacies of using one-sided logic in two-sided markets, most of which are applicable for the CSPs. For example, the prices of one-sided markets reflect the relative costs of products, which means that the price is high for the high-cost product/service, and the price can be low for the low-cost product/service. However, this is not true for CSP for which higher cost of sending a commuter to her/his worksite does not necessarily increases the price of customers since the CSP also charges the worksites; see more discussions on this in the next section. Based on our discussion of two-sideness, we can summarize the condition of two-sidedness of the monopoly platform as follows,

\gap

\noindent\textbf{(A0)} $b^k > 0, t^k > 0$, ensures that network effects exist ($b^k$ represents cross-side network effects. Same-side effects $t^k$ was introduced in section \ref{sec:HotelModel}).

\gap

When the CSP extracts profits from both sides in a market with network effects, the participation of one side will be affected by the decisions of both sides, thus two-sidedness holds. Condition \textbf{(A0)} is the basic feature of a two-sided market, which is applied to all of the models in this paper. In section \ref{sec:Mono_hotel}, we add Hotelling Model to the monopoly platform, which can also be understood as the same-side negative network effects. In section \ref{sec:DuoSing}, we analyze a market with two CSPs, each of which satisfies condition \textbf{(A0)}.

\subsection{Monopoly platform with linear demand specification} \label{sec:Mono_hotel}
In this part, we add the inconvenience cost to the benchmark model mentioned above. %in section \ref{sec:Mono_Bench}.
We introduce the Hotelling Model to the utility function to represent the horizontal differentiation between joining vs. not joining the CSP. A group $k$ agent incurs fixed benefit $U_0^k$ when joining CSP. Group $k$ agents are uniformly distributed on a unit interval $[0,1]$ with the WF CSP at $x=0$. $t^k$ is the rate of inconvenience cost (same-side ``congestion'' effect) when an agent from group $k$ joins the CSP. A group $k$ agent located at $x^k$ experiences an inconvenience cost (same-side ``congestion'' effect) of $t^k x^k$ to join the WF CSP, or a cost of $t^k (1-x^k)$ if s/he does not join the WF CSP. Adding the new utility terms $U_0^k$ and $-t^k x^k$ to equation \eqref{eq:Mono_U_bench}, we obtain the following utility functions:
\hfill\break
\begin{linenomath}
	\begin{equation}
	U^C = U_0^C + b^C q^B - t^C x^C - p^C \quad \quad U^B =  U_0^B + b^B q^C - t^B x^B - p^B \label{eq:mono_UBC}
	\end{equation}
\end{linenomath}

Here are the conditions that ensure the equilibrium prices of the monopoly model are feasible \citep{economides2012network} :

\gap

\noindent\textbf{(A1)} Cross-side positive effects are not strong. When same-side negative effects and cross-side positive effects follow the condition $4 t^B t^C>(b^B+b^C)^2$, the profit function is concave and the equilibrium prices are feasible;

\noindent\textbf{(A2)} The parameter setting satisfies $4t^B t^C-(b^B+b^C)^2 \ge \max\{(U_0^C-f^C)(b^B+b^C) + 2t^C(U_0^B-f^B), $ $ (b^B+ b^C)(U_0^B- f^B)  + 2 t^B(U_0^C- f^C) \}$, which ensures $q^B,q^C \le 1$.

\gap

Equilibrium exists when worksites or commuters are indifferent between choosing and not choosing the CSP. In the Hotelling model, agents from both sides are uniformly distributed on a unit interval. Notice that we assume unit mass for both sides, thus $x^k=q^k$. The demand of two groups are:
\hfill\break
\begin{linenomath}
	\begin{equation}
	q^B = \frac{ b^B (U_0^C-p^C) + t^C (U_0^B- p^B)}{t^B t^C-b^B b^C} \quad\quad q^C = \frac{ b^C(U_0^B-p^B) + t^B(U_0^C- p^C)}{t^B t^C-b^B b^C} \label{eq:NE_mono_demand}
	\end{equation}
\end{linenomath}
Substitute equation \eqref{eq:NE_mono_demand} to profit function, we can write CSP profit as a function of the prices:
\hfill\break
\begin{linenomath}
	\begin{equation}
	R = \frac{ b^B (U_0^C-p^C) + t^C (U_0^B- p^B)}{t^B t^C-b^B b^C}(p^B-f^B) + \frac{ b^C(U_0^B-p^B) + t^B(U_0^C- p^C)}{t^B t^C-b^B b^C}(p^C-f^C) \label{eq_monopolyProfit}
	\end{equation}
\end{linenomath}
Maximize CSP's profit using the first order condition of equation \eqref{eq_monopolyProfit}. The equilibrium prices are
\hfill\break
\begin{linenomath}
	\begin{equation}
	p^B = \frac{-{b^B}^2 f^B - {b^C}^2 U_0^B + (2t^B t^C-b^B b^C)(f^B+U_0^B)+t^B(b^B-b^C)(U_0^C-f^C)}{4t^B t^C-(b^B+b^C)^2} \label{eq:mono_pB}
	\end{equation}
\end{linenomath}
\begin{linenomath}
	\begin{equation}
	p^C = \frac{-{b^C}^2 f^C - {b^B}^2 U_0^C + (2 t^B t^C- b^B b^C )(f^C + U_0^C) + t^C(b^B - b^C) (f^B - U_0^B )}{4t^B t^C-(b^B+b^C)^2} \label{eq:mono_pC}
	\end{equation}
\end{linenomath}
Substitute equation \eqref{eq:mono_pB} and \eqref{eq:mono_pC} into equation \eqref{eq:NE_mono_demand}. Demands on the CSP platform are
\hfill\break
\begin{linenomath}
	\begin{equation}
	q^B = \frac{(U_0^C-f^C)(b^B+b^C) + 2t^C(U_0^B-f^B)}{4t^Bt^C-(b^B+b^C)^2} \quad q^C = \frac{(b^B+ b^C)(U_0^B- f^B)  + 2 t^B(U_0^C- f^C)}{4t^Bt^C-(b^B+b^C)^2}
	\end{equation}
\end{linenomath}
Substitute equation \eqref{eq:mono_pB} and \eqref{eq:mono_pC} into equation \eqref{eq_monopolyProfit}, we get the profit at equilibrium price
\hfill\break
\begin{linenomath}
	\begin{equation}
	R = \frac{(b^B +b^C) (f^B-U_0^B)( f^C  - U_0^C) +  t^C (f^B-U_0^B)^2 +  t^B (f^C-U_0^C)^2  }{4t^B t^C-(b^B+b^C)^2}
	\end{equation}
\end{linenomath}
Equation \eqref{eq:NE_mono_demand} shows the relation between price and participation. When network effects and fix benefits are given, quantity can be expressed as a linear combination of prices, $q^k = f(p^B, p^C)$. This also indicates the key characteristic of a two-sided market, i.e., the participation of one side of the market is affected by the allocation of prices between the two sides. The equilibrium prices of the two sides are symmetric. It's hard to draw conclusions on the distribution of price allocations based on the equilibrium expressions shown in equation \eqref{eq:mono_pB} and \eqref{eq:mono_pC}. A comprehensive analysis on how network effects impact the distribution of prices, quantities and profits are presented in the numerical experiments next.

\subsection{Numerical experiments of monopoly platform}
In the numerical experiment, we use the Starbucks stores in downtown Seattle as an example (Figure \ref{fig:StarbucksDis}) to explain the participation of the two sides. For this, we assign Starbucks commuters and stores (worksites) as the two sides to the CSP based on their preferences, and maximize the profit of the CSP by selecting the optimal price strategies under different network effects. The baseline parameters for this case study (also summarized in Table \ref{tab:Modelparameter} later) are $U_0^B=1.9$, $U_0^C=2.1$, $b^B=0.5$, $b^C=0.7$, $t^B=1.1$, $t^C=1.5$, $f^B=0.73$, $f^C=0.75$, which satisfy conditions \textbf{(A1)} and \textbf{(A2)}, in addition to \textbf{(A0)} as discussed above. Under this setting, CSP's cost of serving commuters ($f^C$) is larger than that of the worksites ($f^B$). A commuter incurs higher benefit when joining the CSP than a worksite ($U_0^C>U_0^B$) for the approximate commute problem studied here. Commuters value the number of worksites more than worksites value commuters ($b^C > b^B$); commuters dislike the participation of other commuters more than that of worksites ($t^C > t^B$). There are 17 Starbucks in total. The number of employees at each Starbucks may vary depends on the size of the store. Let's say that there are 10 employees at each store, thus 170 commuters in total. Given the parameter setting, there are $17*q^B$ worksites choosing the CSP, $170*q^C$ commuters choosing the CSP. We assume that the commuters on the CSP are split evenly among the worksites that choose the CSP (assumption (d)). For the agents that choose the CSP, the number of employees at each worksite is {\tiny $\frac{170*q^C}{17*q^B}=\frac{10*q^C}{q^B}$}. For agents that do not choose the CSP, there are {\tiny $\frac{10*(1-q^C)}{(1-q^B)}$} employees at each worksite. The number of employees at each worksite is actually determined by the ratio $q^C/q^B$. When implementing our model to a real world problem, we need to add an extra constraint so that this ratio is limited within a threshold ($q^C/q^B \approx 1$), which ensures that there are reasonable number of employees at each worksites. We added this constraint in section \ref{sec:competitive_constraints}.

%(The values may not be integers, but this is not a big concern if the number of agents is large. If we round the number of participants ($170*q^C$) to integer, the proportion of participants ($q^C$) will not change much thus will not be affected by the approximation.)

\subsubsection{Demand-price relation}

\begin{figure}[H]
	\center
	\subfloat[$q^B$]{%
		\includegraphics[clip, trim=1cm 1cm 2cm 0cm, width=0.4\textwidth]{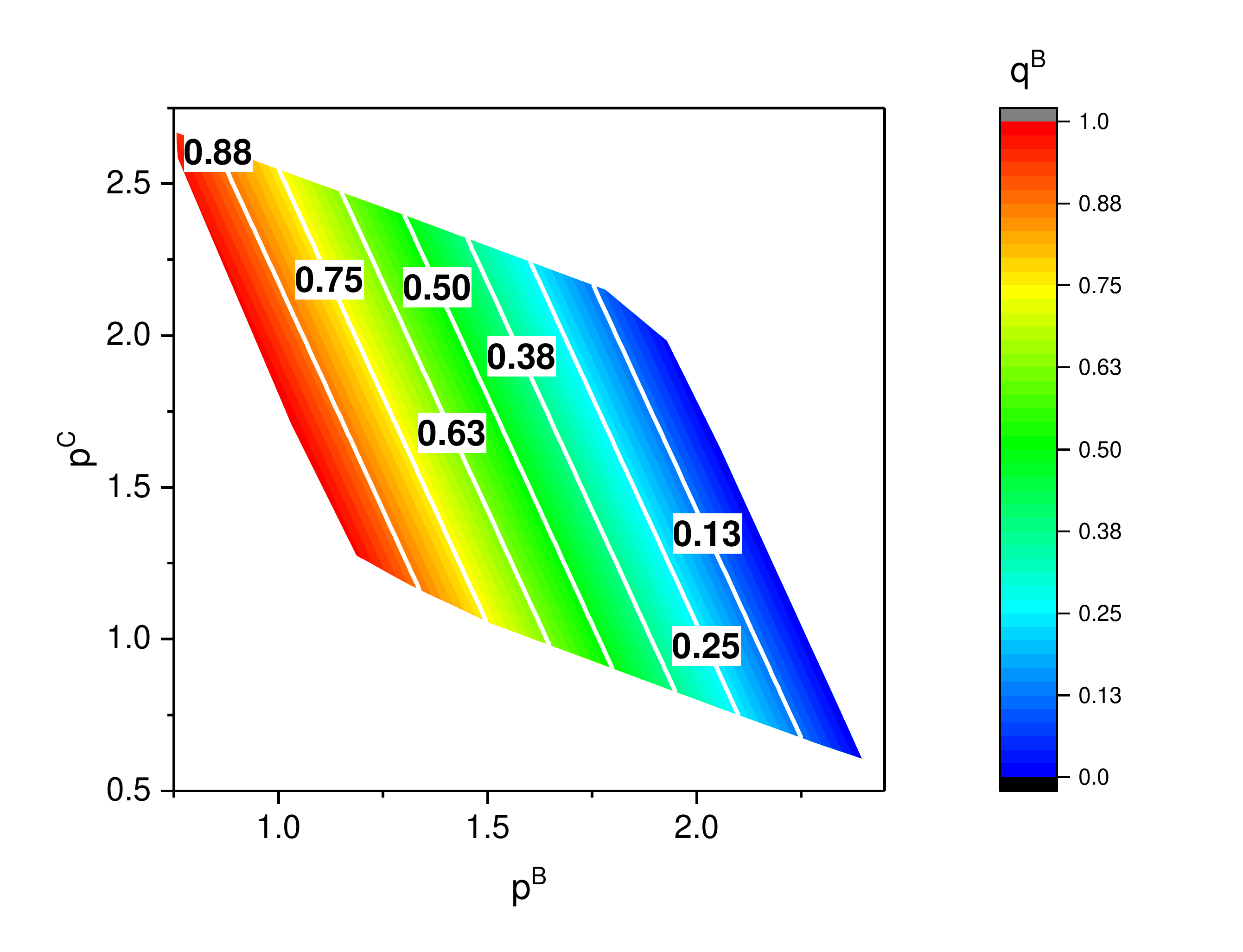}
		\label{fig:qB_ptBpC}%
	}%
	\hspace{0.5cm} %\hfill%
	\subfloat[$q^C$]{%
		\includegraphics[clip, trim=1cm 1cm 2cm 0cm, width=0.4\textwidth]{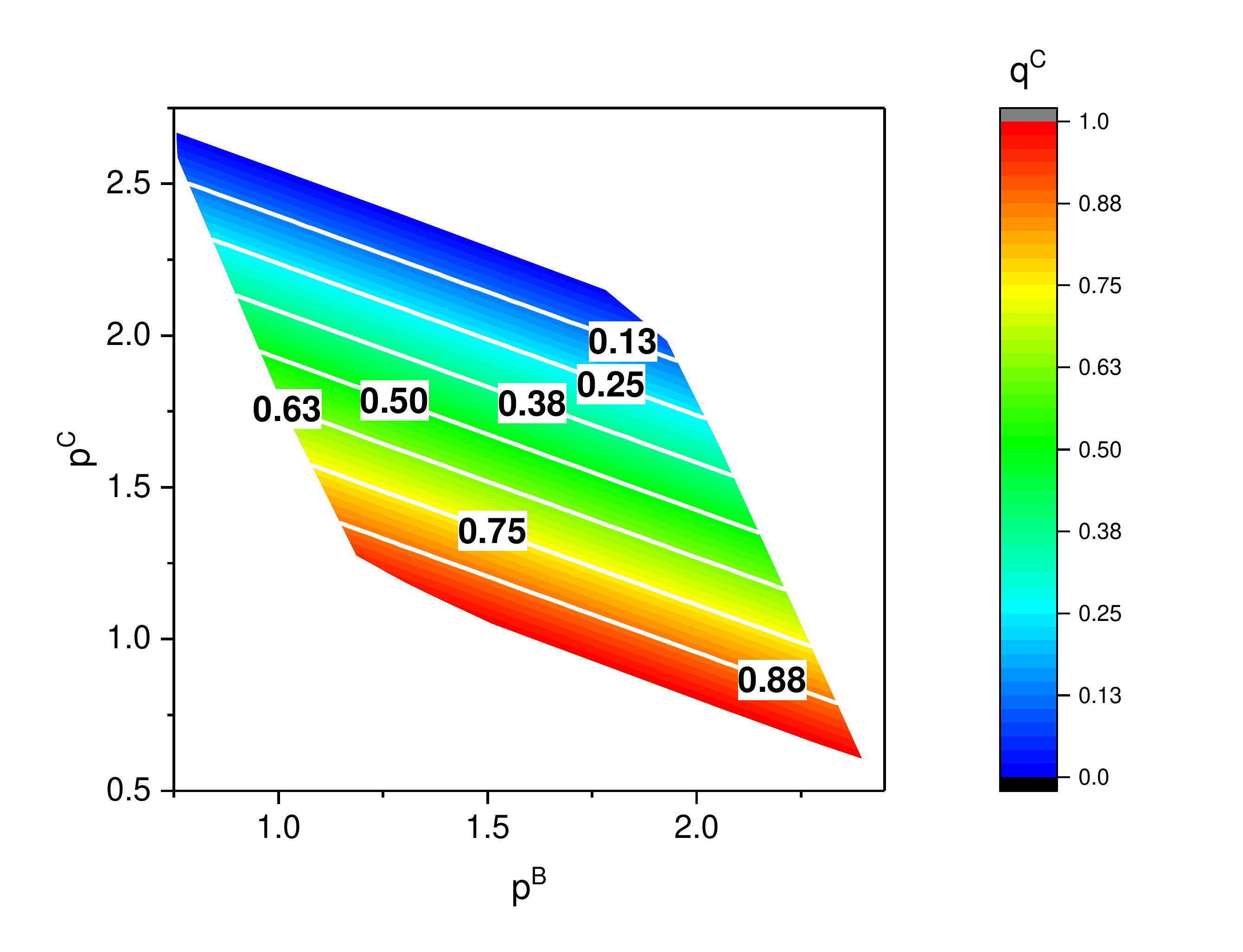}
		\label{fig:qC_pBpC}%
	}%
	\caption{The change of participation as a function of $(p^B,p^C)$  }
	\label{fig:qBqC_pBpC}
\end{figure}

According to equation \eqref{eq:NE_mono_demand}, given the parameters, demand can be written as a linear function of prices. An example of the demand-price relation is shown in Figure \ref{fig:qBqC_pBpC} when using the baseline parameters. Because of the cross-side positive network effects, the choices of Starbucks worksites affect the choices of commuters (vice versa). Therefore, if the CSP increases the price of commuters ($p^C$), less commuters will choose the CSP, the participation of Starbucks worksites will decrease as well. Under the current parameter setting, the price of one side has dominant impact on the participation of the same side, and has minor impact on the participation of the other side. For example, the participation of commuters ($q^C$) decreases when the CSP increases the price for the commuters or the Starbucks worksites, which however decreases faster with the price of commuters ($p^C$).
%More details are unveiled in the following numerical examples.

\subsubsection{Cross-side positive network effects}
\begin{figure}[H]
	\center
	\includegraphics[clip, trim=5cm 2.5cm 4cm 3cm, width=0.7\textwidth]{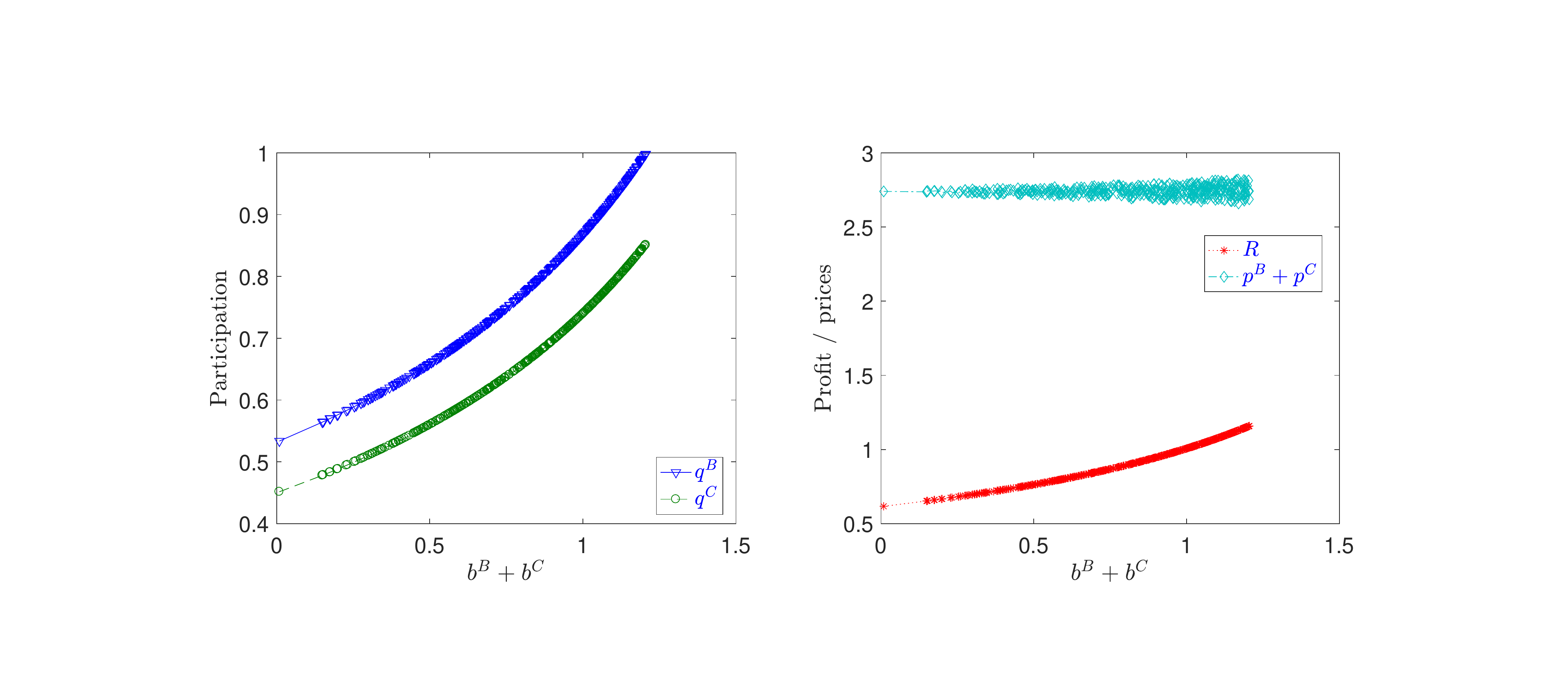}
	\caption{Sensitivity analysis of $b^B+b^C$}
	\label{fig:Mono_bBbC}
\end{figure}

\begin{figure}[H]
	\center
	\includegraphics[clip, trim=5cm 2.5cm 4cm 3cm, width=0.7\textwidth]{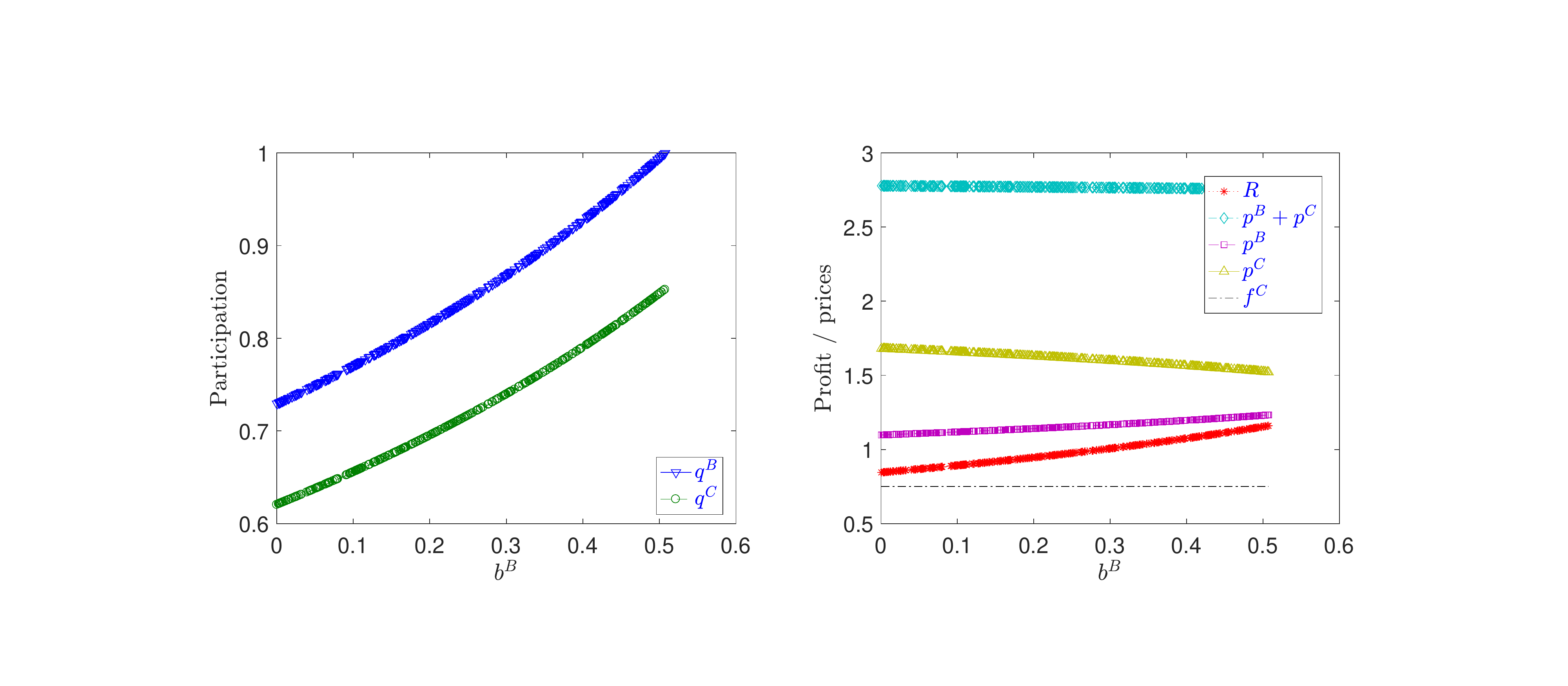}
	\caption{Sensitivity analysis of $b^B$}
	\label{fig:Mono_bB}
\end{figure}

\begin{figure}[H]
	\center
	\includegraphics[clip, trim=5cm 2.5cm 4cm 3cm, width=0.7\textwidth]{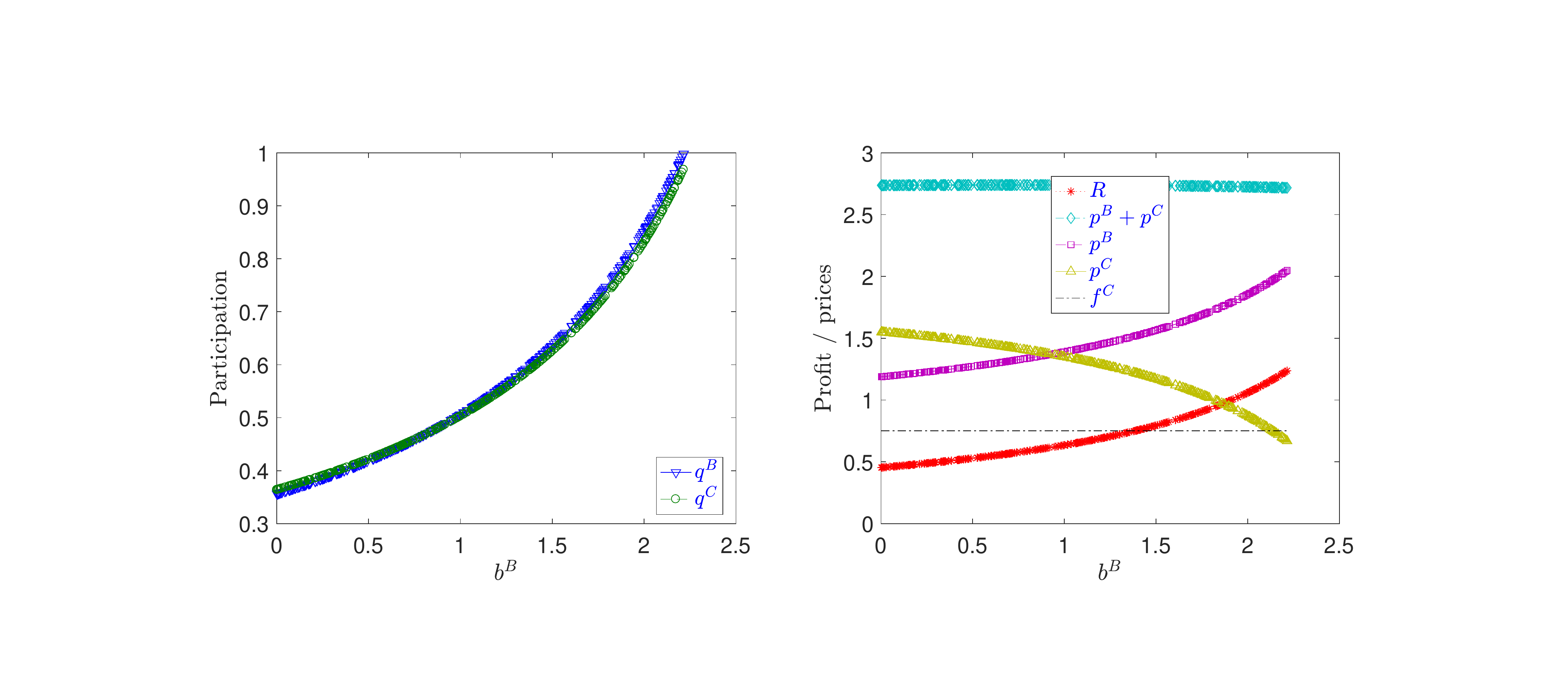}
	\caption{Sensitivity analysis of $b^B$ (when $t^B$ and $t^C$ are adjusted)}
	\label{fig:Mono_bB_LossLeader}
\end{figure}

In order to explore further the cross-side positive effects, we test $b^B$ and $b^C$ while fixing other parameters to the baseline values. Results in Figure \ref{fig:Mono_bBbC} show that when the overall cross-side effect ($b^B+b^C$) increases, the participation ($q^B$ and $q^C$) from both sides increases, the aggregated price ($p^B+p^C$) keeps steady, and the CSP's profit ($R$) increases. Notice that the aggregated price does not change, implying that CSP profit increases as a result of the increased participation from both sides. This tells us that the price allocation, not only the aggregated price, is effective to change the participation and profit of the CSP. This indicates the two-sidedness of the CSP. Without raising the aggregated price, the CSP may make use of the cross-side effects to attract more end-users (both Starbucks stores and employees), thus increasing its profit. The CSP can achieve higher profit when the two sides value the choices of each other (larger cross-side effects). It is common for big companies like Starbucks to provide commuting subsidies to improve employees' satisfaction at work. With CSP, this means that the Starbucks worksites are willing to pay subscription fee to the CSP. In such case, the CSP can charge the worksites with higher prices and take the commuters side as a loss leader so that more worksites and commutes prefer the CSP over other options.
%Based on the relative value of the cross-side effects experienced by the Starbucks worksites and commuters, the CSP can optimize the price allocation between the two side and increase the subscription
Since the equilibrium price structure is affected by both $b^B$ and $b^C$, we test $b^B$ unilaterally in order to capture the change of price structure. In this test, we use the baseline parameters except for $b^B$. Results are shown in Figure \ref{fig:Mono_bB}. When worksites value the number of commuters more ($b^B$ increases), the CSP can make more profit by reducing the price of commuters (thus attracting more participation from both sides) and recouping profit from worksites. By setting lower prices to commuters, more commuters will be willing to join the CSP. Worksites highly value the number of commuters, and as a result they will join CSP even if their price is high.
In Figure \ref{fig:Mono_bB}, increasing cross-side positive effect of worksites reduces the price of commuters. In the monopoly model, all parameters follow the conditions described in \textbf{(A1)} and \textbf{(A2)}. So under baseline parameter setting, $b^B$ cannot exceeds 0.8. Because of these constraints on the parameters, the CSP starts to set lower prices to commuters but not to the extent of subsidizing. With an attempt to show the ``loss leader", we change $t^B,t^C$ to $t^B = 2, t^C=2.2$, while keeping the other parameters the same as the baseline values. We then unilaterally test $b^B$. Results show that CSP will subsidize commuters when $b^B$ keeps increasing (Figure \ref{fig:Mono_bB_LossLeader}). When $b^B$ exceeds 2.2, the price of commuters $p^C$ falls below cost $f^C=0.75$, in which case the commuter side is the ``loss leader". At the same time, the CSP recoups profit by charging worksites much higher prices. Comparing Figure \ref{fig:Mono_bB} and Figure \ref{fig:Mono_bB_LossLeader}, we can see that when worksites highly value the number of commuters, CSP has the intention to reduce the price of commuters, whereas the actual ``subsidizing'' level will depend on the parameter setting of the model.

\subsubsection{Same-side ``congestion'' effects}
\begin{figure}[H]
	\center
	\subfloat[$q^B$]{%
		\includegraphics[clip, trim=1.1cm 1cm 2cm 1cm, width=0.4\textwidth]{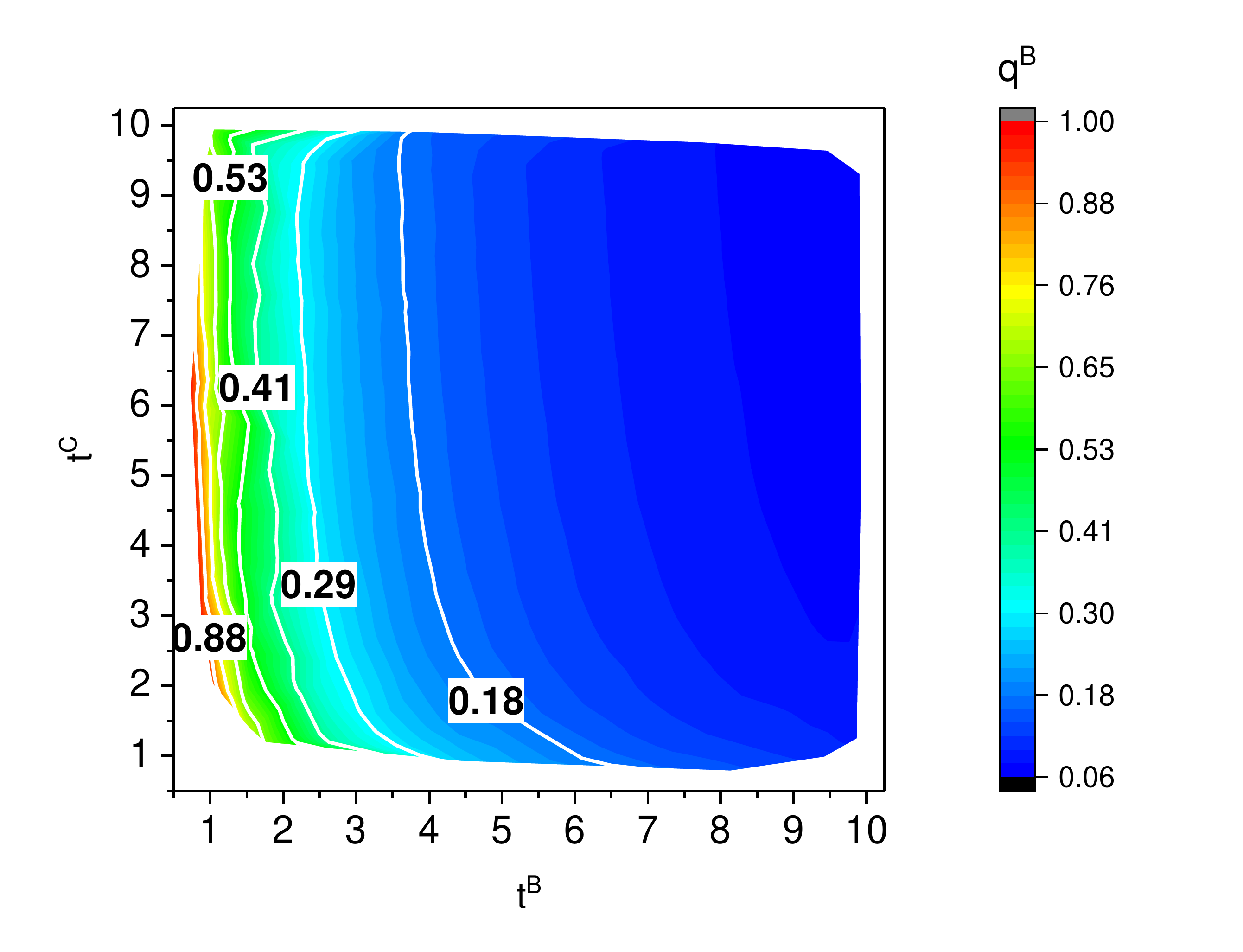}
		\label{fig:qB_tBtC}%
	}%
	\hspace{0.5cm}%\hfill%
	\subfloat[$q^C$]{%
		\includegraphics[clip, trim=1.1cm 1cm 2cm 1cm, width=0.4\textwidth]{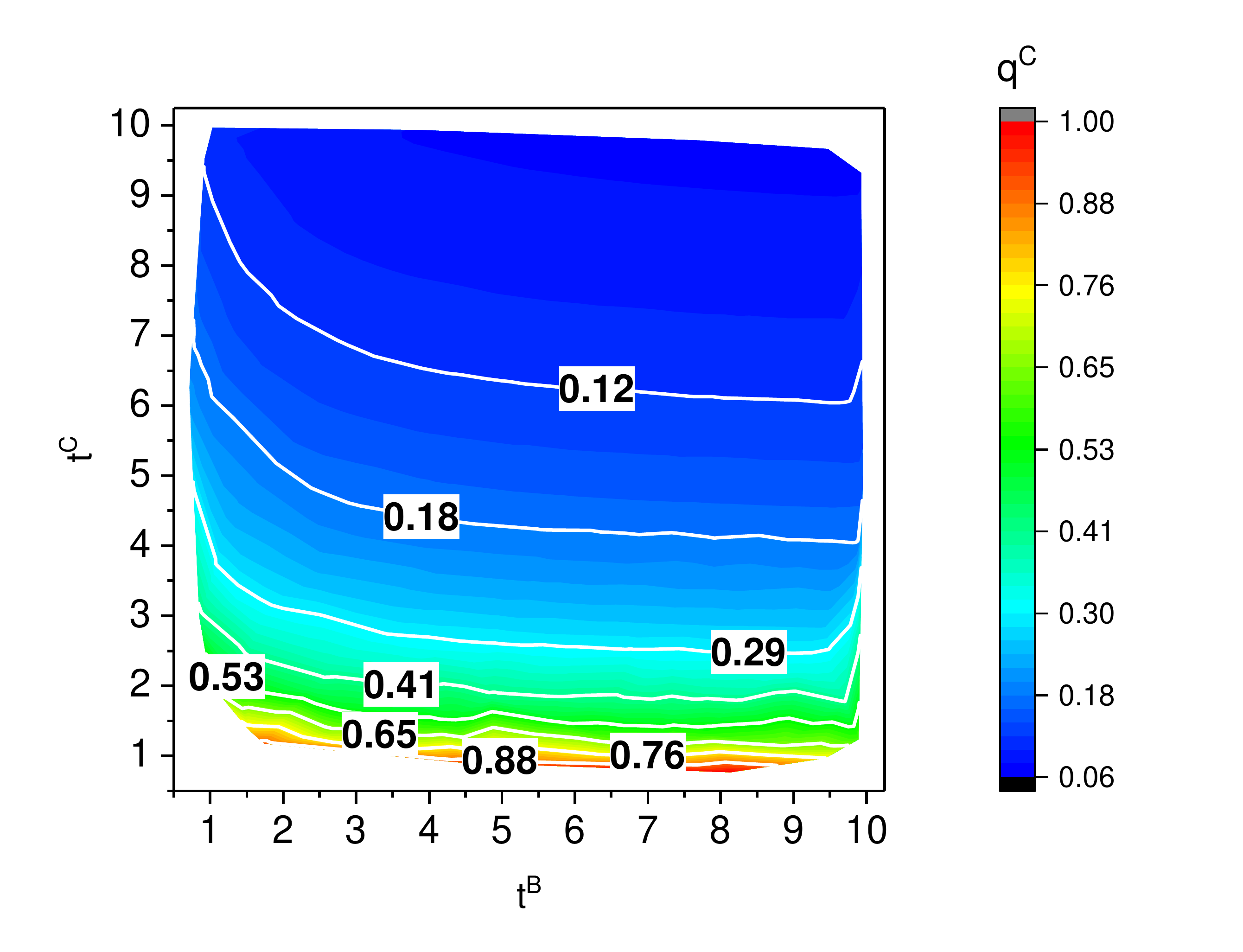}
		\label{fig:qC_tBtC}%
	}%
	\hfill
	\subfloat[$p^B$]{%
		\includegraphics[clip, trim=1.1cm 1cm 2cm 1cm, width=0.4\textwidth]{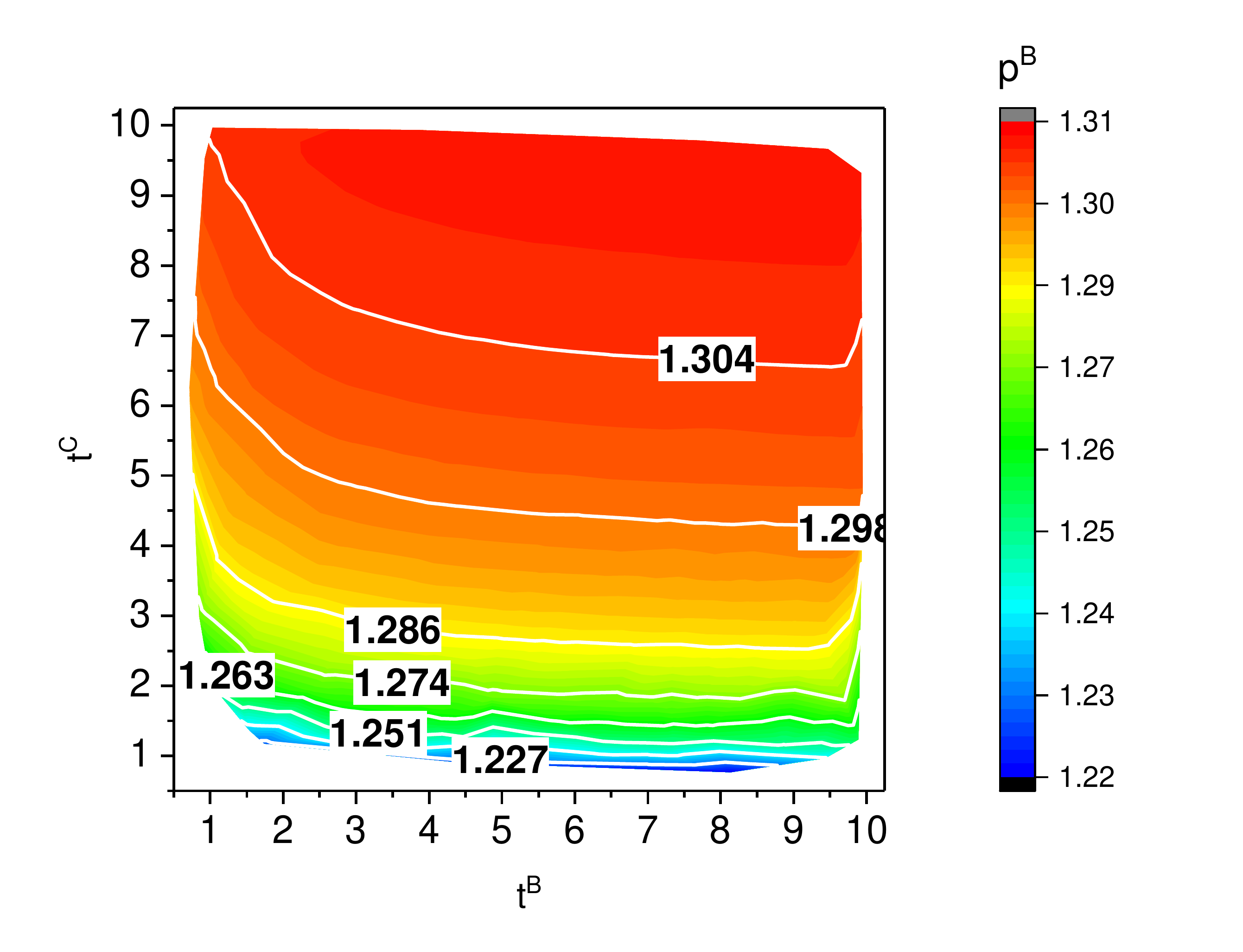}
		\label{fig:pB_tBtC}%
	}%
	\hspace{0.5cm}%\hfill
	\subfloat[$p^C$]{%
		\includegraphics[clip, trim=1.1cm 1cm 2cm 1cm, width=0.4\textwidth]{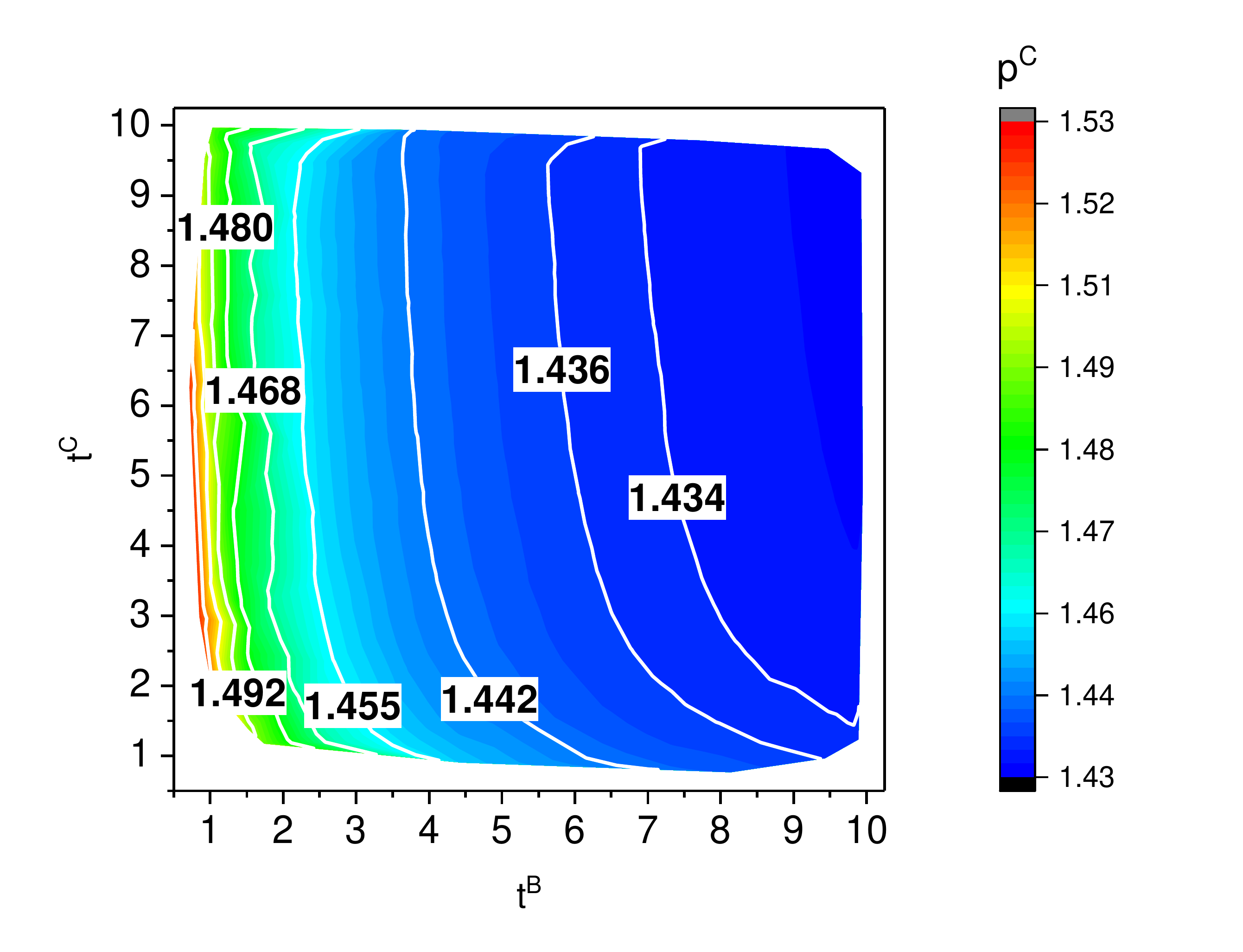}
		\label{fig:pC_tBtC}%
	}%
	\hfill
	\subfloat[$R$]{%
		\includegraphics[clip, trim=1cm 1cm 2cm 1cm, width=0.4\textwidth]{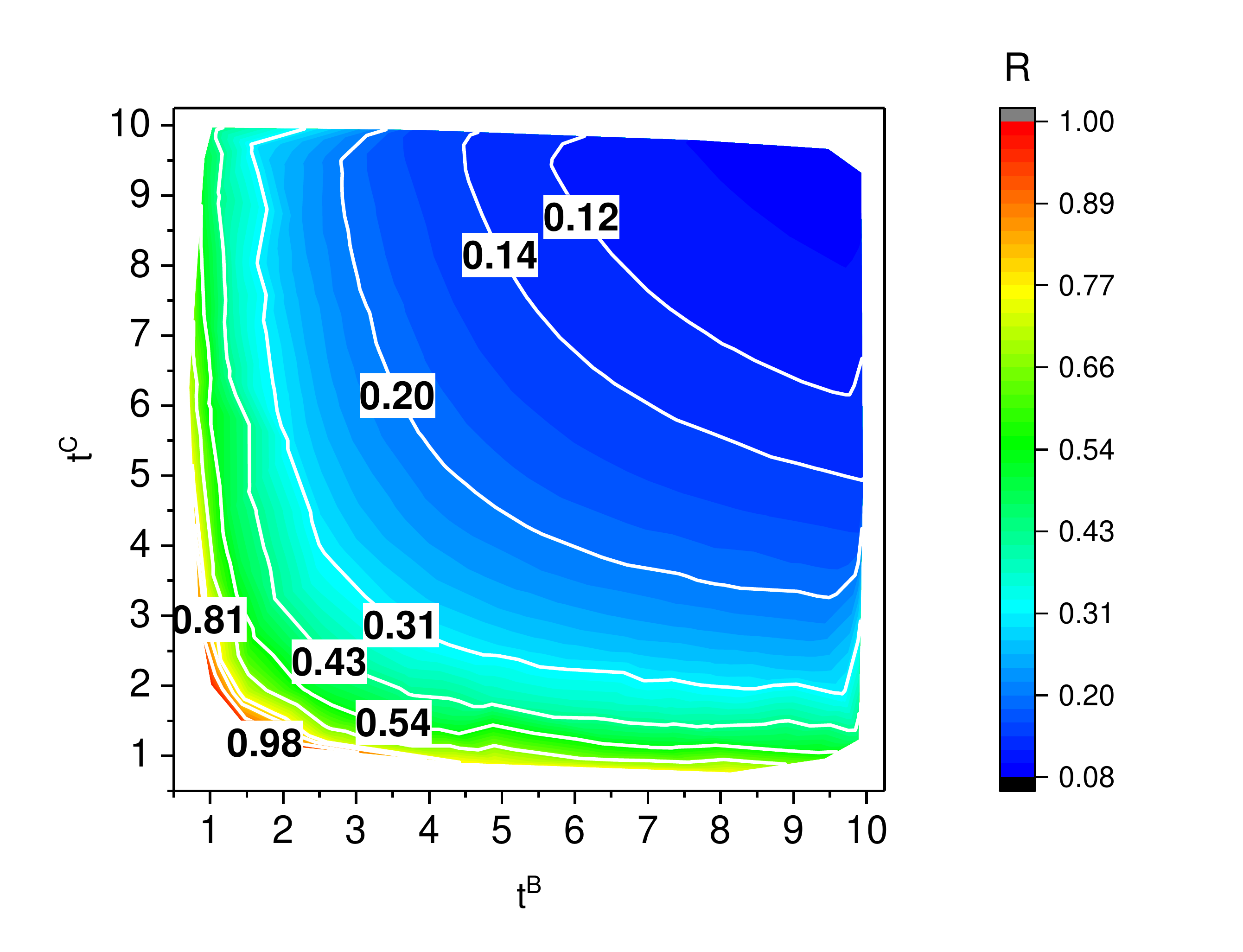}
		\label{fig:Profit_tBtC}
	}%
	\caption{Sensitivity analysis of $t^B,t^C$ }
	\label{fig:results_tBtC}
\end{figure}

We use the baseline parameters and only change the value of $t^C$ and $t^B$ to investigate the same-side ``congestion'' effects. The resulting patterns of participation, prices and profits are presented in Figure \ref{fig:results_tBtC}. Intuitively, when $t^B$ becomes larger, a worksite will experience larger disutility if s/he joins CSP given existing worksites on CSP, in which case the participation of worksites is discouraged. The results shown in Figure \ref{fig:qB_tBtC} agree with our intuition. When each worksite is less discouraged by fellow worksites on CSP ($t^B$ is small), the participation of worksites increases. Also notice that, as $t^B$ gets smaller, $t^C$ will have weaker impact on $q^B$. This means that when $t^B$ is small, the number of worksites on the CSP ($q^B$) is mainly affected by the same-side ``congestion" effect of worksites ($t^B$). On the other hand, if the same-side ``congestion'' effect of worksites ($t^B$) is larger, the same-side ``congestion'' effect of the commuter side ($t^C$) will have a greater impact on the participation of worksites ($q^B$). Thus. when $t^B$ is large, the same-side ``congestion" effect of both groups will discourage the participation of worksites. Commuters have similar behaviors, as shown in Figure \ref{fig:qC_tBtC}. Figure \ref{fig:pB_tBtC} and \ref{fig:pC_tBtC} show that the price allocation changes very slowly with the same-side ``congestion'' effects, with a price variation below 0.08. But it is interesting that the price of worksites is generally more sensitive to the same-side ``congestion'' effect of commuters. When $t^B$ becomes larger given small $t^C$, the CSP will lose many worksites and a few commuters, as shown in Figure \ref{fig:qB_tBtC} and \ref{fig:qC_tBtC}. The CSP fails to maintain profits in such case. Therefore, when the same-side ``congestion'' effect gets too large, more agents from both groups will leave the CSP. The case when $t^C$ becomes larger given small $t^B$ also holds. Ultimately, $t^B,t^C$ will affect the CSP profits. Increasing same-side ``congestion'' effects of either side reduces CSP profits mainly because of the loss of participants, which follows from the discussion of participation and prices. The contour lines in Figure \ref{fig:Profit_tBtC} show that for a range of combination of $t^B,t^C$, i.e., the $t^B,t^C$ that fall onto the same contour line, the CSP can manage to maintain the same profit level (e.g., with a profit of 0.54). In practice, when the participation of both groups is large for the CSP, it could happen that the commuting services of the CSP may degrade if the number of commuters exceeds the capacity of the service. In such scenario, the potential CSP commuters or Starbucks worksites will experience higher same-side ``congestion'' effects, which could lead to dramatic decrease of participation. This implies that when there are increasing participation of the CSP, the CSP needs to control the same-side effects in order to maintain the profit and service quality.

%When a commuter is less discouraged by the number of commuters on the CSP ($t^C$ is small), more commuters will join the CSP and this trend is marginally affected by the same-side ``congestion'' effect of worksites ($t^B$). When commuters face increasing pressure from the same side ($t^C$ gets larger), the same-side ``congestion'' effect of worksites ($t^B$) will increase its impact on the participation of commuters, especially when $t^B$ lies in $[1,3]$.

%The above analytical results show the pattern of participation and the price strategies under different network effects. \textbf{[The below Starbuck-specific analysis should be integrated with the previous text where each analysis was done] }Now let's check the number of participants in the Starbucks example. In Figure \ref{fig:Mono_bBbC}, \ref{fig:Mono_bB}, \ref{fig:Mono_bB_LossLeader}, we can see that participation from both sides increases simultaneously with cross-side network effects, thus $q^C/q^B$ keeps steady ($q^C/q^B \approx 1 $) \textbf{[Figure 4 and Figure 5 show $q^B$ and $q^C$ are a bit different]}. The employees at each worksite ({\tiny $\frac{170*q^C}{17*q^B}$} for CSP subscribers, {\tiny $\frac{170*(1-q^C)}{17*(1-q^B)}$} for non-CSP subscribers) only changes slightly. However, top-left and right-bottom part of Figure \ref{fig:qB_tBtC} \ref{fig:qC_tBtC} shows that there are cases when $q^C/q^B$ deviates away from 1 \textbf{[not immediately clear to the reader]}, which is undesirable. In section \ref{sec:competitive_constraints}, we add constraints to the model so that there is a reasonable number of employees at each worksite ($q^C/q^B \approx 1 $).

\section{Duopoly platforms} \label{sec:DuoSing}
\subsection{General Model for duopoly platforms}
In practice, there are often multiple options for commuting, which indicates there should be multiple CSPs in real life commuting services. Thus, a two-sided market with more than one platform is also valuable for understanding the competition among CSPs. Here we investigate the competition in the proposed two-sided market when there are two CSPs. We still consider worksites and commuters as the two sides. The two CSPs are specified as the WF CSP and the NWF CSP, denoted as $i \in \{ W,N \}$. We consider two cases: (i) worksites and commuters single-home (i.e., a worksite or a commuter can only join one of the two CSPs) ; (ii) worksites multi-home (i.e., a worksite can join both CSPs), commuters single-home (i.e., a commuter can only join one of the two CSPs). We apply the model similar as \cite{armstrong2006competition} to the scenario in this paper. We assume that an agent from group $k$ has an incurred utility $U_0^k$ by joining the market. The cost for platform $i$ to serve a group $k$ agent is $f_i^k$. We consider that each agent from group $C$ (or $B$) values the number of agents from the other group $B$ (or $C$) with whom s/he can interact with. In particular, an agent from group $C$ obtains benefit $\beta_i (q_i^B + Q^B) $ by joining CSP $i$ with $\beta_i$ the cross-side benefit of commuters joining CSP $i$, as s/he will have the potential to interact with $ q_i^B + Q^B$ agents from group $B$. $\beta_i$ denotes the cross-side benefit rate of commuters on CSP $i$. On the other hand, an agent from group $B$ also values the number of agents from group $C$. A group $B$ agent obtains cross-side benefit $\alpha_i ( q_i^C + Q^C) $ by joining CSP $i$. $\alpha_i$ can be interpreted as the cross-side benefit rate of worksites on CSP $i$. CSP $i$ charges a non-negative subscription price $p_i^k$ to agents from group $k$. We assume that when an agent chooses a platform, s/he will be worse off when more other agents from the same group join the same platform, referred to as the same-side negative network effect (same-side “congestion” effect) as shown by the Hotelling model in Section \ref{sec:HotelModel}. The analysis of the same-side effect for the duopoly model is similar to that for the monopoly model, which is omitted here. The utility of a worksite located at $x^B \in [0,1]$, when s/he joins WF CSP or NWF CSP, is given by:
\hfill\break
\begin{linenomath}
	\begin{equation}
	U_W^B = \underbrace{U_0^B}_{\text{fixed benefit}}  - \underbrace{p_W^B}_{\text{subscription price}} - \underbrace{t^B x^B}_{\begin{subarray}{l} \vspace*{2pt} \text{same-side/inconvenience} \\ \text{ costs}\end{subarray}} + \underbrace{\alpha_W (q_W^C + Q^C) }_{\text{cross-side benefits}} \label{EQutilityB}
	\end{equation}
	
	\begin{equation}
	U_N^B = \underbrace{U_0^B}_{\text{fixed benefit}}  - \underbrace{p_N^B}_{\text{subscription price}} - \underbrace{t^B (1-x^B)}_{\begin{subarray}{l} \vspace*{2pt} \text{same-side/inconvenience} \\ \text{ costs}\end{subarray}} + \underbrace{\alpha_N (q_N^C + Q^C)}_{\text{cross-side benefits}} \label{EQutilityB_N}
	\end{equation}
\end{linenomath}
The utility of a commuter located at $x^C \in [0,1]$, when s/he joins WF CSP or NWF CSP, is given by:
\hfill\break
\begin{linenomath}
	\begin{equation}
	U_W^C = \underbrace{U_0^C}_{\text{fixed benefit}}  - \underbrace{p_W^C}_{\text{subscription price}} - \underbrace{t^C x^C}_{\begin{subarray}{l} \vspace*{2pt} \text{same-side/inconvenience} \\ \text{ costs}\end{subarray}} +  \underbrace{\beta_W (q_W^B + Q^B) }_{\text{cross-side benefits}}
	\end{equation}
	
	\begin{equation}
	U_N^C = \underbrace{U_0^C}_{\text{fixed benefit}}  - \underbrace{p_N^C}_{\text{subscription price}} - \underbrace{t^C (1-x^C)}_{\begin{subarray}{l} \vspace*{2pt} \text{same-side/inconvenience} \\ \text{ costs}\end{subarray}} + \underbrace{\beta_N (q_N^B + Q^B) }_{\text{cross-side benefits}} \label{EQutilityC}
	\end{equation}
\end{linenomath}
When a worksite multi-homes, s/he obtains utility:
\hfill\break
\begin{linenomath}
	\begin{equation}
	U_{NW}^B = \underbrace{U_0^B}_{\text{fixed benefit}}  - \underbrace{(p_W^B+p_N^B)}_{\text{subscription price}} - \underbrace{t^B}_{\begin{subarray}{l} \vspace*{2pt} \text{same-side/inconvenience} \\ \text{ costs}\end{subarray}} + \underbrace{(\alpha_W q_W^C + \alpha_N q_N^C)}_{\text{cross-side benefits}}  \label{Eq_utility_multihome}
	\end{equation}
\end{linenomath}
The profit of platform $i$ is:
\hfill\break
\begin{linenomath}
	\begin{equation}
	R_i = \underbrace{(q_i^B + Q^B)(p_i^B - f_i^B)}_{\text
		{profit collected from worksites}} + \underbrace{(q_i^C + Q^C) (p_i^C - f_i^C)}_{\text
		{profit collected from commuters}} \quad \forall i \in \{ W,N \} \label{EQprofits}
	\end{equation}
\end{linenomath}
Based on this general Model, we can analyze duopoly platforms when both sides singlehome in section \ref{sec:BC_single-home}. Section \ref{sec:competitive_constraints} adds demand constraints to the model in section \ref{sec:BC_single-home}. The multi-home case is presented in Appendix \ref{Append_oneside_multihome_platform}.

\subsection{Duopoly platforms when worksites and commuters single-home} \label{sec:BC_single-home}
The following conditions ensure that agents from both groups are single-homing and the equilibrium price is feasible, as formally shown in Lemma \ref{lem:single-home} and Proposition \ref{prop:prices_competitive}:

\gap

\noindent\textbf{(B1)} $U_0^B$ and $U_0^C$ are sufficiently high such that all agents wish to subscribe to at least one CSP;\\
\noindent\textbf{(B2)} $t^B > \alpha_W q_W^C + \alpha_N q_N^C $ and $t^C >  \beta_W q_W^B + \beta_N q_N^B    $ : ensures that the incremental utility from single-home to multi-home is always negative, so that no agent multi-homes at any non-negative prices set by the two CSPs; \\
\noindent\textbf{(B3)} $ 4 t^B t^C > (\alpha^+ + \beta^+)^2 $: ensures that the profits of the CSPs are positive.

\begin{lemma} \label{lem:single-home}
	 Under condition \textbf{(B2)}, no agent multi-homes at any non-negative price set by the two CSPs.
\end{lemma}

\begin{myproof}
	Single-home case can be further divided into sub-cases: (i) all agents from group $k$ prefer one CSP to another, $q_W^k q_N^k=0, q_W^k + q_N^k=1$; (ii) agents from group $k$ choose either CSPs, $q_W^k q_N^k>0, q_W^k + q_N^k=1$. To guarantee all agents single-home, we need to show that group $B$ agents with the lowest utility does not want to multi-home. (The proof of group $C$ single-homing can be derived similarly)
	
	\gap
	
	Case (i): all group $B$ agents prefer the WF CSP to the NWF CSP, so that $U_W^B > U_N^B$. Here, the agents with the lowest utility is located at $x^k=1$, and they are most likely to multi-home. Evaluated at $x^B=1$, the incremental utility of multi-homing with respect to single-homing is $U_{WN}^B - U_W^B\rvert_{x^B=1} = U_0^B - (p_W^B+p_W^C) -t^B + (\alpha_W q_W^C + \alpha_N q_N^C) - \big[U_0^B - p_W^B - t^B + \alpha_W q_W^C \big] = -p_N^B + \alpha_N q_N^C$. We assume $U_W^k > U_N^k$, thus $U_W^B \rvert_{x^B=1} > U_N^B \rvert_{x^B=1}$ holds, from which yields $-p_W^B - t^B + \alpha_W q_W^C > -p_N^B + \alpha_N q_N^C$. From condition (A2), we know that $t^B > \max \{\alpha_W q_W^C , \alpha_N q_N^C \}$, so $-p_W^B - t^B + \alpha_W q_W^C$ is negative. Hence the incremental utility of multi-homing for group $B$ is also negative. Similarly, we can prove that the incremental utility of multi-homing for group $C$ is also negative under condition (A2).
	
	\gap
	
	Case (ii): agents from group $k$ are most likely to multi-home when they are indifferent between the two CSPs, namely, $U_W^B\rvert_{x^B=x^*} = U_N^B\rvert_{x^B=x^*}$. This is the lowest utility an agent experiences by choosing single-home. Evaluate at location $x^*$, the incremental utility from multi-homing with respect to single-homing is $U_{WN}^B - \frac{1}{2}(U_W^B\rvert_{x^B=x^*} + U_W^B\rvert_{x^B=x^*})  = U_0^B - (p_W^B+p_W^C) -t^B + (\alpha_W q_W^C + \alpha_N q_N^C) - \frac{1}{2}\big[2U_0^B - p_W^B - p_N^B - t^B + \alpha_W q_W^C + \alpha_N q_N^C \big] = \frac{1}{2} \big[ - p_W^B - p_N^B - t^B + \alpha_W q_W^C + \alpha_N q_N^C \big]$, given condition (A2), $t^B > \alpha_W q_W^C + \alpha_N q_N^C$, so the incremental utility is negative.
\end{myproof}

%[\textbf{JB: this proposition should have two parts: (a) under (B1)-(B3), an equilibrium solution exists; (b) if we further assume the prices are the same, then we can derive simple forms of the results}. \textbf{The current description is a bit confusing}]
\begin{prop} \label{prop:prices_competitive}
	(i) Under condition \text{(B1)-(B3)}, an equilibrium exists and all agents single-home. In other words, \text{(B1)-(B3)} are the sufficient conditions for equilibrium existence.  \\
	(ii) For simplicity, we only show the formulation of equilibrium when the two CSPs have the same pricing strategy, $p_W^B=p_N^B$ and $p_W^C=p_N^C$. Half of the agents from each group will join each platform, i.e., $q_W^B=q_W^C=q_N^B=q_N^C=0.5$. If $f_W^B + t^B  + \psi_W^B > \frac{\beta^+}{2}$ and $f_W^C + t^C  + \psi_W^C > \frac{\alpha^+}{2}$, the equilibrium prices are:
	\begin{linenomath}
		\begin{equation}
		p_W^B = f_W^B + t^B - \frac{\beta^+}{2} + \psi_W^B \ge 0
		\quad \quad
		p_W^C =   f_W^C + t^C- \frac{\alpha^+}{2}  + \psi_W^C \ge 0 \label{EQunconstrain_p}
		\end{equation}
	\end{linenomath}
	\noindent	where
	\begin{linenomath}
		\begin{equation}
		\psi_W^B = \frac{ 2(\beta^+ - \alpha^+)\beta^- t^B + ({\beta^+}^2 - 4t^B t^C)\alpha^-}{8t^B t^C - 2 \alpha^+ \beta^+}
		\quad \quad
		\psi_W^C = \frac{2(\alpha^+ - \beta^+)\alpha^- t^C + ({\alpha^+}^2 - 4t^Bt^C)\beta^-}{8t^B t^C - 2 \alpha^+ \beta^+}
		%\frac{\partial p_W^B}{\partial \alpha^+} &= \frac{({\beta^+}^2--4t^B t^C)(\alpha^- \beta^+ + 2 \beta^- t^B)}{2(4t^B t^C - \alpha^+ \beta^+)^2} < 0 \nonumber \\
		\end{equation}
	\end{linenomath}

	\noindent Each CSP makes profit:
	\begin{linenomath}
		\begin{equation}
		R_i = \frac{t^B + t^C - \frac{\beta^+ + \alpha^+}{2} + \psi_W^B + \psi_W^C}{2} > 0  \quad \forall i \in \{ W,N \}  \label{EQunconstrain_RW}
		\end{equation}
	\end{linenomath}
\end{prop}

\begin{myproof}
	First, under condition (\textbf{B2}), all agents single home as shown in Lemma \ref{lem:single-home}. We know that $q_N^k + q_W^k =1$. According to Nash equilibrium, an agent experiences the same utility from either CSP, i.e., $U_W^B=U_N^B$ and $U_W^C=U_N^C$. From equation \eqref{EQutilityB}-\eqref{EQutilityC} we get the following relations:	
	\begin{linenomath}
		\begin{equation}
		\noindent U_0^B  - p_W^B - t^Bx^B + \alpha_W (q_W^C + Q^C) = U_0^B  - p_N^B - t^B  (1-x^B)  + \alpha_N (q_N^C + Q^C) \label{eq:DUO_NE_B}
		\end{equation}
		\begin{equation}
		U_0^C  - p_W^C - t^C x^C + \beta_W (q_W^B + Q^B)  = U_0^C  - p_N^C - t^C (1-x^C)  + \beta_N (q_N^B + Q^B) \label{eq:DUO_NE_C}
		\end{equation}
	\end{linenomath}
	Set $x^k = q_W^k$. We obtain the participation of each group on a CSP. Set $\alpha^+ = \alpha_N + \alpha_W $, $\beta^+ = \beta_N + \beta_W$, $\alpha^- = \alpha_N - \alpha_W $, $\beta^- = \beta_N - \beta_W$, the number of group $k$ agents joining the WF CSP can be written as:
	\begin{align}
		q_W^B = \frac{1}{2} + \frac{\alpha^+( p_N^C - p_W^C) + 2t^C (p_N^B - p_W^B)}{  4 t^B t^C - \alpha^+\beta^+}  - \frac{ t^C \alpha^- +   \frac{\alpha^+  \beta^-}{2}   }{4 t^B t^C - \alpha^+\beta^+}   \label{eq:Duo_qWB_p} \\
		q_W^C =    \frac{1}{2} + \frac{\beta^+( p_N^B - p_W^B) + 2t^B ( p_N^C - p_W^C) }{ 4 t^B t^C - \alpha^+\beta^+}  - \frac{ t^B \beta^- +   \frac{\alpha^-  \beta^+}{2}   }{4 t^B t^C - \alpha^+\beta^+}   \label{eq:Duo_qWC_p}
	\end{align}
	Condition \textbf{(B3)} ensures that $4 t^B t^C - \alpha^+\beta^+ > 0$ and the profit function is concave with respect to prices. Substitute the demand functions, equation \eqref{eq:Duo_qWB_p} and  \eqref{eq:Duo_qWC_p}, into the profit function \eqref{EQprofits}. Assume $p_W^B=p_N^B$ and $p_W^C=p_N^C$. Based on the first order condition of profit function \eqref{EQprofits} over prices, we can obtain the optimal prices.
	\begin{align}
	\frac{\partial R_i}{\partial p_i^k} =0 \qquad \forall i \in \{ W,N\}, k \in \{B,C\} \\
	s.t. \qquad  p_W^B=p_N^B,  p_W^C=p_N^C \nonumber
	\end{align}
\end{myproof}
Note that we assume $p_W^B=p_N^B$ and $p_W^C=p_N^C$ in Proposition \ref{prop:prices_competitive} for more concise expressions of equilibrium prices. We do not have such assumptions when generating numerical results in section \ref{sec:numerical_competitive} and \ref{sec:numerical_competitive_constraints}.

%[\textbf{JB: for Figures 9-11 and maybe 13-15 too, if we divide each figure into 4 quadrants, what can we say about each quadrant?}]
%w
\subsubsection{Numerical experiments for duopoly model (single-homing)} \label{sec:numerical_competitive}
 To avoid redundancy, we present the results for the duopoly model selectively. The same-side effects of the duopoly model is similar as the monopoly model, so the numerical experiments for the same-side effects are omitted here. Also, the analysis for the cross-side effects of the two groups are similar, so we only analyze the cross-side effects of the worksites. We use the same Starbucks example in this subsection. The baseline parameters are set as $\alpha_N=0.7$, $\alpha_W = 0.6$, $\beta_N=0.5$, $\beta_W = 0.8$, $t^B = 1.1$, $t^C=1.2$, $f_W^B= 0.7$, $f_N^B=0.73$, $f_W^C=0.73$, $f_N^C=0.75$, which satisfy condition \textbf{(B2)} and \textbf{(B3)}. We further assume \textbf{(B1)}, which implies that the conditions of Lemma \ref{lem:single-home} and Proposition \ref{prop:prices_competitive} hold, i.e., all agents single home and an equilibrium solution holds for the duopoly model. Under this parameter setting, worksites experience more cross-side benefits when joining the NWF CSP than the WF CSP ($\alpha_N > \alpha_W$). In practice, this may be due to the reduced cost when a worksite sets fixed working hours for their employees. Commuters obtain more cross-side benefits from the number of worksites on the WF platform ($\beta_W > \beta_N$). This is because commuters tend to value more flexible working hours. Commuters dislike the participation of other commuters more than that of worksites ($t^C > t^B$). For CSPs, the service cost of each commuter ($f_i^C$) is higher than the per-worksite cost ($f_i^B$). The WF CSP spends less than the NWF CSP to serve customers from the same group ($f_W^k < f_N^k$). For the Starbucks example, we assume that employees who choose CSP $i$ are evenly split among the worksites that also choose CSP $i$ (assumption (d)). For the WF CSP subscribers, there are {\tiny $\frac{170*q_W^C}{17*q_W^B}=\frac{10*q_W^C}{q_W^B}$} employees at each Starbucks worksite. For the NWF CSP subscribers, there are {\tiny $\frac{10*q_N^C}{q_N^B} = \frac{10*(1-q_W^C)}{(1-q_W^B)}$} employees at each Starbucks worksite. For the duopoly model, the number of employees at a worksite is also affected by the ratio $q_W^C/q_W^B$. We will first show the analytical results when $q_W^C/q_W^B$ is not constrained in this section, and then show the results when $q_W^C/q_W^B$ is constrained (so that $q_W^C \approx q_W^B$) in section \ref{sec:competitive_constraints}.

\gap

\begin{figure}[H]
	\center
	\subfloat[$q_W^B$]{%
		\includegraphics[clip, trim=0.5cm 1cm 2cm 0.8cm, width=0.4\textwidth]{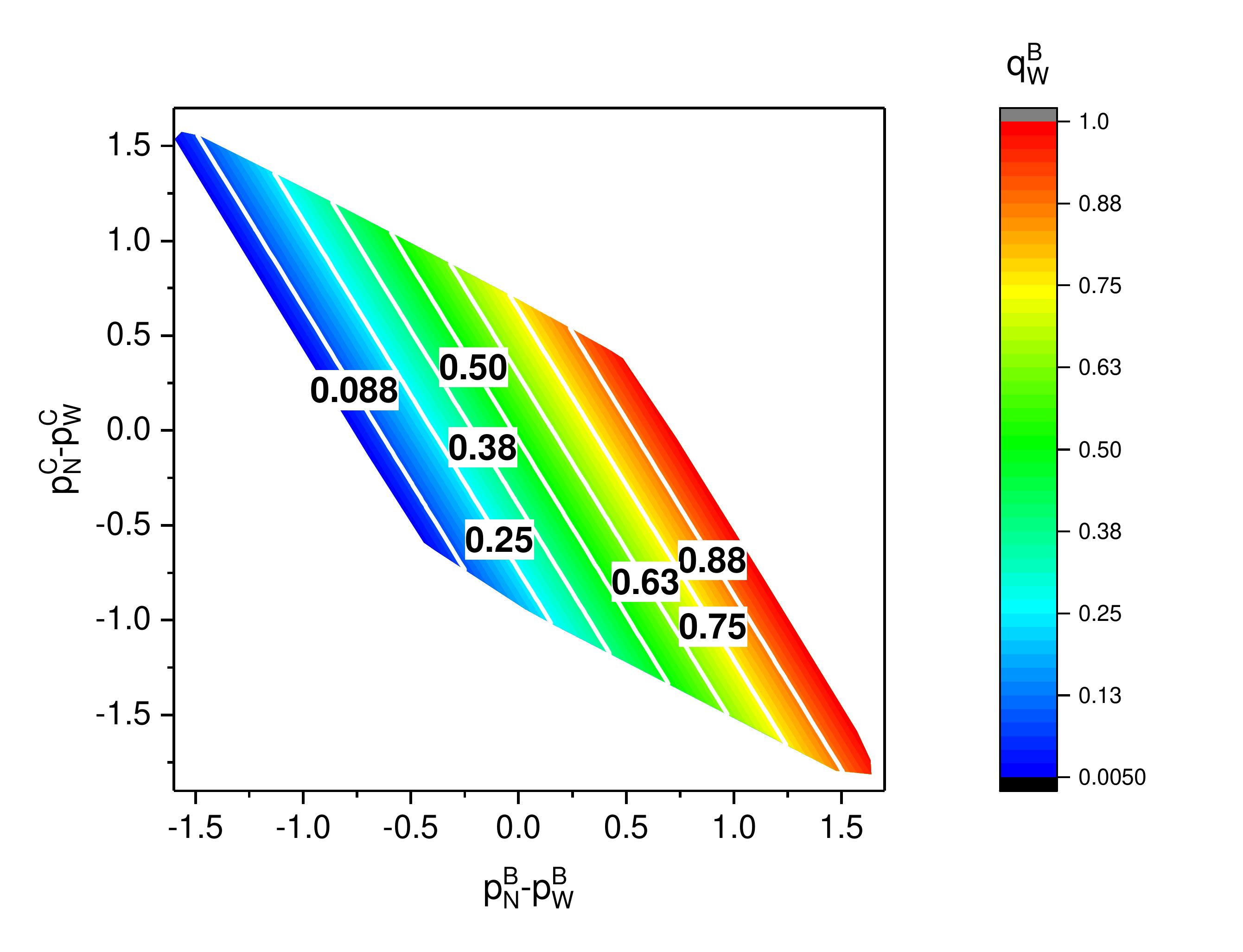}
		\label{fig:qWB_pWBpWCpNBpNC}%
	}%
	\hspace{0.5cm}%\hfill%
	\subfloat[$q_W^C$]{%
		\includegraphics[clip, trim=0.5cm 1cm 2cm 0.8cm, width=0.4\textwidth]{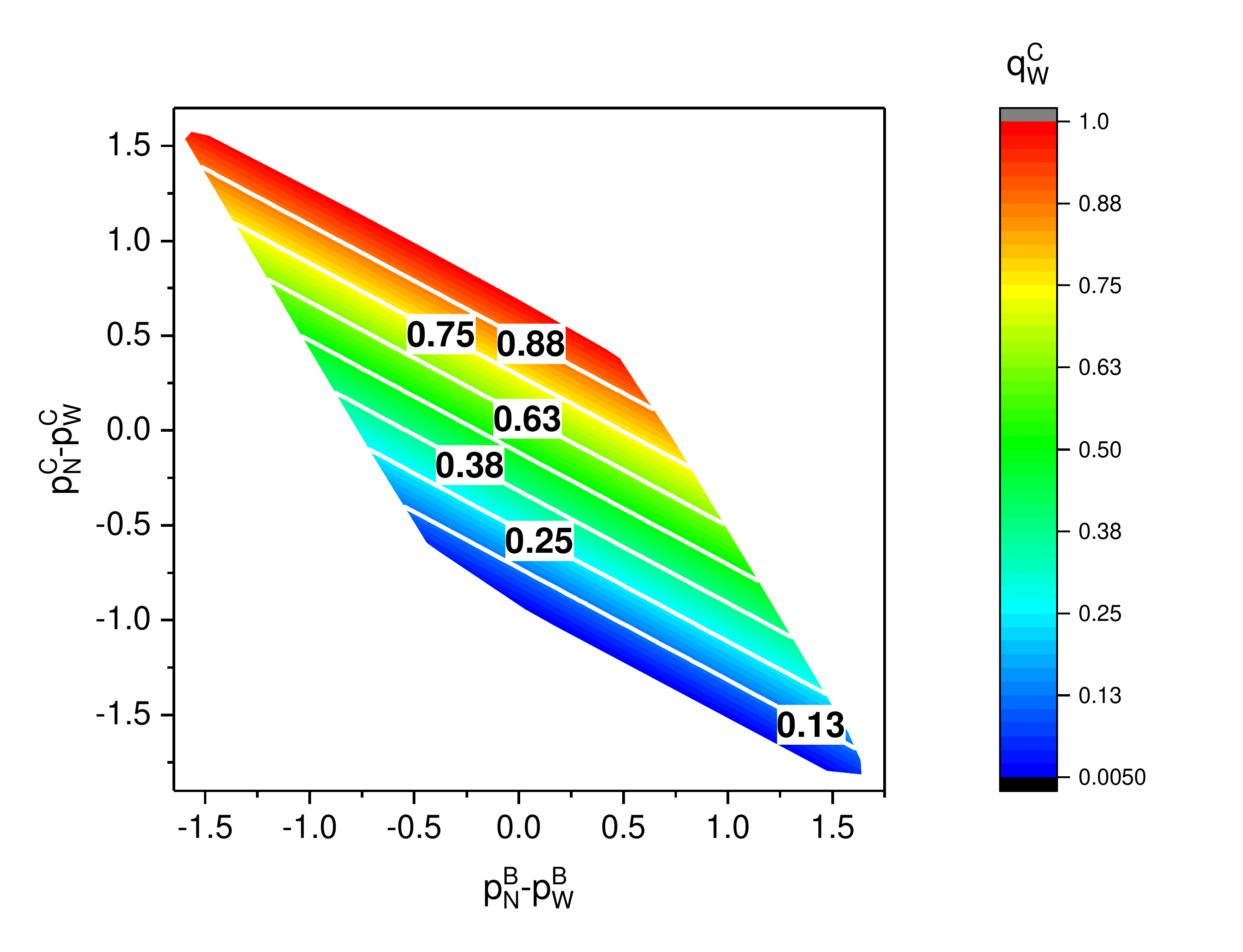}
		\label{fig:qWC_pWBpWCpNBpNC}%
	}%
	\caption{The change of participation as a function of $(p_N^B-p_W^B,p_N^C-p_W^C)$  }
	\label{fig:qWBqWC_pWBpWCpNBpNC}
\end{figure}

According to equation \eqref{eq:Duo_qWB_p} and \eqref{eq:Duo_qWC_p} , participation (demand) can be expressed as a linear function of $(p_N^C - p_W^C)$ and $(p_N^B - p_W^B)$ when we hold the other parameters constant. Using the baseline parameters, we obtain the demand-price relation as shown in Figure \ref{fig:qWBqWC_pWBpWCpNBpNC}. The number of agents from either group on the WF CSP increases when the WF CSP sets lower prices than the NWF CSP. With the current parameters, $p_N^B-p_W^B$ has larger impact than $p_N^C-p_W^C$ on the participation of worksites on the WF CSP ($q_W^B$). Similarly, $p_N^C-p_W^C$ has larger impact than $p_N^B-p_W^B$ on the participation of commuters on the WF CSP ($q_W^C$). This indicates that the participation of group $k$ on CSP $i$ is affected, but not at the same level, by the prices of both groups on the two CSPs. Therefore, CSPs should be more strategic to allocate the prices between the two groups in order to be more competitive in the market.

$\alpha_N$ and $\alpha_W$ are the cross-side benefit rates of worksites on the two CSPs. We define $\alpha^+$ (positive, $\alpha^+ = \alpha_N + \alpha_W$) representing the overall cross-side benefit rate of worksites from the two CSPs, and $\alpha^-$ (can be negative, $\alpha^- = \alpha_N - \alpha_W$) is the difference of cross-side benefit rate of worksites between the NWF CSP and the WF CSP. When $\alpha^- > 0$, worksites value higher of the commuters on the NWF CSP; when $\alpha^- < 0$, worksites value higher of the commuters on the WF CSP. $\alpha^+,\alpha^-$ are linear combinations of $\alpha_N$ and $\alpha_W$. Here we test the sensitivity of $\alpha^+,\alpha^-$ in order to see how the cross-side benefits of the two CSPs affect participation, prices and CSP profits. Notice that the impact of $\beta^+$ and $\beta^-$ can be analyzed in a similar way, which is omitted here. %[\textbf{JB: have we done the analysis of $\beta^+$ and $\beta^-$? What are the main findings? We might need to provide a brief, high-level summary here or in the table later}]

\gap

\begin{figure}[H]
	\center
	\subfloat[$q_W^B$]{%
		\includegraphics[clip, trim=1cm 1cm 2cm 0.8cm, width=0.4\textwidth]{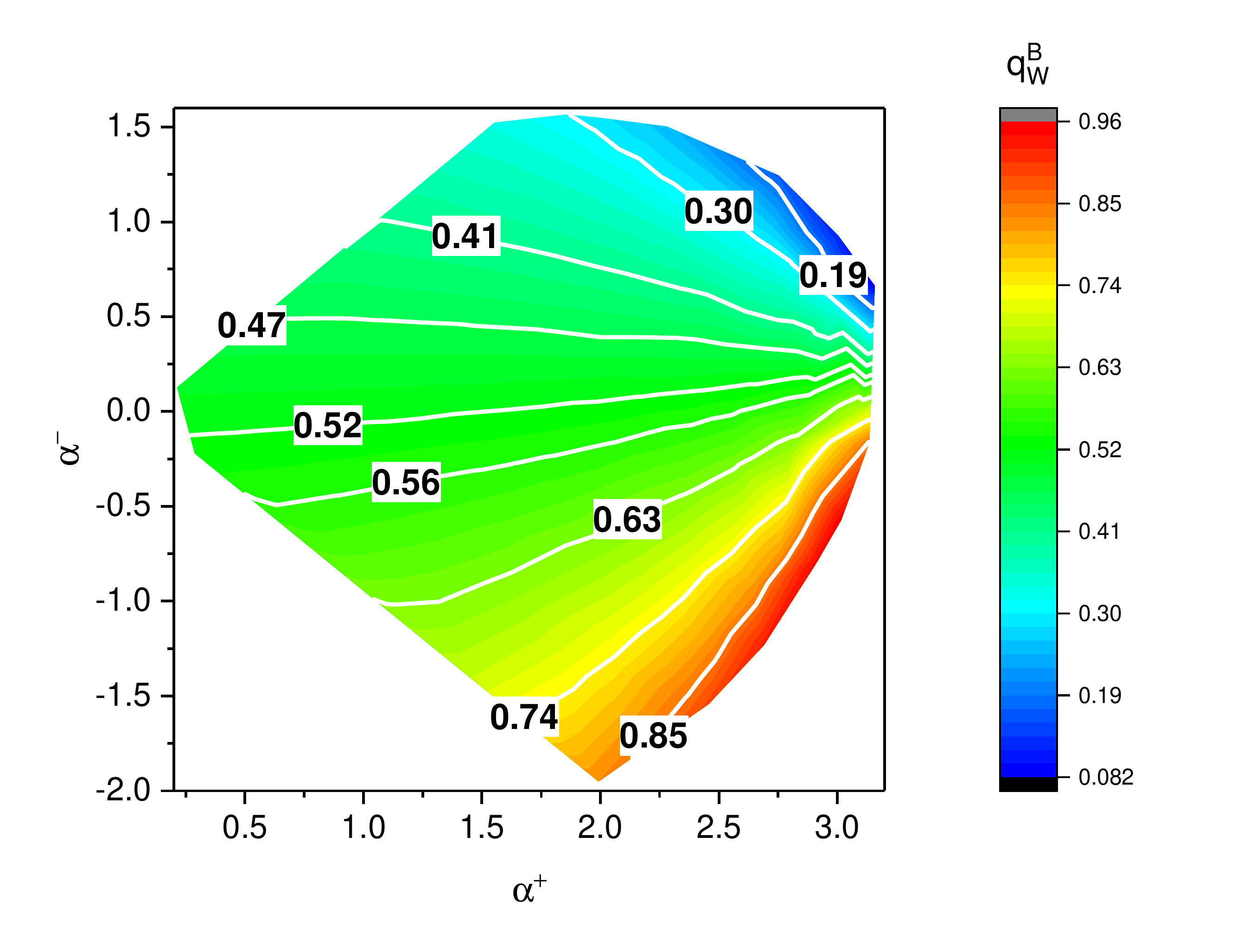}
		\label{fig:qWB_ALPHAalpha}
	}%
	\hspace{0.5cm}%\hfill%
	\subfloat[$q_W^C$]{%
		\includegraphics[clip, trim=1cm 1cm 2cm 0.8cm, width=0.4\textwidth]{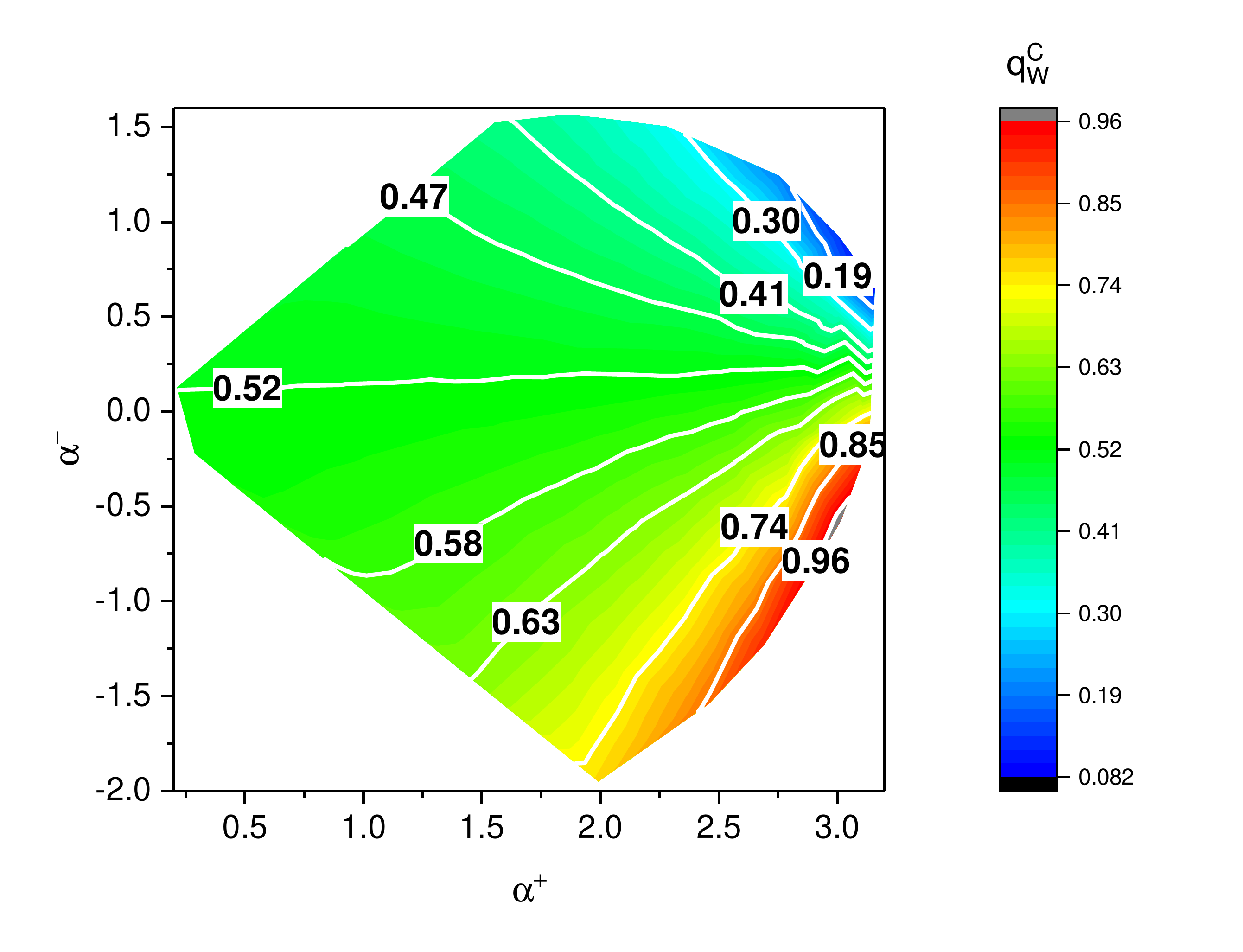}
		\label{fig:qWC_ALPHAalpha}
	}%
\iffalse
	\hfill
	\subfloat[$U_W^B$]{%
		\includegraphics[clip, trim=1cm 1cm 2cm 0.8cm, width=0.4\textwidth]{DuoSing_UWB_ALPHAalpha_color.pdf}
		\label{fig:UWB_ALPHAalpha}
	}%
	\hspace{0.5cm}%\hfill%
	\subfloat[$U_W^C$]{%
		\includegraphics[clip, trim=1cm 1cm 2cm 0.8cm, width=0.4\textwidth]{DuoSing_UWC_ALPHAalpha_color.pdf}
		\label{fig:UWC_ALPHAalpha}
	}%
\fi
	\caption{The change of participation  with $\alpha^+, \alpha^-$}
	\label{fig:q_ALPHAalpha}
\end{figure}

Fixing other parameters as the baseline values, we change the values of $\alpha^+,\alpha^-$ unilaterally. Figure \ref{fig:qWB_ALPHAalpha} and \ref{fig:qWC_ALPHAalpha} show that when $\alpha^- \approx 0$, the participation of worksites / commuters does not change much with $\alpha^+$. Under this scenario, worksites/commuters are indifferent toward the two CSPs and the overall level of cross-side benefit rate ($\alpha^+$) marginally affects the participation of worksites/commuters on the WF CSP. When the cross-side benefit rate on the WF CSP exceeds that of the NWF CSP ($\alpha^- < 0$), the WF CSP becomes more attractive to both worksites and commuters. When $\alpha^- $ is negative and fixed, if $\alpha^+$ increases, the WF CSP becomes more attractive to both sides. Note that in this case, $\alpha_N < \alpha_W$ and $\alpha_N$ increases together with $\alpha_W$. But even $\alpha_N$ increases, the participation on the NWF CSP is still decreasing (since the participation on the WF CSP keeps increasing), showing that $\alpha_W$ becomes the dominant factor to the change of participation ($q_W^B$ and $q_W^C$). The same is true when $\alpha^-$ is larger.
%On the other hand, when the cross-side benefit rate of the WF CSP is smaller than that of the NWF CSP ($\alpha^- > 0$), participation of both worksites and commuters will decrease with the overall cross-side benefit rate ($\alpha^+$). This shows that if $\alpha^-$ is positive and fixed, $\alpha_N$ will be the dominant factor of participation, and the increase of $\alpha_N$ reduces participation on the WF CSP despite the fact that $\alpha_W$ increases together with $\alpha_N$ (when $\alpha^-$ is fixed, $\alpha_N$ and $\alpha_W$ increase together).\textbf{[What does this tell us? Can we say the larger cross-side benefit is dominant in deciding the participation of the two CSPs from both sides??]}
Therefore, the larger cross-side benefit is dominant in deciding the participation of the two CSPs from both groups. Similar participation of the two groups indicates that the cross-side benefit effects lead the participation of the two groups to change in the same direction for the  duopoly model, which is consistent with the findings from the monopoly model.
%\textbf{[Why showing the utility figures?]} The utility of joining WF CSP from both sides can be explained from Figure \ref{fig:UWB_ALPHAalpha} and \ref{fig:UWC_ALPHAalpha}. When the overall cross-side benefit rate is large ($\alpha^+$ is large), both worksites and commuters experience high utility. However, the participation is also dependent on the relation between $\alpha_W$ and $\alpha_N$, as shown in Figure \ref{fig:qWB_ALPHAalpha} and \ref{fig:qWC_ALPHAalpha}.
%\textbf{[What is the purpose of the example here??]}

% Sep. 09, 2019: Comment the following paragraph; a bit redundant and unclear [JB]
%Starbucks worksites tend to subsidize the type of commuting service that is preferred by their employees, which means the worksites value the choices of commuters ($\alpha_W, \alpha_N$).
%When $\alpha^+$ increases, the utilities of Starbucks worksites and commuters increase (Figure \ref{fig:UWB_ALPHAalpha}, \ref{fig:UWC_ALPHAalpha}).
%The commuters benefit from the WF CSP because shorter commuting time, more flexible time schedule. Higher employee satisfaction leads to higher work productivity, in which way worksites will also benefit from choosing WF CSP. In this case, the worksites will value the employees on the WF CSP more ($\alpha_W > \alpha_N$), participation on the WF CSP ($q_W^C, q_W^B$) increases.

\gap

Although we do not specify the number of employees on each worksite, Figure \ref{fig:qWB_ALPHAalpha} and \ref{fig:qWC_ALPHAalpha} show similar patterns of participation from the two sides on the WF CSP. The numerical tests show that $q_W^B$ is always close to $q_W^C$, and more particularly $-0.0675 \le q_W^C-q_W^B \le 0.0854 $ holds. This means that under equilibrium, the fraction of commuters choosing a CSP is close to the fraction of worksites on the same CSP, indicating there is a relative reasonable range of the number of employees working for each worksite.

\begin{figure}[H]
	\center
	\subfloat[$p_W^B$]{%
		\includegraphics[clip, trim=1cm 1cm 2cm 0.8cm, width=0.4\textwidth]{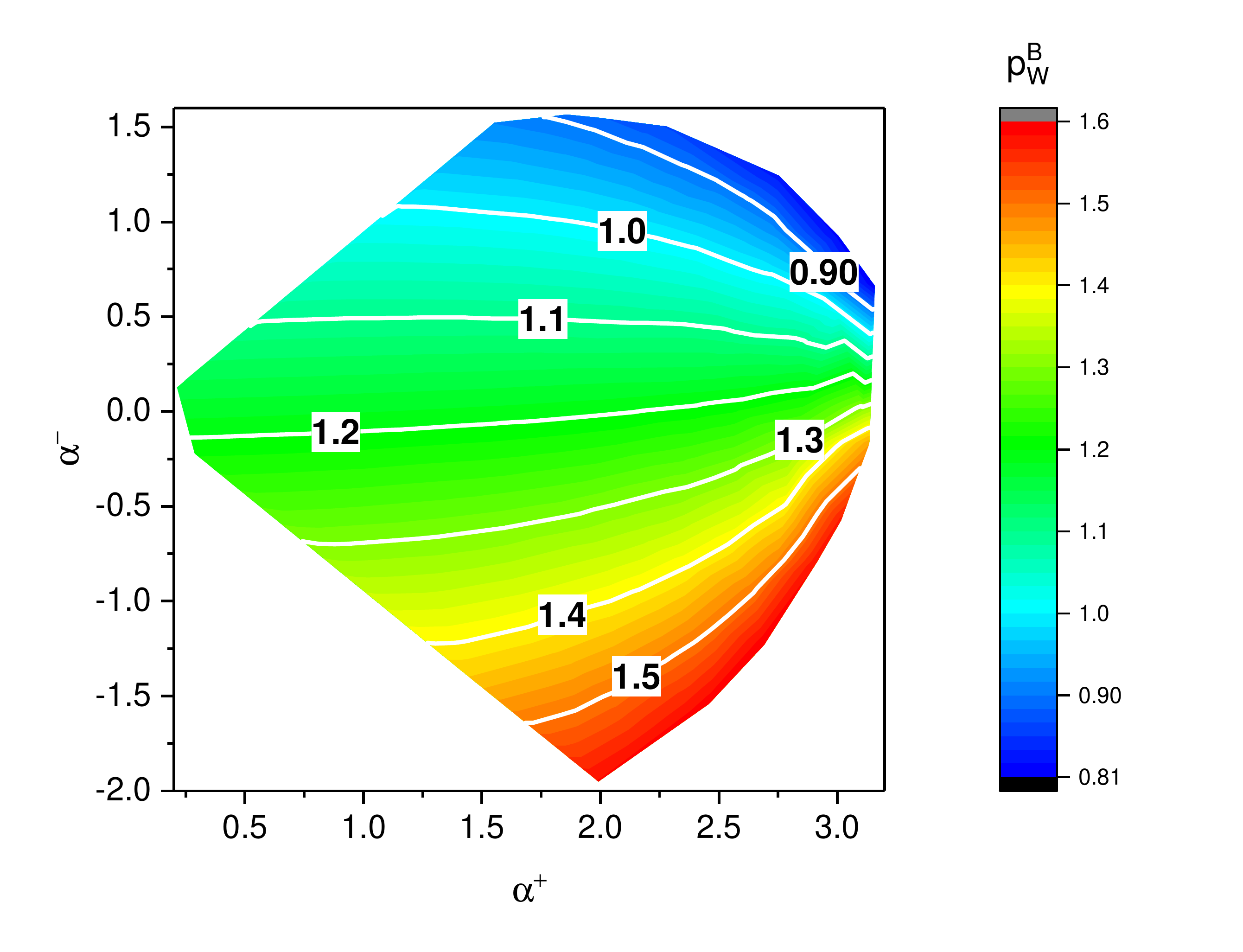}
		\label{fig:pWB_ALPHAalpha}
	}%
	\hspace{0.5cm}%\hfill%
	\subfloat[$p_N^B$]{%
		\includegraphics[clip, trim=1cm 1cm 2cm 0.8cm, width=0.4\textwidth]{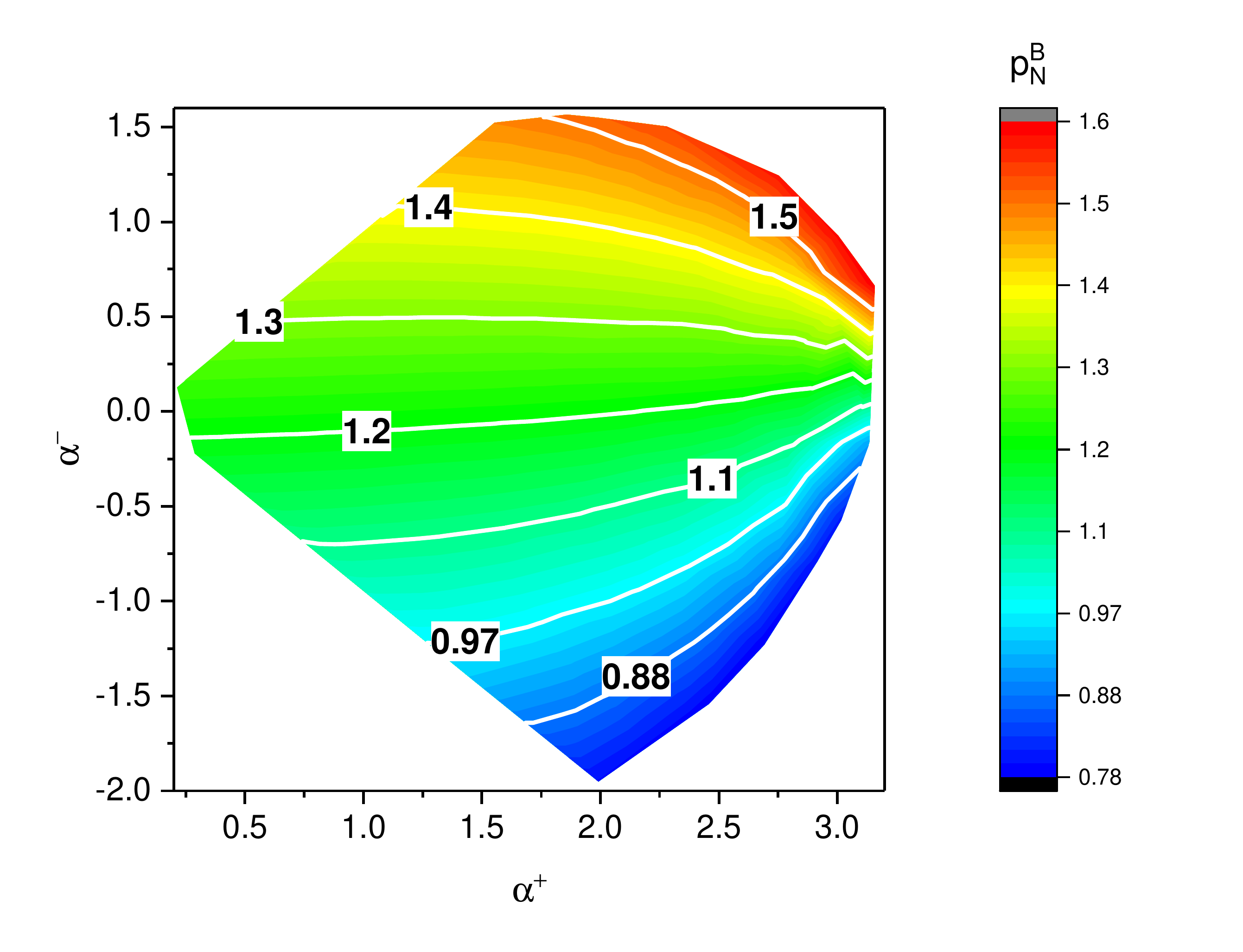}
		\label{fig:pNB_ALPHAalpha}
	}%
	\hfill%
	\subfloat[$p_W^C$]{%
		\includegraphics[clip, trim=1cm 1cm 2cm 0.8cm, width=0.4\textwidth]{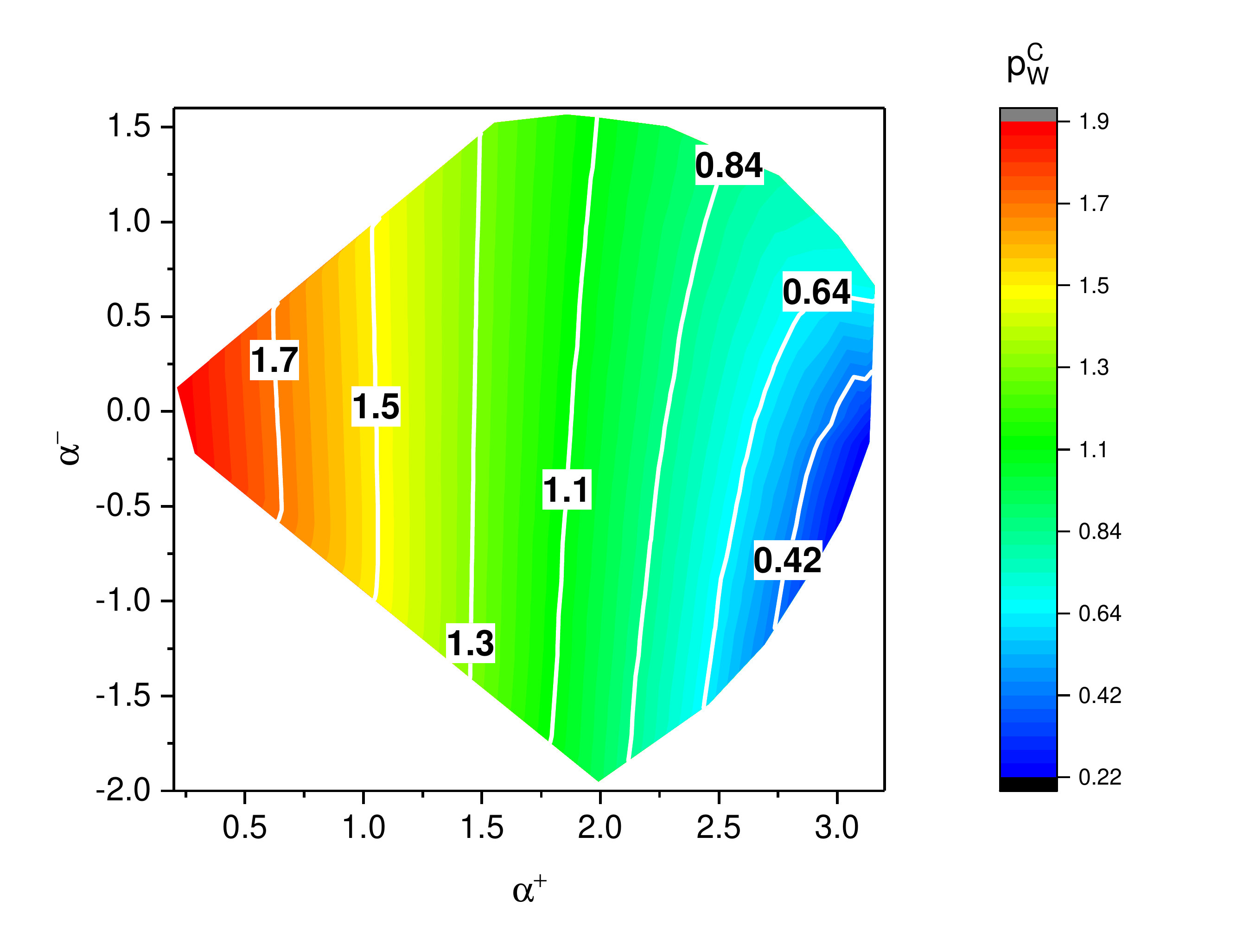}
		\label{fig:pWC_ALPHAalpha}
	}%
	\hspace{0.5cm}%\hfill%
	\subfloat[$p_N^C$]{%
		\includegraphics[clip, trim=1cm 1cm 2cm 0.8cm, width=0.4\textwidth]{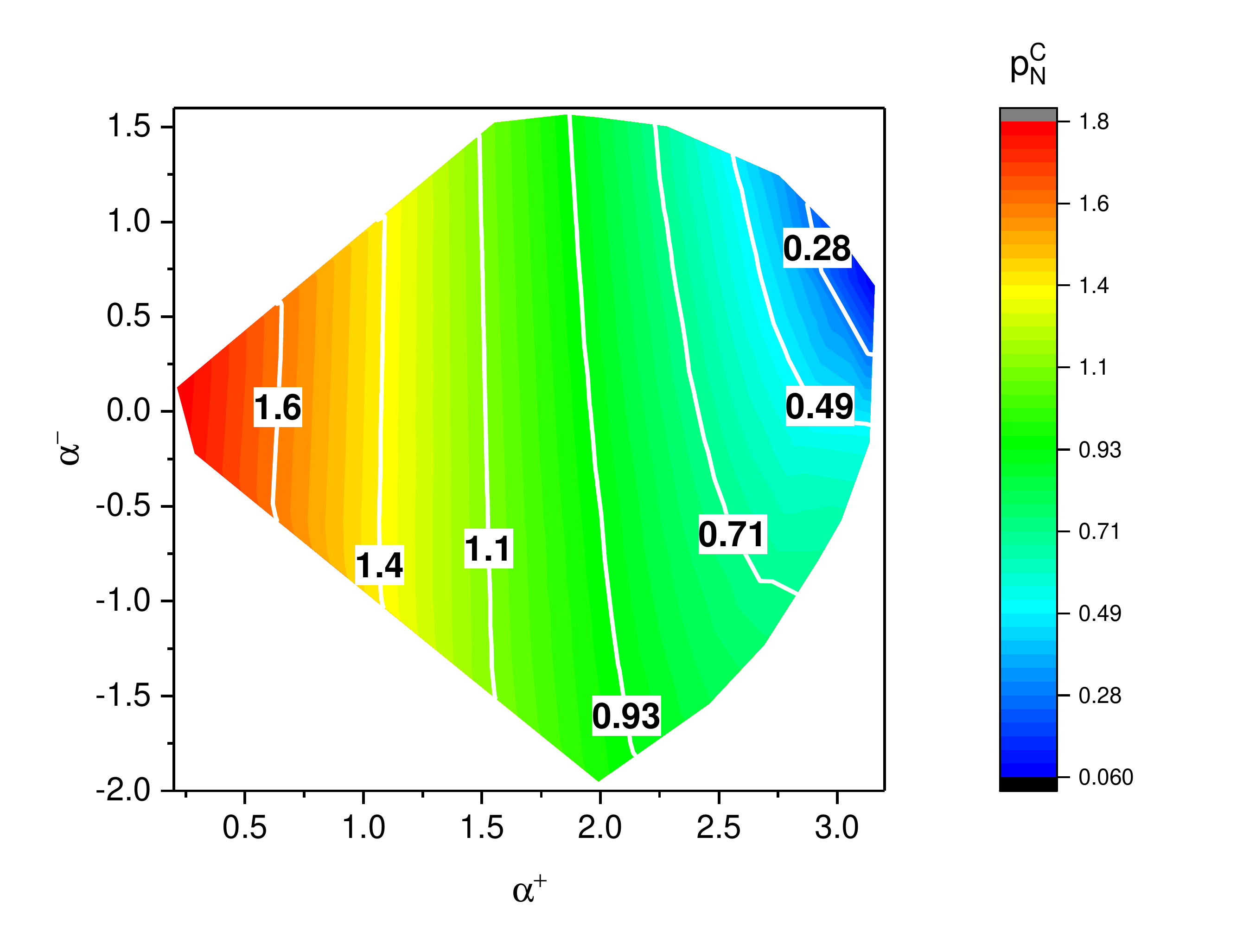}
		\label{fig:pNC_ALPHAalpha}
	}%
	\caption{The change of prices with $\alpha^+, \alpha^-$}
	\label{fig:UR_ALPHAalpha}
\end{figure}

When we change $\alpha^+, \alpha^-$, each CSP adjusts its price structure to maximize profit. The price structure patterns are shown in Figure \ref{fig:UR_ALPHAalpha}. We can see that the prices of worksites are mainly affected by $\alpha^-$, while the prices of commuters are mainly affected by $\alpha^+$. By comparing Figure  \ref{fig:qWB_ALPHAalpha} and Figure \ref{fig:pWB_ALPHAalpha}, we notice that for worksites, prices have the similar pattern as that of participation. The reason is that $\alpha_W$ and $\alpha_N$ are the cross-side benefit of worksites, and the relative value of these two parameters ($\alpha^-$) can be understood as the cross-side benefit discrepancy of the two CSPs from the perspective of worksites.
%Thus, $\alpha^-$ controls over the competition for worksites between the two CSPs.
When $\alpha^-<0$, worksites get higher cross-side benefits on the WF CSP. Knowing that worksites care less about price and care more about the number of commuters on the WF CSP, the WF CSP sets higher price to worksites (bottom-right part of Figure \ref{fig:pWB_ALPHAalpha}). In the meantime, worksites are less attracted to the number of commuters on the NWF CSP. The NWF CSP tries to set lower price to worksites to encourage participation (bottom-right part of Figure \ref{fig:pNB_ALPHAalpha}), but the participation is still low (bottom-right part of Figure \ref{fig:qWB_ALPHAalpha} and \ref{fig:qWC_ALPHAalpha}, the high participation on the WF CSP also means that the participation on the NWF CSP is low). The case when $\alpha^->0$ can be understood in a similar way. From commuters' perspective, the overall cross-side effect of worksites ($\alpha^+$) matters more. Because $\alpha^+$ measures how much the worksites group value commuters. If the worksites value commuters more, meaning $\alpha^+ > \beta^+ = 1.3$, then the CSPs will set lower prices to commuters to attract both groups on board. When $\alpha^+$ is much larger than $\beta^+$, the CSP will even take the commuters group as a loss leader. In section \ref{sec:Mono_Bench}, we define the threshold of subsidization/loss leader as $p_i^k < f_i^k$. Here is an example when the commuters group is a loss leader in the duopoly model. In Figure \ref{fig:pWC_ALPHAalpha}, $p_W^C < f_W^C = 0.73$ when $\alpha^+$ exceeds 2.4, in which case commuters group is a loss leader to the WF CSP.
%\textbf{[What happens if it does exceed 2.4??]}
Otherwise, if the worksites value commuters less, meaning $\alpha^+ < \beta^+ = 1.3$, CSPs will increase the prices of commuters and set lower prices to worksites. This shows that the aggregated cross-side benefit of worksites ($\alpha^+$) in the duopoly model has similar effect as the cross-side benefit of worksites ($b^B$) in the monopoly model (also $\beta^+$ is similar as $b^C$). Therefore, the price patterns in Figure \ref{fig:UR_ALPHAalpha} are consistent with the findings in the monopoly model. For the Starbucks example, if the worksites on the WF CSP value the commuters more ($\alpha^- <0$, $\alpha_W > \alpha_N$), they are willing to pay higher subscription fee on the WF CSP. The reverse is also true when the worksites on the NWF CSP value the commuters more ($\alpha^- >0$, $\alpha_N > \alpha_W$). In both cases, the CSP with higher cross-side benefit can set higher prices to the worksites, and offer discount price to the commuters to increase participation. %Figure \ref{fig:pW_ALPHAalpha} and \ref{fig:pN_ALPHAalpha} shows the change of aggregated prices on two CSPs. We can see that $\alpha^+$ is the dominant factor to the aggregated prices. This indicates that overall cross-side effects encourage participation, in which case a platform would be willing to reduce the aggregated price. The aggregated price of platform $W$ decrease with $\alpha^-$ when $\alpha^+$ is fixed. If $\alpha^+$ is fixed, $\alpha^-$ will be the only factor that affects the competition between two CSPs. Still keep $\alpha^+$ constant and reduce $\alpha^-$, platform $W$ could raise aggregated price without sacrificing participation with the help of increasing $\alpha_W$ ($\alpha_N$ is decreasinG). On the contrary, when $\alpha_+$ is fixed, platform $N$ wants $\alpha_N$ to be relatively larger, so that she can take advantage of larger $\alpha_N$ and increase aggregated price, reflected as $p_N^B+p_N^C$ increase with $\alpha^-$.

\iffalse
\caption{The change of prices  with $\alpha^+, \alpha^-$}
\end{figure}
\begin{figure}[!ht]
\ContinuedFloat
\fi

\begin{figure}[H]
	\center
	\subfloat[$R_W$]{%
		\includegraphics[clip, trim=1cm 1cm 2cm 0.8cm, width=0.4\textwidth]{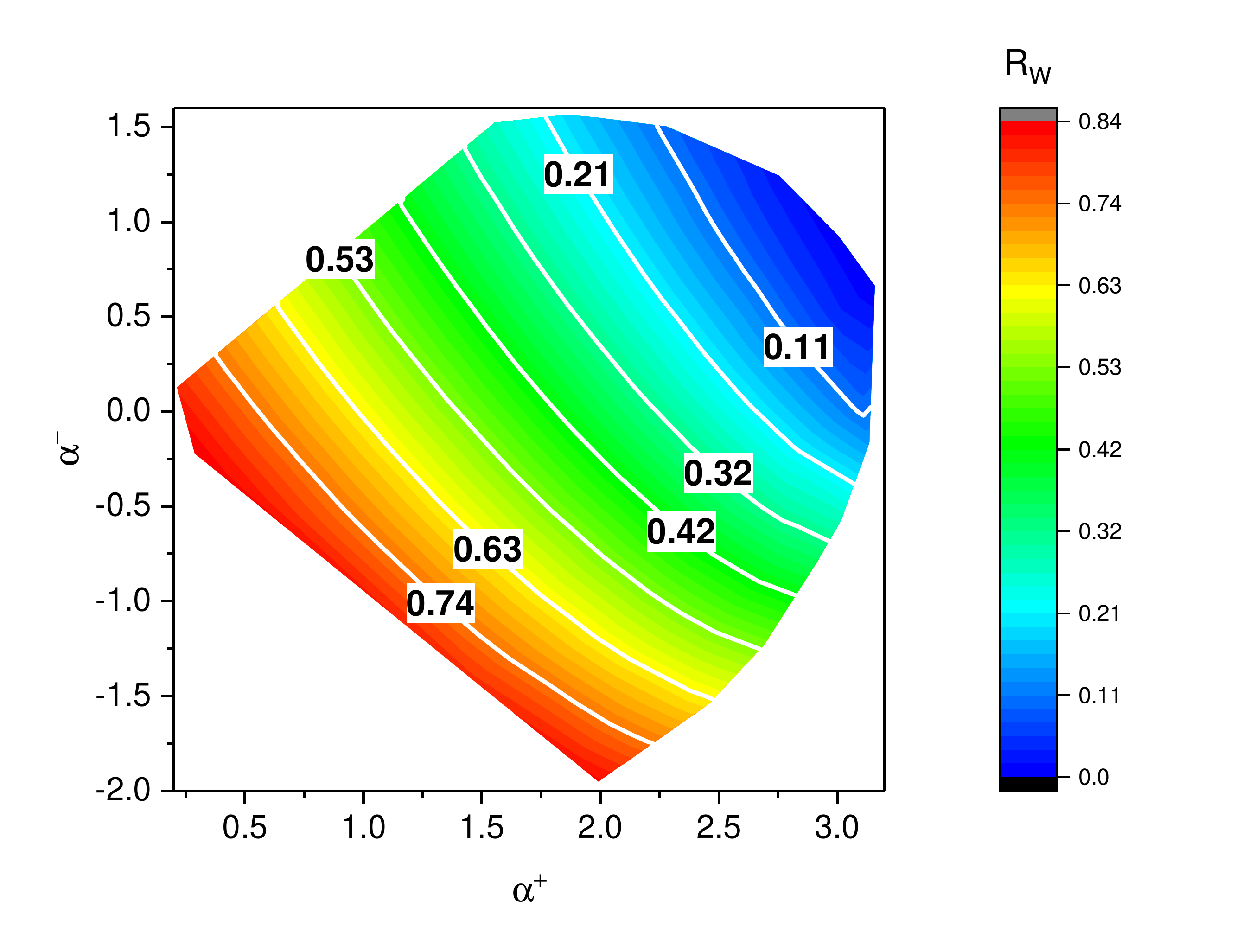}
		\label{fig:RW_ALPHAalpha}
	}%
	\hspace{0.5cm}%\hfill%
	\subfloat[$R_N$]{%
		\includegraphics[clip, trim=1cm 1cm 2cm 0.8cm, width=0.4\textwidth]{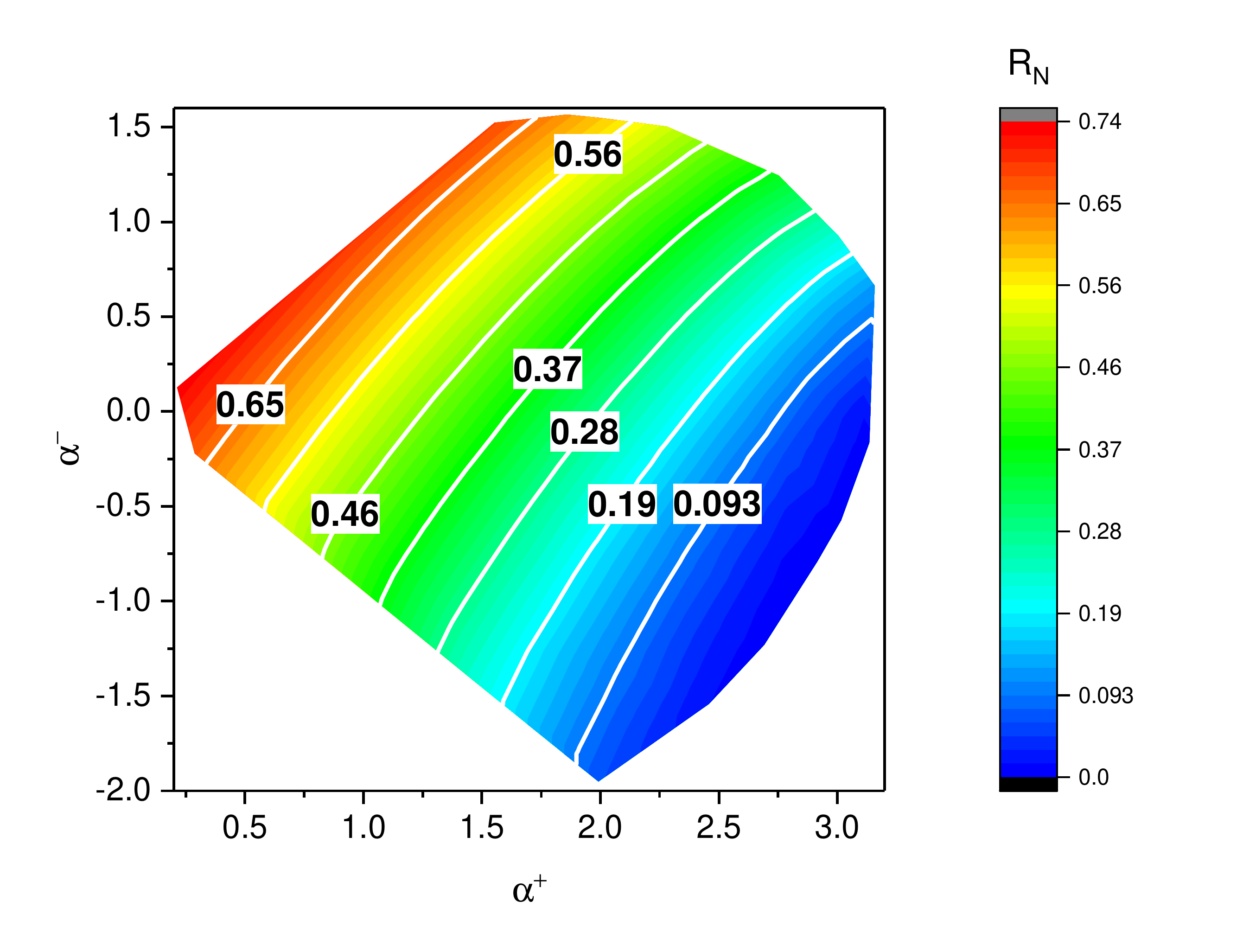}
		\label{fig:RN_ALPHAalpha}
	}%
	\caption{The change of CSP profit with $\alpha^+, \alpha^-$ }
	\label{fig:R_ALPHAalpha}
\end{figure}

Figure \ref{fig:R_ALPHAalpha} shows how profits on the two CSPs change with $\alpha^+, \alpha^-$. When $\alpha^-<0$ ($\alpha_W>\alpha_N$), the WF CSP can make higher profit. Similarly, the NWF CSP tends to make higher profit when $\alpha^->0$ ($\alpha_N > \alpha_W$). In contrast, large overall cross-side benefit rate ($\alpha^+$) is not always desired by CSPs. When $\alpha^-$ is fixed, profits on both CSPs decrease with $\alpha^+$. Referring to Figure \ref{fig:q_ALPHAalpha} and \ref{fig:UR_ALPHAalpha}, we can explain why profits decrease with $\alpha^+$. Take the WF CSP as an example, when $\alpha^->0$ and $\alpha^+$ is large (the top-right part of Figure \ref{fig:RW_ALPHAalpha} and \ref{fig:RN_ALPHAalpha}), the WF CSP charges low prices for both groups (the top-right part of Figure \ref{fig:pWB_ALPHAalpha} and \ref{fig:pWC_ALPHAalpha}), but few agents from either groups join WF CSP (the top-right part of Figure \ref{fig:qWB_ALPHAalpha} and \ref{fig:qWC_ALPHAalpha} ). Thus, the WF CSP gains low profit because of the low participation and low prices. Consider another situation, when $\alpha^- < 0$ and $\alpha^+$ is large (the bottom-right part of Figure \ref{fig:RW_ALPHAalpha} and \ref{fig:RN_ALPHAalpha}). The WF CSP can attract lots of participants from both groups (the bottom-right part of Figure \ref{fig:qWB_ALPHAalpha} and \ref{fig:qWC_ALPHAalpha}), and even takes advantage of high $\alpha_W$ to charge worksites high price (the bottom-right part of Figure \ref{fig:pWB_ALPHAalpha}), but it still makes low profit. The reason is that, the WF CSP subsidizes commuters too much and fails to recoup enough profits from worksites. From Figure \ref{fig:pWC_ALPHAalpha}, we can see that in this case the price of commuters decreases with the overall cross-side benefit ($\alpha^+$). The commuters are charged with only 0.42 ($0.42 = p_W^C < f_W^C = 0.73$) when $\alpha^+ \approx 2.7$, which implies more commuters on WF CSP means more profit loss. Similar analysis can also applies to the NWF CSP. At the bottom-right part of Figure \ref{fig:RW_ALPHAalpha} and \ref{fig:RN_ALPHAalpha}, we notice that the profit of NWF CSP is lower than that of WF CSP. This means that large $\alpha^+$ intensifies the competition between the two CSPs. One of the CSP attracts lots of participants by making use of cross-side network effects and over subsidizing, while the other platform fails to attract consumers even if it sets very low prices. Neither of the CSPs makes good profits in this case. The profit patterns also distinguish the CSPs from the one-sided market. Even the highest participation from both sides (the bottom-right part of Figure \ref{fig:qWB_ALPHAalpha}, \ref{fig:qWC_ALPHAalpha} and \ref{fig:RW_ALPHAalpha}) cannot ensure high profit on the WF CSP, although its profit is higher than that of the NWF CSP. The most desirable cases for the WF CSP are on the bottom-left part of Figure \ref{fig:RW_ALPHAalpha}, when the CSP charges medium prices to worksites and commuters without taking any of them as loss leaders (the bottom-left part of Figure \ref{fig:pWB_ALPHAalpha}, \ref{fig:pWC_ALPHAalpha}), and is also able to have more than 50\% of the customer share from the both sides (the bottom-left part of Figure \ref{fig:qWB_ALPHAalpha}, \ref{fig:qWC_ALPHAalpha}).

% Also notice that, on the WF CSP, the price variation of worksites (Figure \ref{fig:pWB_ALPHAalpha}), is less than the price variation of commuters (Figure \ref{fig:pWC_ALPHAalpha}).

Similar to the monopoly model, the participation of the two groups are similar when we change the cross-side network effects (Figure \ref{fig:qWB_ALPHAalpha}, \ref{fig:qWC_ALPHAalpha}). However, it is still possible that $q_W^C/q_W^B$ deviate from 1 under some parameter settings. In the following section, we will add demand constraints to the duopoly model when both sides single-home.

\subsection{Duopoly platforms with demand constraints (single-homing) } \label{sec:competitive_constraints}
In section \ref{sec:BC_single-home}, we assume that each worksite does not specify the number of employees. Practically, a worksite loses profit if there are too many or too few employees. Thus, it is important to ensure a reasonable range of the number of employees at each worksite. Notice that the results in section \ref{sec:Mono_hotel} and \ref{sec:BC_single-home} already show similar participation for worksites and commuters, since the two groups are modeled to benefit from the cross-side effects. In order for the model to be more practical, here we impose stricter demand constraints for worksites. In particular, we assume that employees who choose a CSP are evenly split among the worksites on the same CSP. We can then add the demand constraints to the duopoly model in section \ref{sec:BC_single-home} as:
\begin{align}
q_i^C = q_i^B \pm \eta \quad i \in \{W,N\} \label{eq:constraints}
\end{align}
where $\eta$ reflects the demand flexibility of each worksite (either on the WF CSP or the NWF CSP). Here we still work with the Starbucks example with 17 stores (worksites) in downtown Seattle. We assume that, ideally, each worksite has 10 employees. But this number may change because of the discrepant preferences of CSPs from the two groups, as represented in the above demand constraints using $\eta$ . For example, when we set $\eta=0.05$ and $q_W^B=0.4$ (i.e., $40\%$ of worksites choose the WF CSP), the actual percentage of commuters choosing the WF CSP is $35\% \sim 45\%$. Because we assume unit quantity for both worksites and commuters, in this case, $60\%$ of worksites choose the NWF CSP, and $55\% \sim 65\%$ of commuters choose the NWF CSP. It is obvious that the number of employees at each worksite converges to 10 when $\eta \rightarrow 0$.

% (For the case when the worksites have different number of employees, we can add demand constraint to each of the worksite to meet the demand requirement.)

% Assumption (d) assume that the worksites on a CSP have even number of employees. In this case, the number of employees at each worksite is constrained to be no less than 19 and no larger than 21.

\subsubsection{Numerical experiments for duopoly platforms with demand constraints}
\label{sec:numerical_competitive_constraints}
The baseline parameters are the same as the previous duopoly model (section \ref{sec:numerical_competitive}), $\alpha_N=0.7$, $\alpha_W = 0.6$, $\beta_N=0.5$, $\beta_W = 0.8$, $t^B = 1.1$, $t^C=1.2$, $f_W^B= 0.7$, $f_N^B=0.73$, $f_W^C=0.73$, $f_N^C=0.75$, which satisfy condition \textbf{(B2)} and \textbf{(B3)}. Further assume \textbf{(B1)} holds, the duopoly model with demand constraint will have an equilibrium solution with all agents single-home. This section shows the results when the above demand constraints (i.e., assumption (d)) are added to the duopoly model, and make a comparison with the duopoly model without demand constraints in Section \ref{sec:numerical_competitive}. We only present the results of the duopoly model with demand constraints when $\eta=0.05$ and $\eta=0.01$ in the following discussions. For the Starbucks example, this means that $q_W^C/q_W^B = (q_W^B \pm \eta)/q_W^B \approx 1$. So that the number of employees at each worksite is $\frac{170*q_W^C}{17*q_W^B} \approx \frac{170}{17} = 10$.

\begin{figure}[H]
	\center
	\subfloat[$q_W^B (\eta=0.05)$]{%
		\includegraphics[clip, trim=0.5cm 0.5cm 2cm 0.5cm, width=0.37\textwidth]{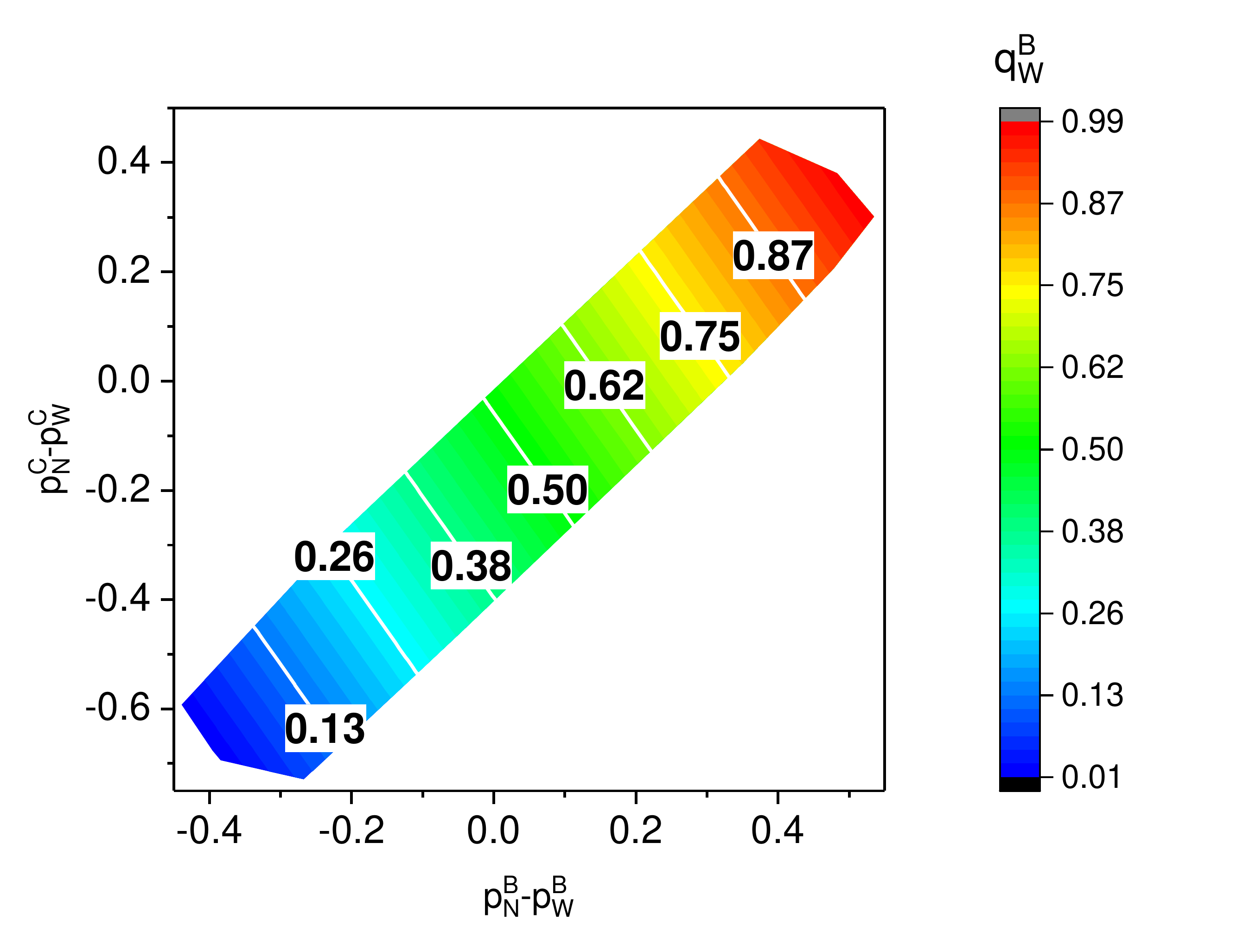}
		\label{fig:_qWB_p_demandCon}%
	}%
	\hspace{0.5cm}%\hfill%
	\subfloat[$q_W^C (\eta=0.05)$]{%
		\includegraphics[clip, trim=0.5cm 0.5cm 2cm 0.8cm, width=0.37\textwidth]{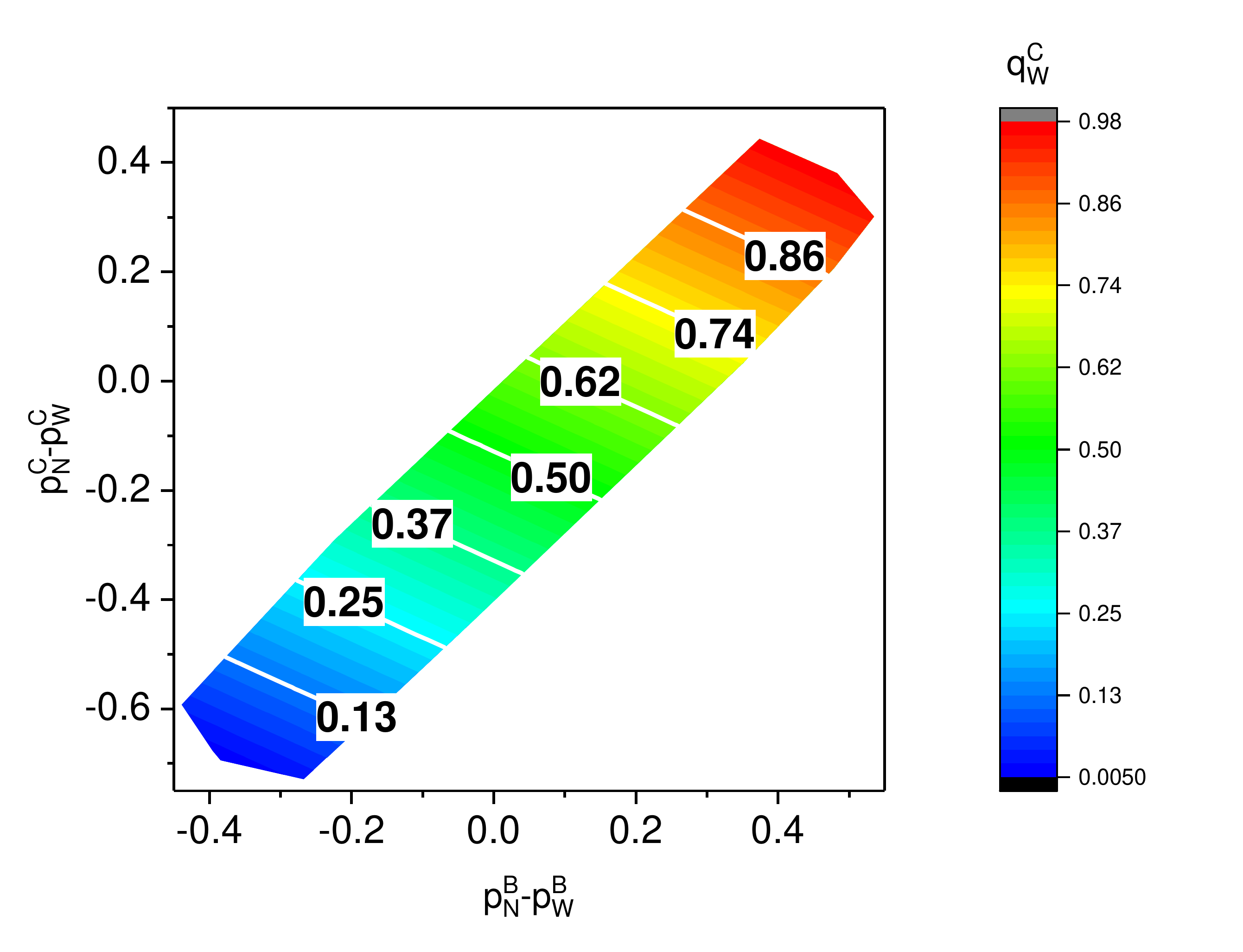}
		\label{fig:qWC_p_demandCon}%
	}%
	\hfill%
	\subfloat[$q_W^B (\eta=0.01)$]{%
		\includegraphics[clip, trim=0.5cm 0.5cm 2cm 0.8cm, width=0.37\textwidth]{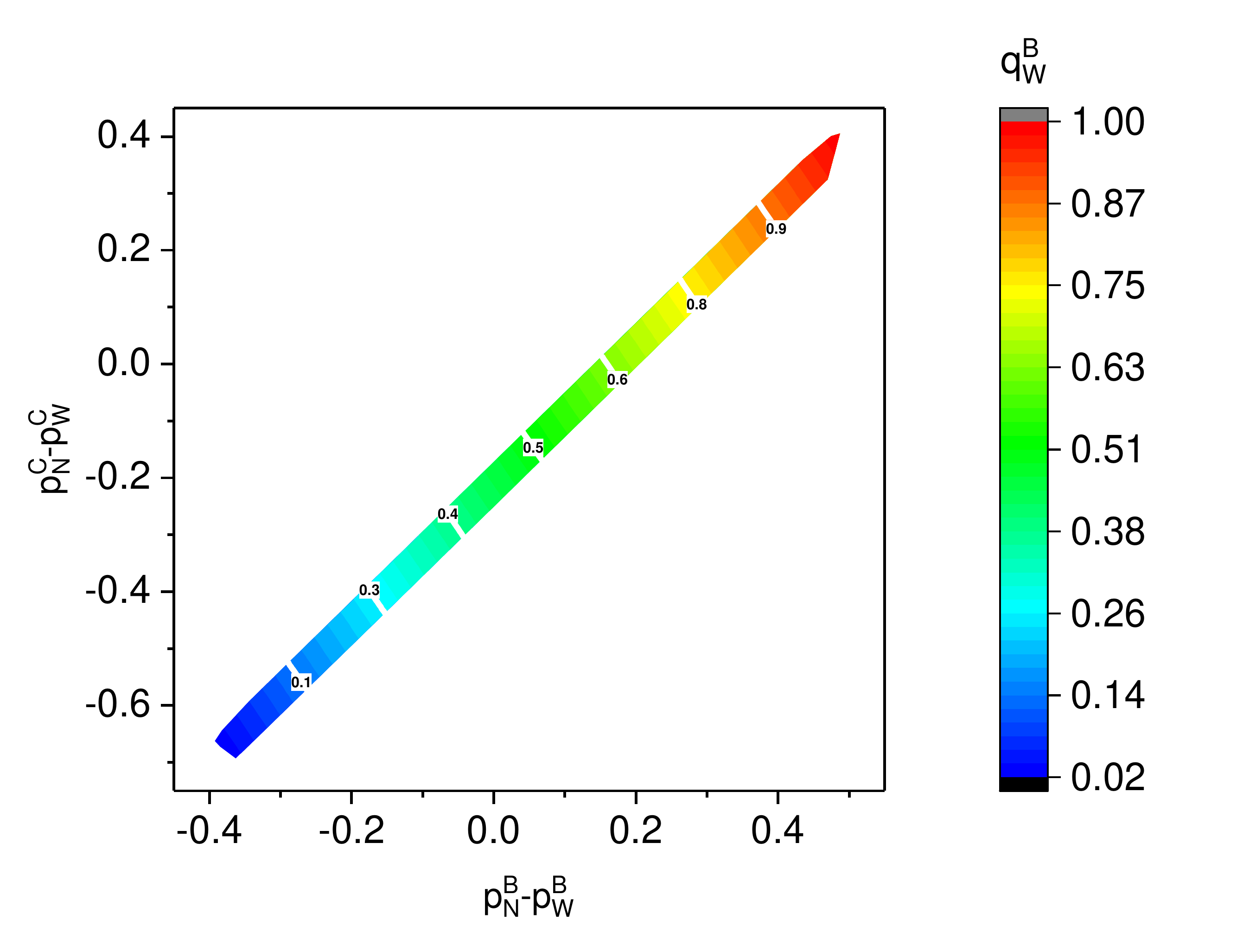}
		\label{fig:_qWB_p_demandCon}%
	}%
	\hspace{0.5cm}%\hfill%
	\subfloat[$q_W^C (\eta=0.01)$]{%
		\includegraphics[clip, trim=0.5cm 0.5cm 2cm 0.8cm, width=0.37\textwidth]{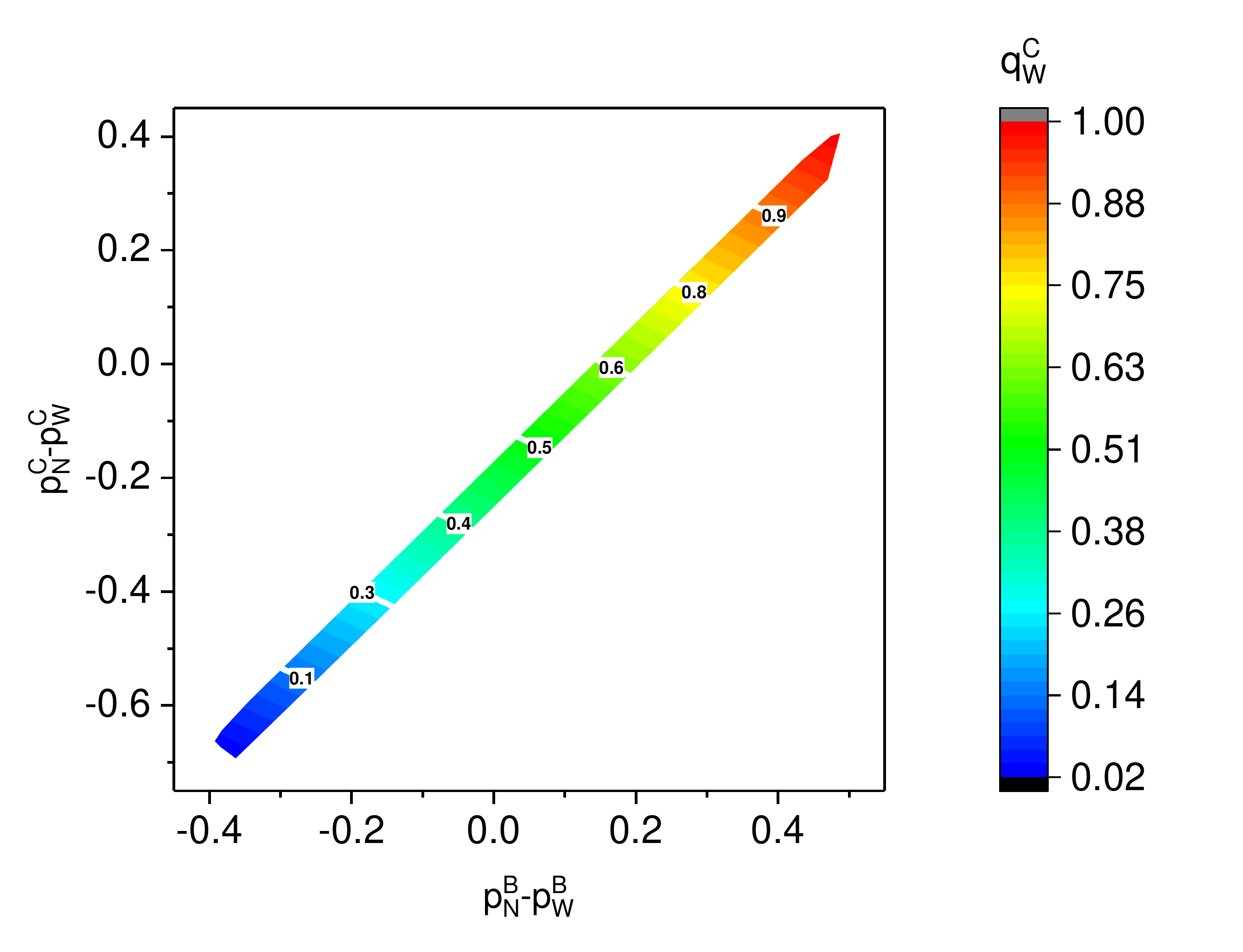}
		\label{fig:qWC_p_demandCon}%
	}%
	\caption{Demand-price relation of the duopoly platforms (with demand constraints) }
	\label{fig:qWBqWC_p_demandCon}
\end{figure}

Under equilibrium, the relation between prices and participation is shown in Figure \ref{fig:qWBqWC_p_demandCon}. The number of group $k$ agents joining the WF CSP has the same expression as equation \eqref{eq:Duo_qWB_p} and  \eqref{eq:Duo_qWC_p} besides the constraints from equation \eqref{eq:constraints}. From Figure \ref{fig:qWBqWC_p_demandCon}, we can see that the demand constraints are very effective in forcing the participation of worksites and commuters to be similar on the same CSP. The demand constraints filter the data points in the top-left and bottom-right of Figure \ref{fig:qWBqWC_pWBpWCpNBpNC} and reserve the data points along the diagonal line. When we reduce $\eta$ from $0.05$ to $0.01$, the demand constraints become stricter, the feasible region of the demand-price relation shrinks toward the diagonal line more dramatically. Thus, the proposed demand constraints effectively select price allocations to force the participation of the two sides to be similar on the same CSP at the same time. Results shown in Figure \ref{fig:qWBqWC_p_demandCon} are consistent with the results in the duopoly model (Figure \ref{fig:qWBqWC_pWBpWCpNBpNC}) and the monopoly model (Figure \ref{fig:qBqC_pBpC}). The participation of worksites on the WF CSP ($q_W^B$) is more sensitive to the relative prices of worksites on the two CSPs ($p_N^B-p_W^B$), and less sensitive to the relative prices of commuters on the two CSPs ($p_N^C-p_W^C$). The participation of worksites on the WF CSP ($q_W^B$) increases when the WF CSP set lower prices to the two groups than the NWF CSP. The same is also true for the commuters side.

%A discussion of the cross-side positive network effect has already been presented in section \ref{sec:numerical_competitive}.

\begin{figure}[H]
	\center
	\subfloat[$q_W^B (\eta=0.05)$]{%
		\includegraphics[clip, trim=1cm 0.5cm 2cm 0.8cm, width=0.4\textwidth]{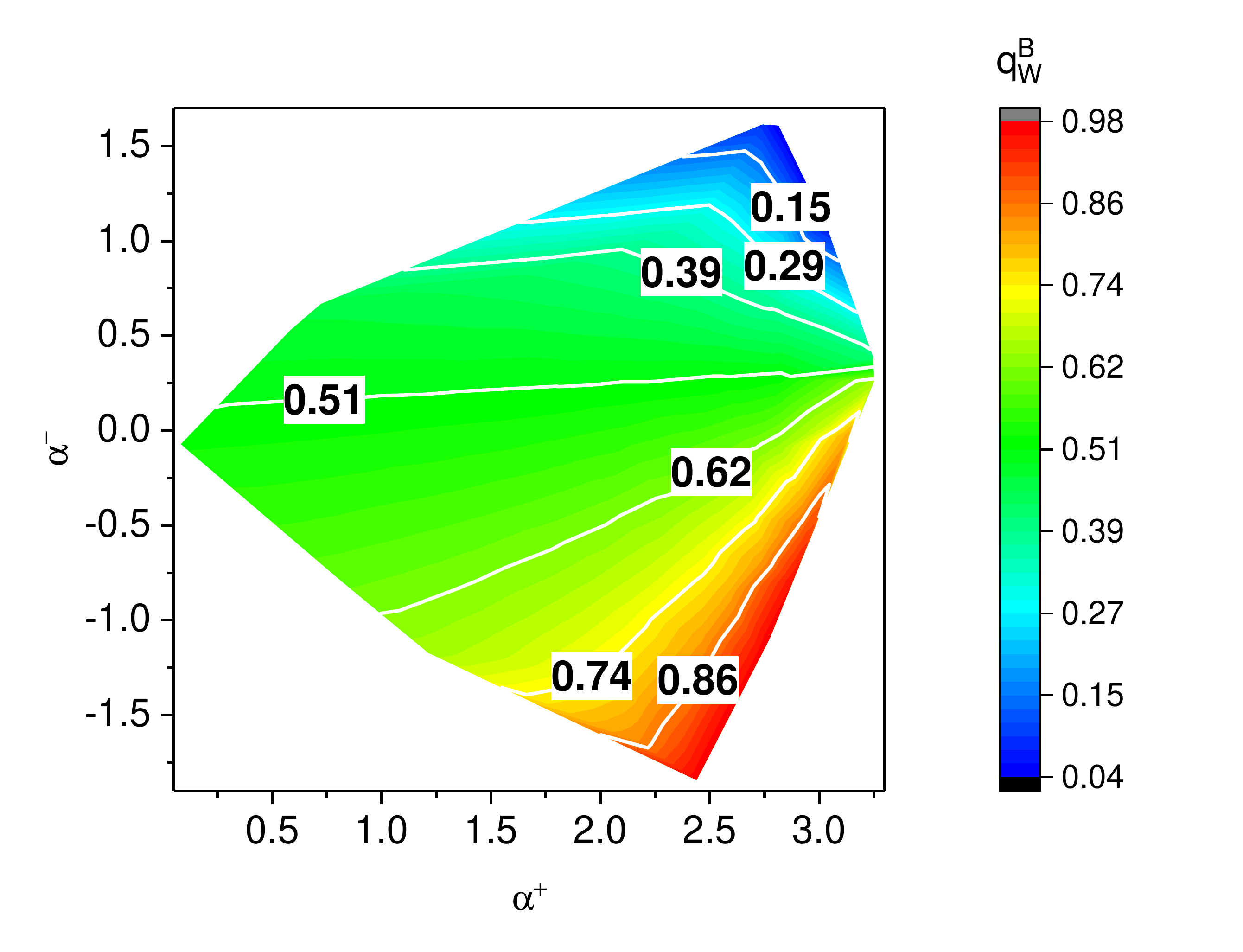}
		\label{fig:qWB_ALPHAalpha_dcon}
	}%
	\hspace{0.5cm}%\hfill%
	\subfloat[$q_W^C (\eta=0.05)$]{%
		\includegraphics[clip, trim=1cm 0.5cm 2cm 0.8cm, width=0.4\textwidth]{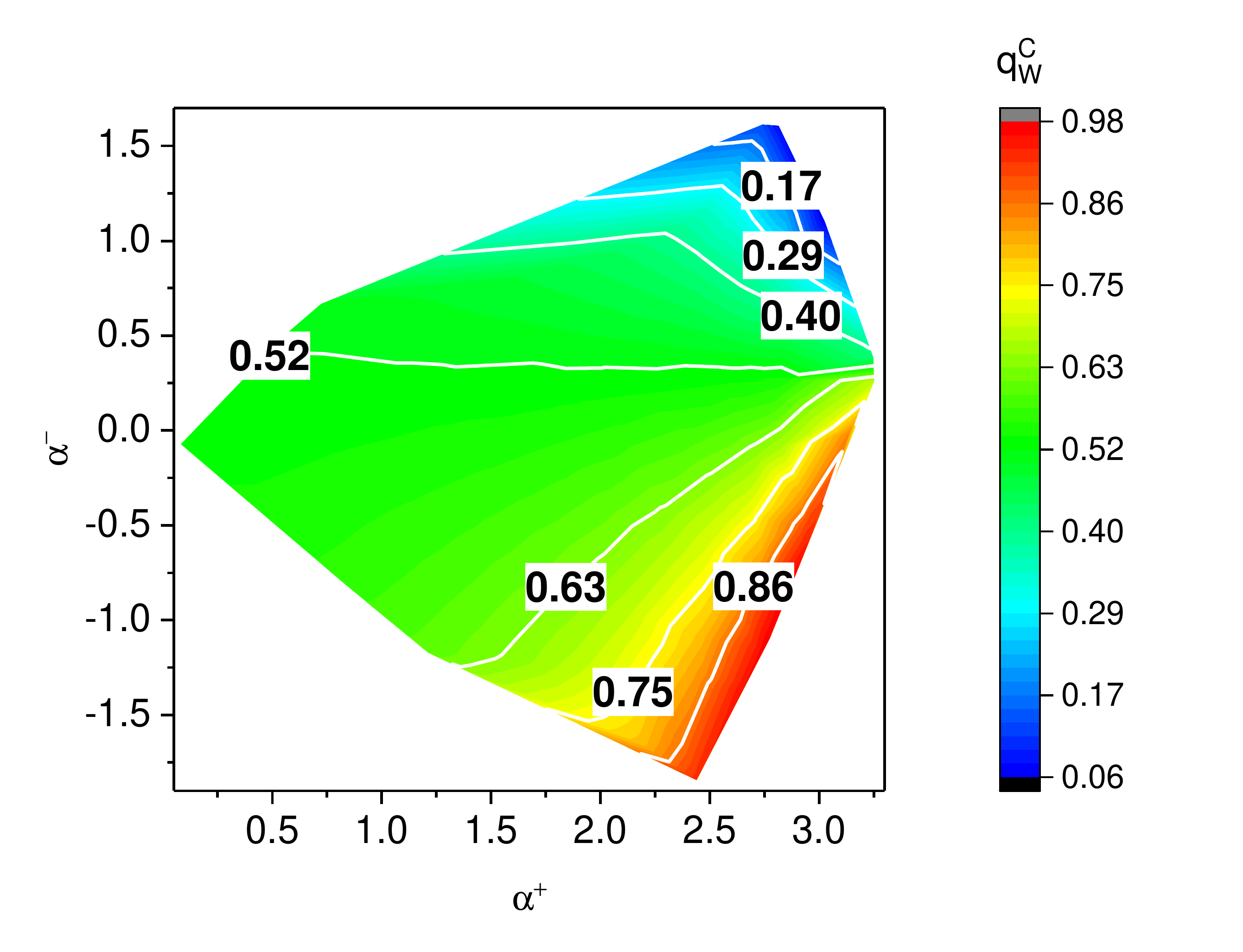}
		\label{fig:qWC_ALPHAalpha_dcon}
	}%
	\hfill
	\subfloat[$q_W^B (\eta=0.01)$]{%
		\includegraphics[clip,  trim=1cm 0.5cm 2cm 0.8cm, width=0.4\textwidth]{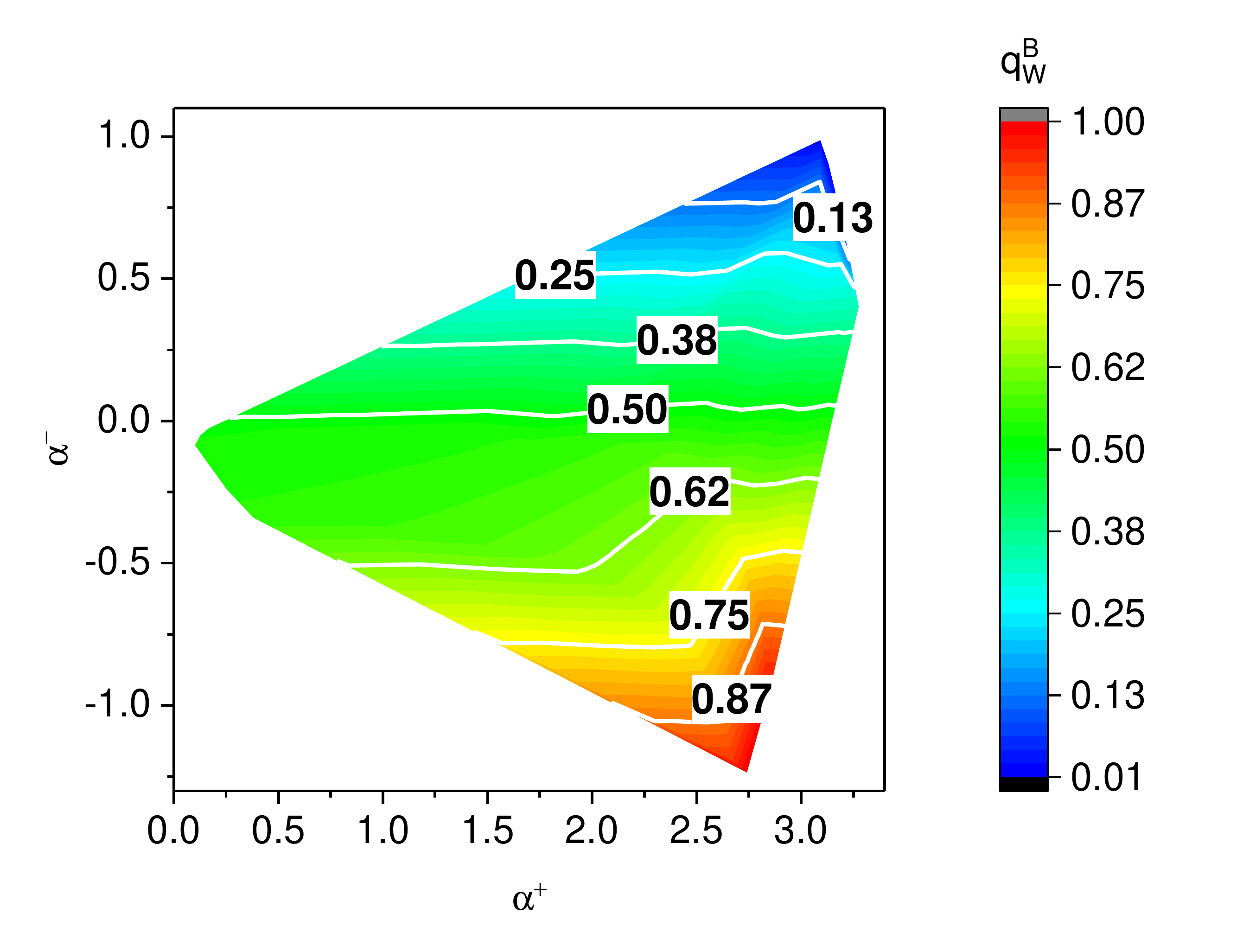}
		\label{fig:qWB_eta01_ALPHAalpha_dcon}
	}%
	\hspace{0.5cm}%\hfill%
	\subfloat[$q_W^C (\eta=0.01)$]{%
		\includegraphics[clip, trim=1cm 0.5cm 2cm 0.8cm, width=0.4\textwidth]{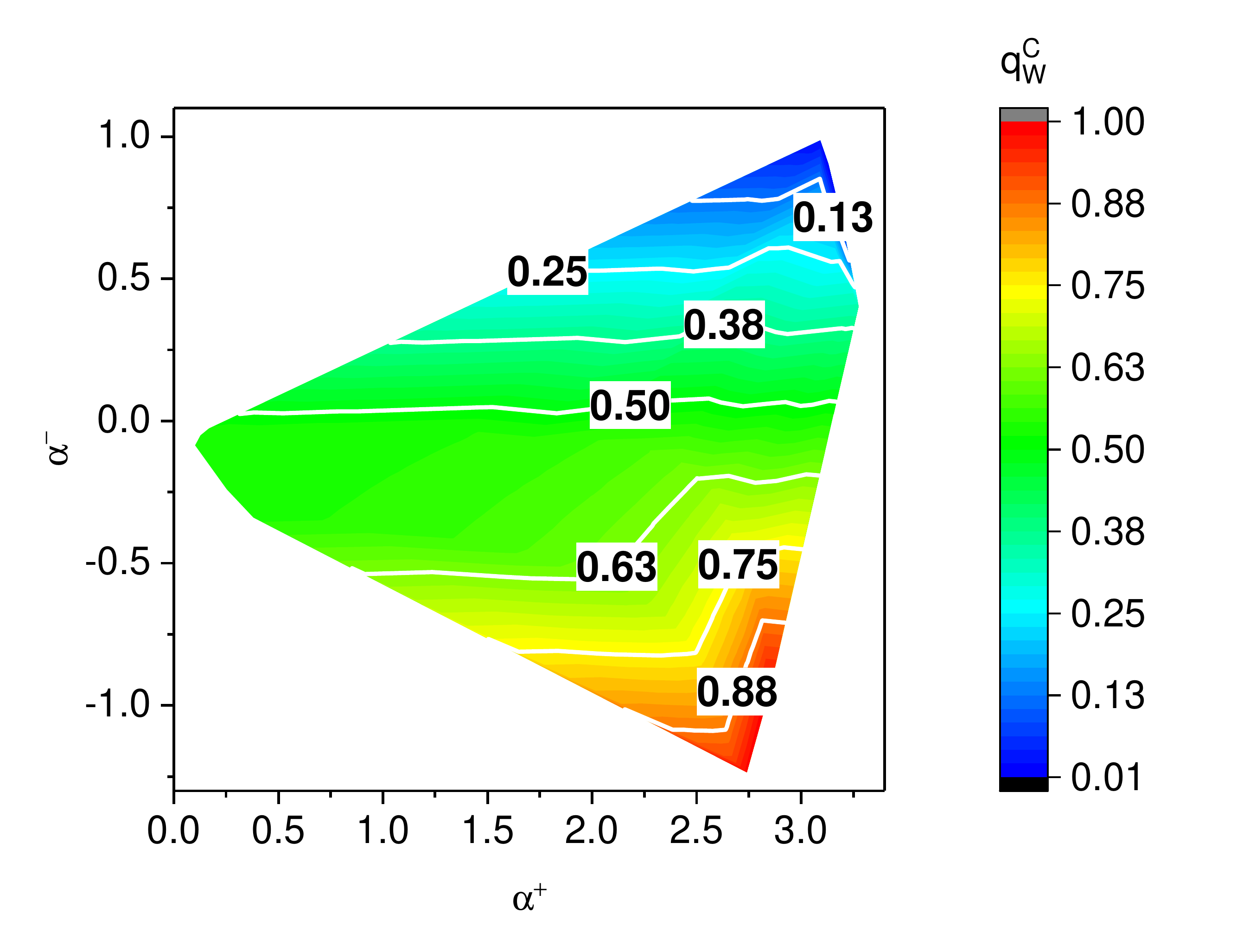}
		\label{fig:qWC_eta01_ALPHAalpha_dcon}
	}%
	\caption{The change of participation and with $\alpha^+, \alpha^- \quad (\eta=0.05 , 0.01)$ }
	\label{fig:q_ALPHAalpha_dcon}
\end{figure}

We continue to investigate what changes demand constraints bring to the duopoly model regarding the cross-side effects in terms of $\alpha^+$ and $\alpha^-$ (same can be done for $\beta^+$ and $\beta^-$). Generally, the participation patterns shown in Figure \ref{fig:q_ALPHAalpha_dcon} are consistent with Figure \ref{fig:q_ALPHAalpha}. When $\alpha^- = 0$ ($\alpha_W=\alpha_N$), the participation of worksites on the WF CSP ($q_W^B$) is marginally affected by the aggregated cross-side benefits ($\alpha^+$). When $\alpha^- > 0$ ($\alpha_W<\alpha_N$), the participation of worksites on the WF CSP ($q_W^B$) decreases with $\alpha^+$. When $\alpha^- <0$ ($\alpha_W>\alpha_N$), the participation of worksites on the WF CSP ($q_W^B$) increases with $\alpha^+$. The participation of commuters on the WF CSP ($q_W^C$) shows similar patterns. It is interesting to see that demand constraints weaken the influence of the aggregated cross-side benefit ($\alpha^+$) and strengthen the influence of the relative cross-side benefit ($\alpha^-$). Such effects are stronger when the demand constraints are stricter. When the demand constriants are less strict, i.e., $\eta=0.05$, the participation of worksites on the WF CSP ($q_W^B$) changes slightly with $\alpha^+$, shown in Figure \ref{fig:qWB_ALPHAalpha_dcon} and \ref{fig:qWC_ALPHAalpha_dcon} when $\alpha^+ < 2$. When we add stricter demand constraints to the worksites, i.e., $\eta=0.01$, the influence of the aggregated cross-side benefits ($\alpha^+$) is further weakened, and only a small range of aggregated cross-side benefits ($2.25<\alpha^+<2.75$) affects the participation. As shown in Figure \ref{fig:qWB_eta01_ALPHAalpha_dcon} and \ref{fig:qWC_eta01_ALPHAalpha_dcon}, the relative value of cross-side benefits ($\alpha^-$) largely decides the participation of the two sides. %Under the same $\alpha^-$, the participation of worksites ($q_W^B$) changes slightly when the aggregated cross-side benefit ($\alpha^+$) increases.
%In fact, the participation of worksites ($q_W^B$) changes less with $\alpha^+$ in order to be similar to the participation of commuters ($q_W^C$). As mentioned before, the aggregated cross-side benefit of worksites ($\alpha^+$) affects worksites more, and affects commuters less. The participation of worksites ($q_W^B$) changes less with the aggregated cross-sided benefit ($\alpha^+$) because the participation of commuters does not change much with $\alpha^+$, and the demand constraints force the participation of the two sides to be equal. \textbf{[but why???]} In order words, the stricter the demand constraints, the smaller change in participation from either side due to the aggregated cross-side benefit of worksites ($\alpha^+$).
However, the participation does change with $\alpha^+$ when the absolute value of $\alpha^-$ is large (the top or bottom part of Figure \ref{fig:qWB_eta01_ALPHAalpha_dcon} and \ref{fig:qWC_eta01_ALPHAalpha_dcon}).
%Both $\alpha^-$ and $\alpha^+$ affect participation and the former has larger impact. This is consistent with the findings in Figure \ref{fig:q_ALPHAalpha}. $\alpha^+$ compromises more to the demand constraints but $\alpha^-$ compromises less to the demand constraints, which explains for the dominant effects of $\alpha^-$ on participation. In such case, the participation of Starbucks worksites and commuters are mainly affected by the relative cross-side benefit ($\alpha^-$), and marginally affected by the overall cross-side benefit ($\alpha^+$).
%\textbf{[What is the purpose of this paragraph??]}
In Figure \ref{fig:q_ALPHAalpha}, the larger cross-side effect on the two CSPs determines the participation, in Figure \ref{fig:q_ALPHAalpha_dcon}, the dominant role of the larger cross-side effect is weakened. As a result, demand constraints limits the expansion of the CSP with larger market share.
%When $\alpha^- <0$ , the WF CSP can attract more customers from the both sides.
Let's assume that $\alpha^-$ is fixed. With demand constraints, the WF CSP cannot attract more customers when $\alpha_W$ increases (notice $\alpha_N$ is also increasing because $\alpha^-$ is fixed; see the bottom part of Figure \ref{fig:qWB_eta01_ALPHAalpha_dcon}, \ref{fig:qWC_eta01_ALPHAalpha_dcon}). In contrast, without demand constraints, the WF CSP can increase its user base when the cross-side benefit on the WF CSP increases, even when $\alpha_N$ increases at the same time (bottom part in Figure \ref{fig:qWB_ALPHAalpha}, \ref{fig:qWC_ALPHAalpha}).

% (If the WF CSP sets lower prices to commuters, its profit will be lower. Because the participation of commuters is mostly controlled by the demand constraints and merely increases when the price is low.)

\begin{figure}[H]
	\center
	\subfloat[$p_W^B $]{%
		\includegraphics[clip, trim=1cm 0.5cm 2cm 0.8cm, width=0.4\textwidth]{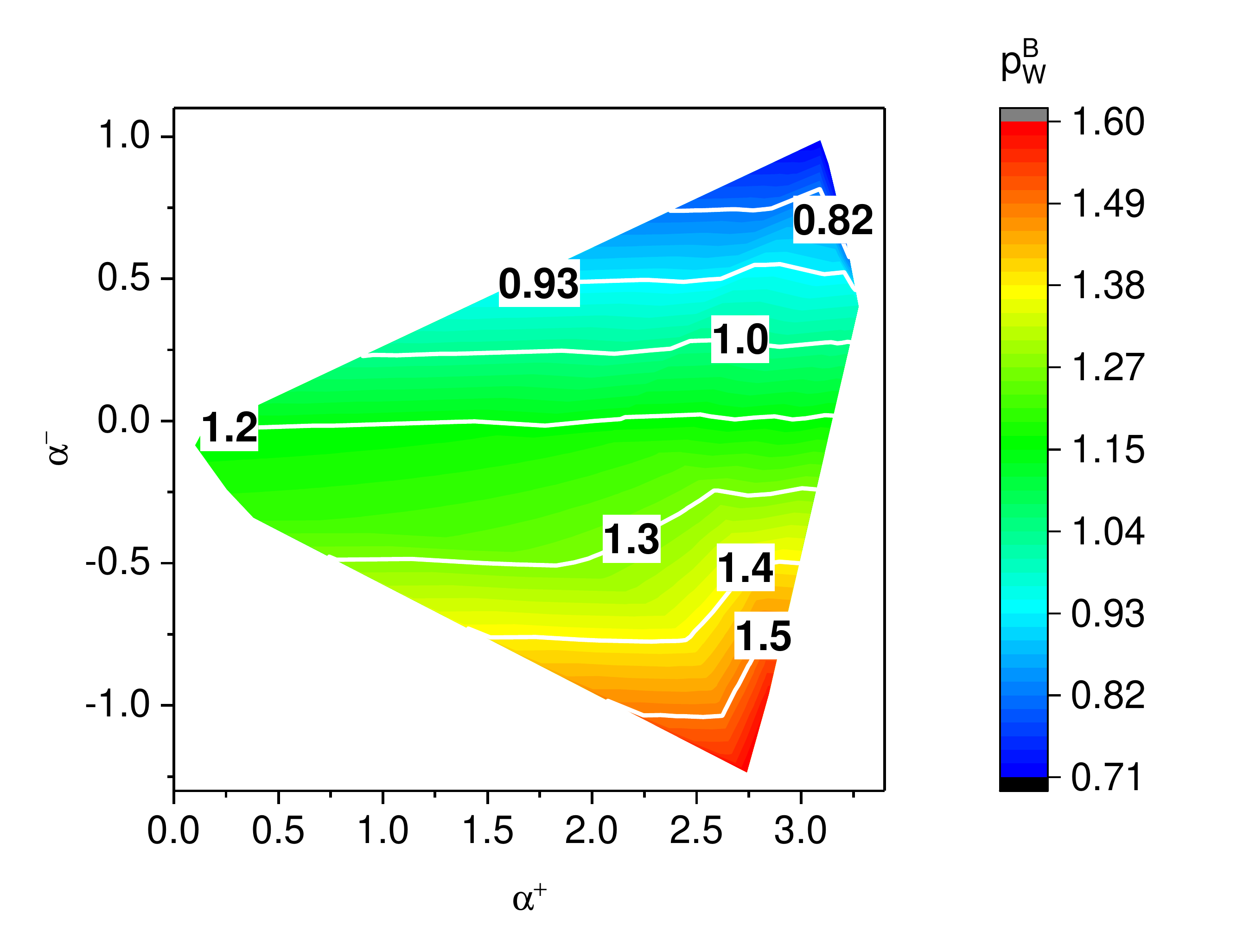}
		\label{fig:pWB_ALPHAalpha_eta}
	}%
	\hspace{0.5cm}%\hfill%
	\subfloat[$p_N^B$]{%
		\includegraphics[clip, trim=1cm 0.5cm 2cm 0.8cm, width=0.4\textwidth]{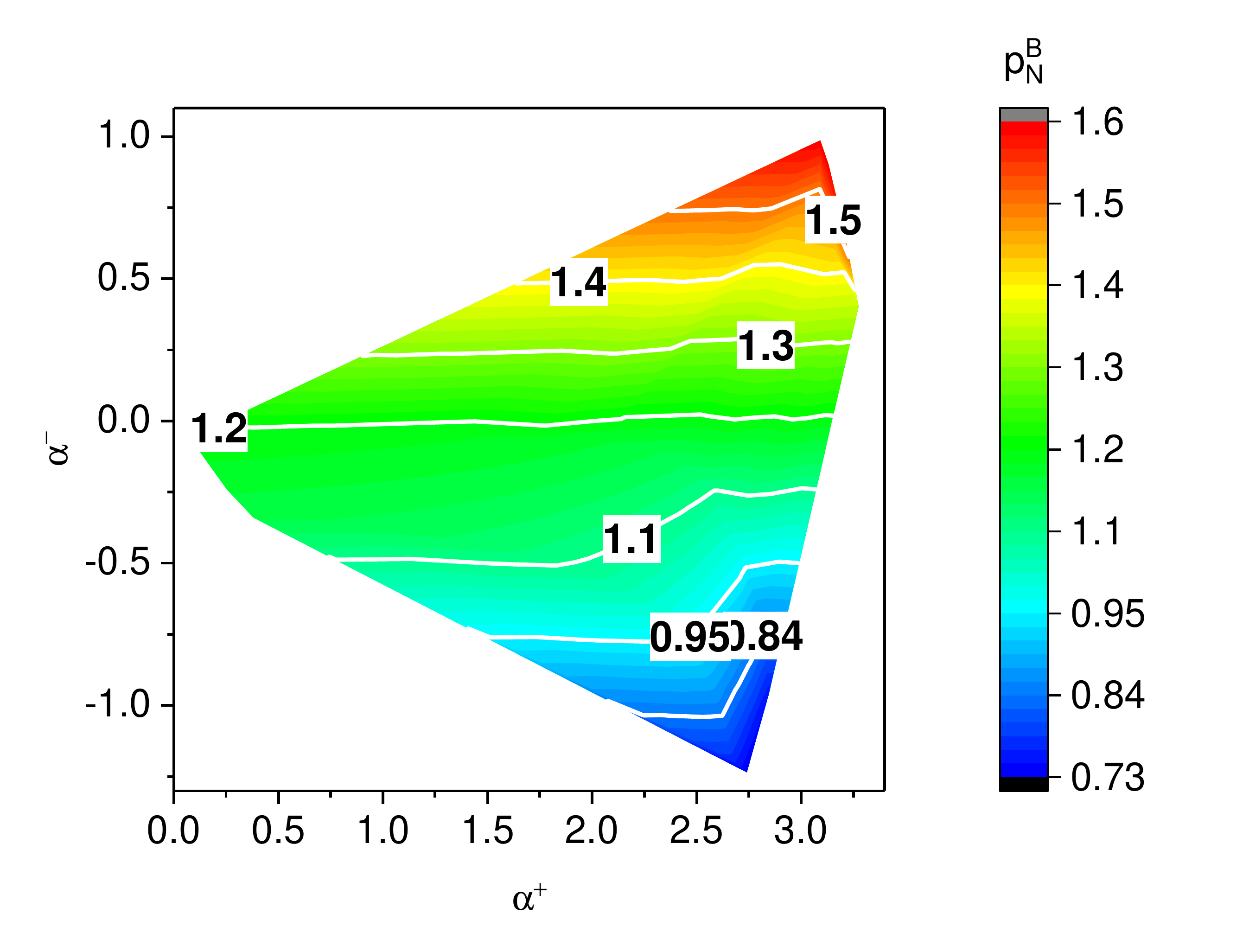}
		\label{fig:pNB_ALPHAalpha_eta}
	}%
	\hfill%
	\subfloat[$p_W^C$]{%
		\includegraphics[clip, trim=1cm 0.5cm 2cm 0.8cm, width=0.4\textwidth]{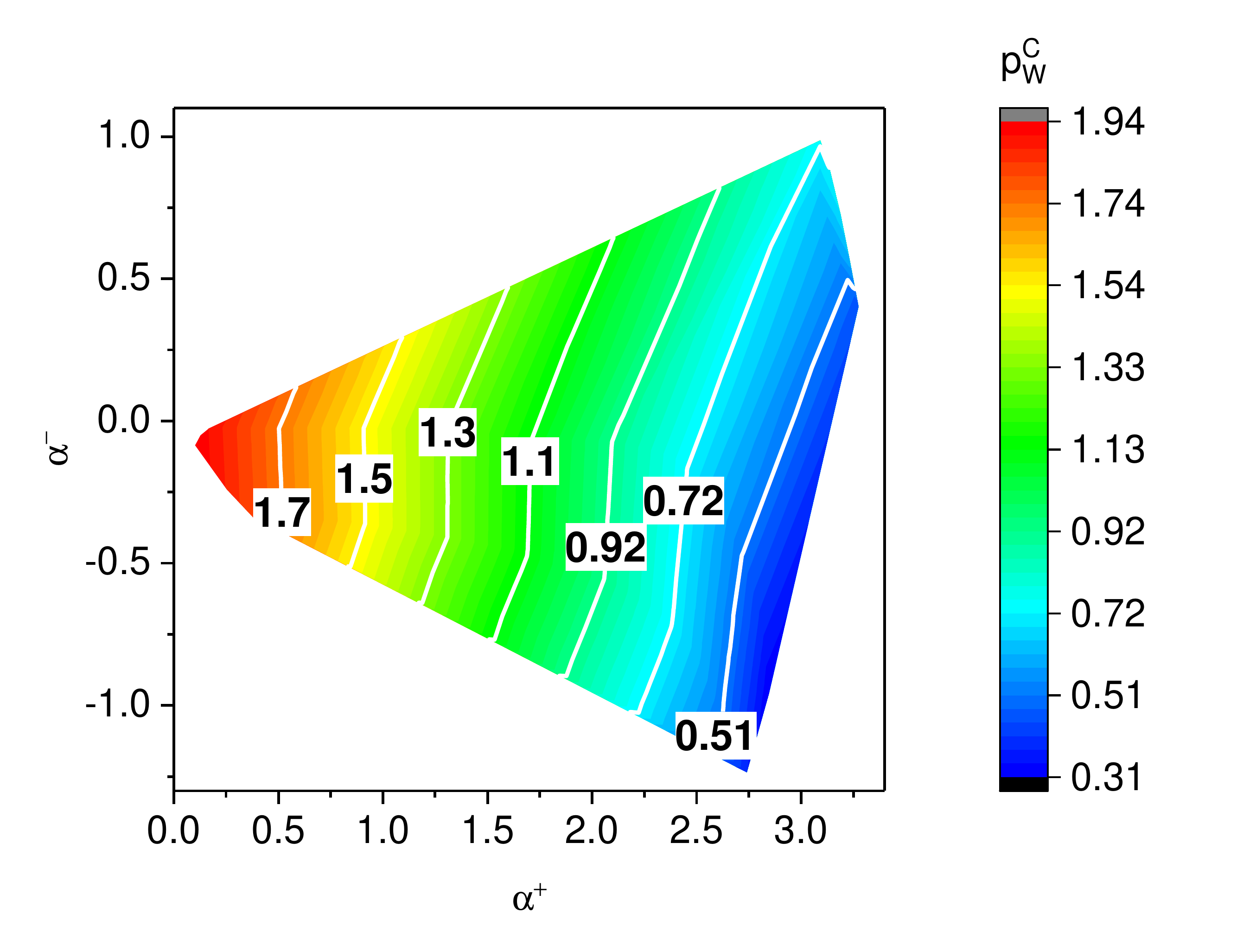}
		\label{fig:pWC_ALPHAalpha_eta}
	}%
	\hspace{0.5cm}%\hfill%
	\subfloat[$p_N^C$]{%
		\includegraphics[clip, trim=1cm 0.5cm 2cm 0.8cm, width=0.4\textwidth]{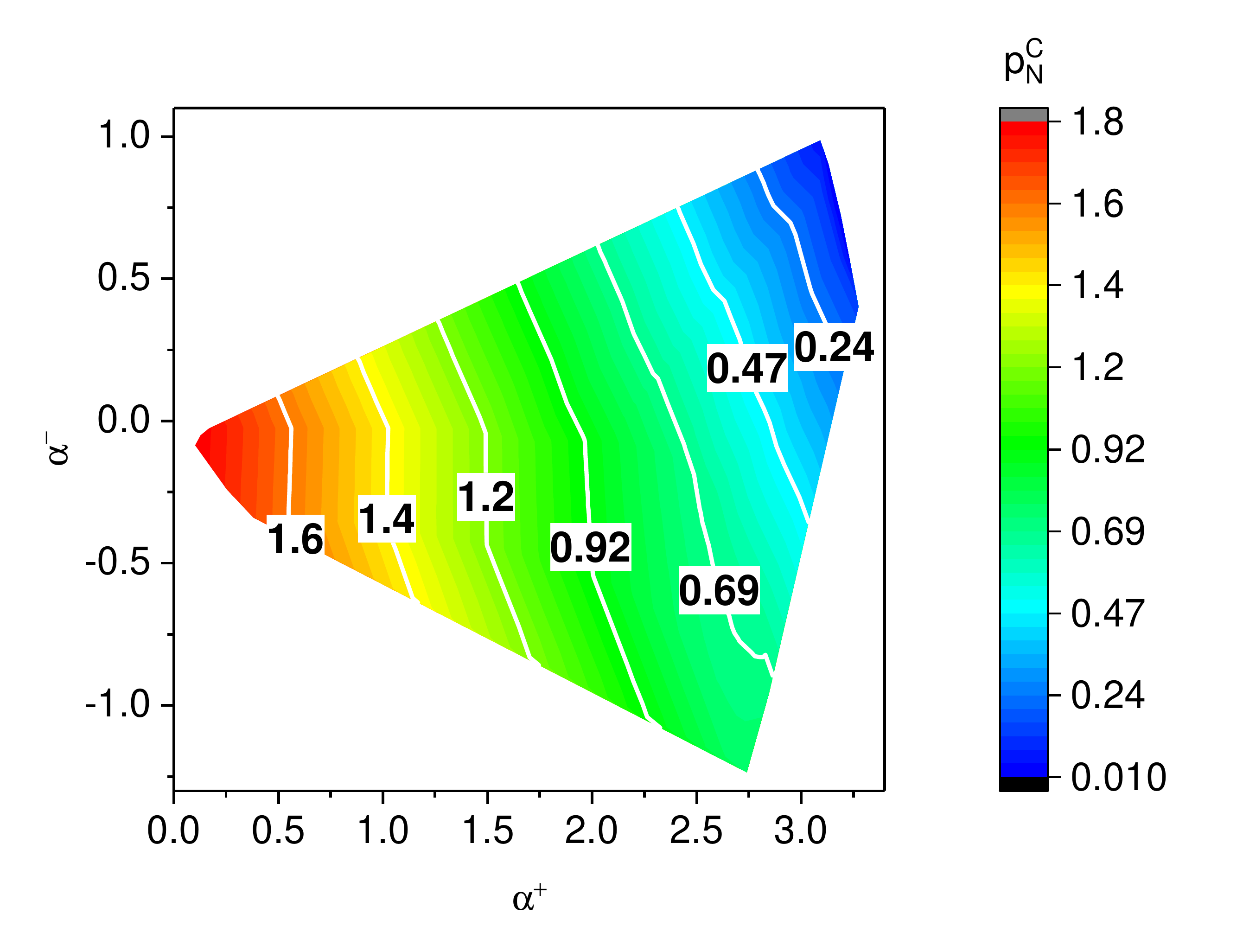}
		\label{fig:pNC_ALPHAalpha_eta}
	}%
	\caption{The change of prices with $\alpha^+, \alpha^- \quad (\eta=0.01)$}
	\label{fig:p_ALPHAalpha_con}
\end{figure}

%\gap

The demand constraints change the price patterns in a similar way, weakening the effects of the aggregated cross-side benefit ($\alpha^+$) and strengthening the effects of the relative cross-side benefit ($\alpha^-$), as shown in Figure \ref{fig:p_ALPHAalpha_con}. The prices ($p_W^B$ and $p_N^B$) of worksites shown in Figure \ref{fig:p_ALPHAalpha_con} are mainly affected by the relative cross-side effects ($\alpha^-$). When $\alpha^- <0$ ($\alpha_W > \alpha_N$ ), worksites experience higher cross-side benefits from the WF CSP, thus choosing the WF CSP even if the price is high. Under such scenario, the NWF CSP fails to attract worksites even if it sets very low prices. Commuters are almost only affected by the aggregated cross-side benefits ($\alpha^+$) in Figure \ref{fig:UR_ALPHAalpha}. When we add the demand constraints, the relative cross-side benefit ($\alpha^-$) starts to have stronger impact on the prices of commuters ($p_W^C$ and $p_N^C$). When $\alpha^+$ is fixed, the price of commuters on the WF CSP ($p_W^C$) increases slowly with $\alpha^-$ (Figure \ref{fig:pWC_ALPHAalpha_eta}).
%When $\alpha^+$ is fixed and $\alpha^-$ increases, worksites experience relatively higher cross-side benefit by choosing the NWF CSP, so the participation of worksites on the WF CSP ($q_W^B$) will decrease, as shown in Figure \ref{fig:qWB_eta01_ALPHAalpha_dcon}. This is different from the duopoly model without demand constraints where the price of commuters on the WF CSP ($p_W^C$ ) changes only slightly with $\alpha^-$ (Figure \ref{fig:pNC_ALPHAalpha}). This can be explained as follows. When we add demand constraints, the participation of commuters on the WF CSP ($q_W^C$) are forced to be similar as the participation of worksites ($q_W^B$), which in turn forces the participation of commuters to decrease with $\alpha^-$. Knowing that the participation of commuters on the WF CSP is forced to decrease, the WF CSP can set higher prices to commuters ($p_W^C$ increases with $\alpha^-$ when $\alpha^+$ is fixed) to maximize its profits.
We can also observe that with demand constraints, under the same relative cross-side benefit ($\alpha^-$ is fixed and negative), the WF CSP is less likely to take advantage of high $\alpha_W$ to increase participation from either side (bottom part of Figure \ref{fig:qWB_eta01_ALPHAalpha_dcon}, \ref{fig:qWC_eta01_ALPHAalpha_dcon}). Also, the WF CSP is less likely to take the commuters side as a loss leader and recoup profit by charging worksites with high prices (bottom part of Figure \ref{fig:pWB_ALPHAalpha_eta}, \ref{fig:pWC_ALPHAalpha_eta}). This is also true for the NWF CSP. Therefore, the competition between the two CSPs is reduced to some extent due to the demand constraints.

\begin{figure}[H]
	\center
	\subfloat[$R_W $]{%
		\includegraphics[clip, trim=1cm 0.5cm 2cm 0.8cm, width=0.4\textwidth]{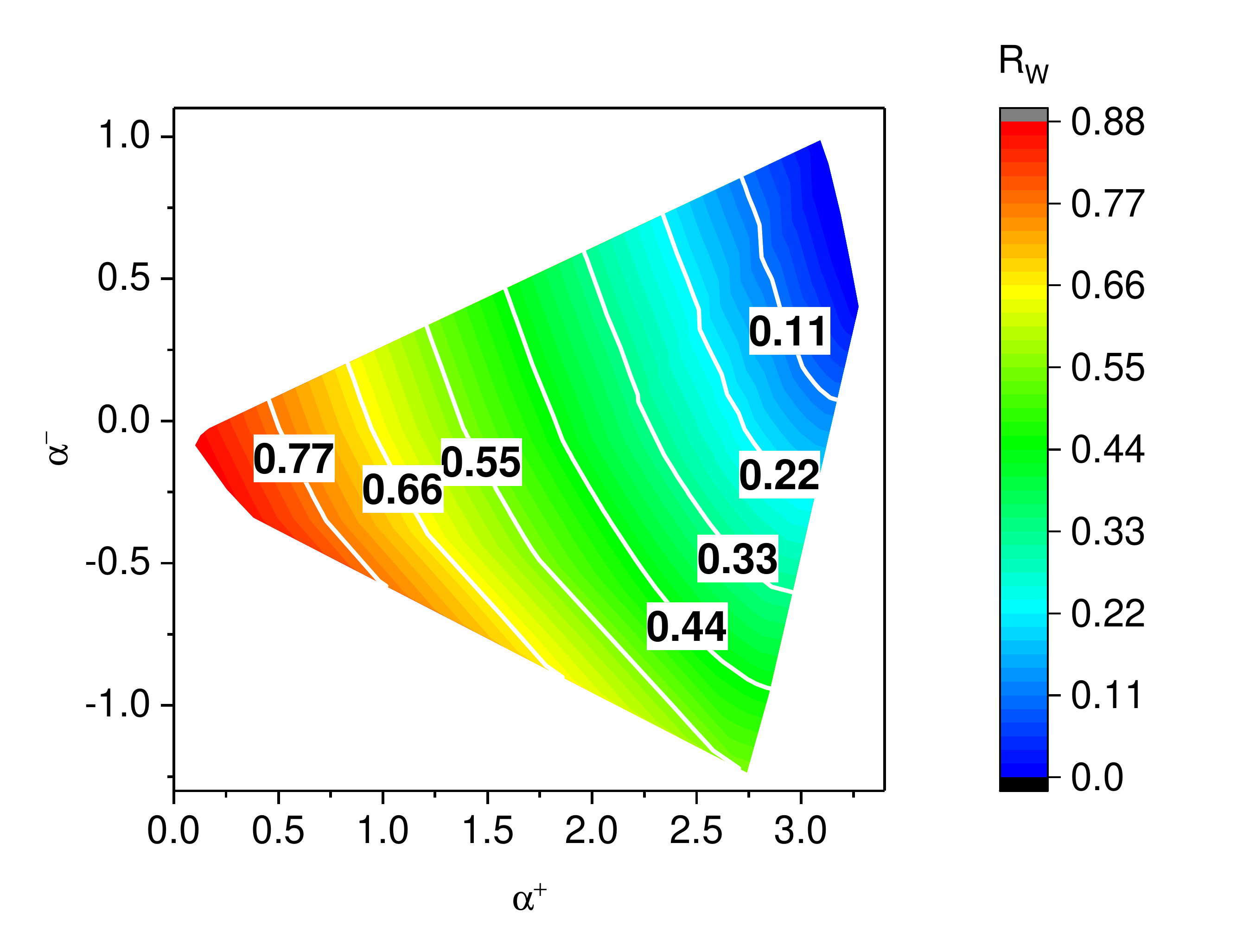}
		\label{fig:RW_ALPHAalpha_eta}
	}%
	\hspace{0.5cm}%\hfill%
	\subfloat[$R_N $]{%
		\includegraphics[clip, trim=1cm 0.5cm 2cm 0.8cm, width=0.4\textwidth]{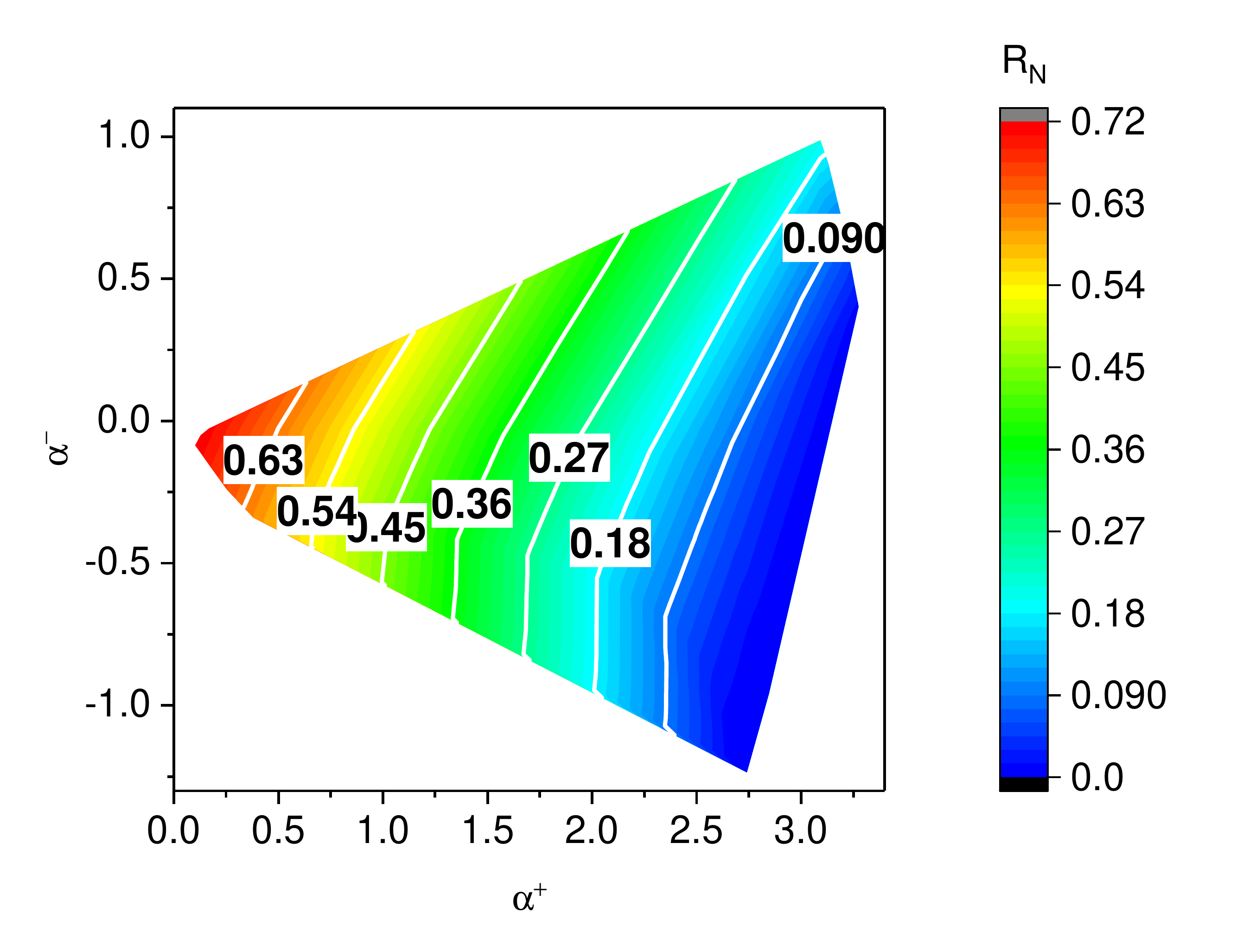}
		\label{fig:RN_ALPHAalpha_eta}
	}%
	\caption{The change of CSP profit with $\alpha^+, \alpha^- \quad (\eta=0.01)$ }
	\label{fig:R_ALPHAalpha_con}
\end{figure}

The profit patterns in Figure \ref{fig:R_ALPHAalpha_con} are consistent with Figure \ref{fig:R_ALPHAalpha}. The feasible region in Figure \ref{fig:R_ALPHAalpha_con} shrinks because of the demand constraints. The profit patterns in Figure \ref{fig:R_ALPHAalpha_con} preserve most of the regions with medium profits, while the cases when there are extreme profits (either very high or very low) are reduced. In reality, mostly likely a worksite has demand constraints, under which the competition between two CSPs are milder than the results shown in Figure \ref{fig:R_ALPHAalpha}. In Figure \ref{fig:R_ALPHAalpha_con}, under most price strategies, the two CSPs share the market and make reasonable profits. The cases when one CSP makes very high profits while the other CSP struggles with low profits only happens when the cross-side benefits fall in certain ranges, i.e., when $\alpha^+$ is very small or very large in Figure \ref{fig:RW_ALPHAalpha_eta} and \ref{fig:RN_ALPHAalpha_eta}. The cases when neither of the CSPs makes decent profit are also reduced. This further confirms that the demand constraints reduce the competition between the two CSPs.

\section{Discussions and implications} \label{sec:discussions}

In this paper, we have presented and analyzed the envisioned CSP in the planning level, especially the numerical results of the monopoly model, and the duopoly models with and without demand constraints. Here we summarize the major findings of the research, based on which to discuss how to build CSPs and associated TDM strategies in practice. First a summary of the numerical experiments for the three models is shown in Table \ref{tab:ResultsSummary}; the baseline parameters used are summarized in Table \ref{tab:Modelparameter}.

\gap

The above analysis of the two-sided market theory on commuting services leads to some interesting findings. In \textbf{the monopoly model}, when a CSP changes the price allocation between the two sides, the participation from both sides will be affected. Actually, even if the CSP only changes the price of one side, e.g., worksites, the participation of both sides will change (Figure \ref{fig:qBqC_pBpC}). Given the specific parameter settings in this paper, the participation of one side (e.g., worksites) is more sensitive to the price of the same side (worksites), and less sensitive to the price of the other side (commuters); see Figure \ref{fig:qBqC_pBpC}. Since the cross-side positive network effects bring more benefits to one side if the participation of the other side increases (see equation \eqref{eq:Mono_U_bench}), increasing cross-side network effects from one side or both sides raise the participation of both sides (Figure \ref{fig:Mono_bBbC}, \ref{fig:Mono_bB}, \ref{fig:Mono_bB_LossLeader}). Furthermore, when worksites highly value the number of commuters on a CSP, the CSP may reduce the price charged on commuters and recoup its profit from the worksites (Figure \ref{fig:Mono_bB_LossLeader}). Worksites are attracted by the commuters on the CSP and will not be discouraged by the high price. The reverse will be also true if commuters value highly the number of worksites on the CSP. Under some specific parameter settings, the CSP is willing to subsidize the commuters to maintain or increase the overall profit (Figure \ref{fig:Mono_bB_LossLeader}). The same-side negative effects discourage an agent to join the CSP when many agents from the same side have already chosen the same CSP (Figure \ref{fig:results_tBtC}). Thus, high level of same-side effects reduce participation and profit on a CSP. However, the CSP can 

\begin{landscape}
	\begin{table}[H]
		\caption{Summary of Major Results} \label{tab:ResultsSummary}
		\small
		\begin{tabular}{| p{.10\textwidth} | p{.21\textwidth} | p{0.20\textwidth} | p{0.19\textwidth} | p{.15\textwidth} |p{.15\textwidth} |p{.15\textwidth} | }
			\hline
			& \multicolumn{3}{|c|}{Cross-side positive effects} & \multicolumn{3}{|c|}{Same-side negative effects} \\ \hline
		& Participation & Price & Profit & Participation & Price & Profit        \\ \hline
			Monopoly model & Increasing cross-side network effects from one
		group raise the participation of both groups. & When the worksites value the commuters more, the CSP takes the commuter group as a loss leader (low price) and recoups profits from worksites (high price). The reverse is also true. & When the cross-side effects of one group or both groups increase, the CSP can achieve higher profit. & Participation decreases with increase of the same-side effects of either group. The same-side effect from the same group is the dominant factor.  & Prices change very slowly with the same-side effects. The price of one group is affected more by the same-side effect on the other group. & The profit decreases with the same-side effects. But the CSP is able to maintain profit under some combinations of the same-side effects.  \\ \hline
		Duopoly model without demand constraints & The larger cross-side effect ($\alpha_W$ or $\alpha_N$) on the two CSPs dominants participation. When worksites value commuters more on a CSP, the CSP can attract more participation from both groups by allocating prices, which is similar to the monopoly model.  & A CSP sets higher price for worksites if the cross-side effect of worksites ($\alpha_W,\alpha_N$) is higher. If the overall cross-side effect of worksites ($\alpha^+$) is high, the price of commuters can be low, similar to the results from the monopoly model.  & When a CSP attract more than 50\% agents from both groups and do not subsidize either groups, it can gain high profits. When one group is regarded as a loss leader, neither of the CSPs can make high profits. & \multicolumn{3}{|p{.45\textwidth}|}{The results are similar to the monopoly model.}\\ \hline
		Duopoly model with demand constraints & \multicolumn{2}{|p{.4\textwidth}|}{Similar to the findings in the duopoly model without demand constraints. However, the impact of the relative cross-side effect (i.e., $\alpha^-$) is amplified, while the impact of the aggregated cross-side effect (i.e., $\alpha^+$) is weakened. The dominant role of the larger cross-side effect is weakened. }  &  The competition between the two CSPs are milder. There are fewer cases for extreme profits (either very high or very low). & \multicolumn{3}{|p{.45\textwidth}|}{Same as the duopoly model} \\
		\hline
	\end{tabular}
\end{table}

\begin{table}[H]
	\caption{Summary of Baseline Parameters} \label{tab:Modelparameter}
	\small
	\begin{tabular}{| p{.35\textwidth} | p{.02\textwidth} | p{0.02\textwidth} | p{0.02\textwidth} | p{.02\textwidth} |p{.02\textwidth} |p{.02\textwidth} |p{.03\textwidth} |p{.03\textwidth} | p{.03\textwidth} |p{.03\textwidth} |p{.03\textwidth} |p{.03\textwidth} |p{.03\textwidth} |p{.03\textwidth} |p{.03\textwidth} |p{.03\textwidth} |p{.10\textwidth} |}
		\hline
		& $U_0^B$ & $U_0^C$ & $b^B$ & $b^C$ & $t^B$ & $t^C$ &  $f^B$ & $f^C$ & $\alpha_N$ & $\alpha_W $ & $\beta_N$ & $\beta_W $ & $f_W^B $ & $f_N^B$ & $f_W^C$ & $f_N^C$ & $\eta$ \\ \hline
		Monopoly model &1.9 &2.1 & 0.5& 0.7& 1.1& 1.5& 0.73& 0.75 & ---& ---&--- &--- &--- &--- &--- &--- & ---\\ \hline
		Duopoly models &  ---  &  ---  &  ---  &  ---  & 1.1& 1.2& --- & --- & 0.7 & 0.6 & 0.5 & 0.8 & 0.7 & 0.73 & 0.73 & 0.75 &--- \\ \hline
		Duopoly model with demand constraints &\multicolumn{16}{|p{.45\textwidth}|}{Same as the duopoly model }  & 0.05 / 0.01 \\
		\hline
	\end{tabular}
\end{table}

\end{landscape}

maintain its profit under some combinations of the same-side effects, shown by the contour line in Figure \ref{fig:Profit_tBtC}.
	
%\textbf{[We should add a table here summarizing the major results of the above analysis in Section 4 and Section 5: three models, key parameters (cross-side benefit and same-side congestion], result measures (participation, price, profit), key findings, etc]}

\gap

\textbf{The duopoly model} inherits the main characteristics of the monopoly model, and helps us understand the competition between the two CSPs. The participation of one side is still affected by the price allocation of both sides (Figure \ref{fig:qWBqWC_pWBpWCpNBpNC}), but with more complex patterns. For example, the participation of worksites is more sensitive to the relative price of worksites ($p_N^B - p_W^B$), and less sensitive to the relative price of commuters ($p_N^C-p_W^C$). The same can also apply to commuters. Generally, the cross-side benefits of worksites on the two CSPs ($\alpha_W$ and $\alpha_N$) encourage participation (Figure \ref{fig:q_ALPHAalpha}), which is consistent with the monopoly model. However, the actual participation pattern is more complex. The higher cross-side benefit has the dominant effects on participation. For example, if $\alpha^- < 0$ and is fixed ($\alpha_W > \alpha_N$), $\alpha_W$ becomes the major factor of participation; thus the participation of worksites on the WF CSP ($q_W^B$) increases with $\alpha^+$ (notice that both $\alpha_W$ and $\alpha_N$ are increasing because $\alpha^-$ is fixed). The same is true when $\alpha^->0$ ($\alpha_W < \alpha_N$). The price of worksites is mainly affected by the relative value of the cross-side benefits ($\alpha^-$; Figure \ref{fig:pWB_ALPHAalpha}, \ref{fig:pNB_ALPHAalpha}),  while the price of commuters is mainly affected by the aggregated cross-side benefits ($\alpha^+$; Figure \ref{fig:pWC_ALPHAalpha}, \ref{fig:pNC_ALPHAalpha}). Remember that higher cross-side benefits attracts more worksites on a CSP. Therefore, the CSP with higher cross-side benefits can attract more worksites even if it sets high prices on worksites; see the right-bottom part of Figure \ref{fig:pWB_ALPHAalpha}.
%\textbf{[Rong, please revise the following sentences to make them clear, concise, and to the point]}
%The prices of commuters on the two CSPs mainly change with the aggregated cross-side benefit. When $\alpha^+$ is large and $\alpha^- < 0$ ($\alpha_W > \alpha_N$), commuters are subsidized on the WF CSP, which means the CSP sets low price to commuters (right-bottom part of Figure \ref{fig:pWC_ALPHAalpha}). Because of the competition between the two CSPs, the NWF CSP also need to set lower price to commuters (right-bottom part of Figure \ref{fig:pNC_ALPHAalpha}). Thus, when either of the CSPs is subsidizing the commuters, the other CSP will also be forced to set lower price to commuters. Notice that the subsidization is mainly caused by the relative cross-side benefit ($\alpha^-$), so the price of commuters is merely affected by $\alpha^-$. Thus, the price of commuters is almost only affected by $\alpha^+$.
A CSP makes more profit when the participation is high and the subsidization level is relatively low. However, there exist ``bad'' competitions where neither of the CSPs makes high profit. For example, in the right-bottom part of Figure \ref{fig:R_ALPHAalpha}, the NWF CSP has low profit because of low participation from both sides. Although the WF CSP successfully attracts participation from both sides, its profit is still low (slightly higher than that of the NWF CSP) because of the subsidization to the commuters side.

\gap

When \textbf{demand constraints} are added to the duopoly CSPs, the demand constraints force the participation of commuters to be similar as that of worksites. As a result, the feasible region of demand-price relation is shrank to around the diagonal line (Figure \ref{fig:qWBqWC_p_demandCon}).
%\textbf{[Rong, please revise the following sentences to make them clear, concise, and to the point; if it is similar to the cases without demand constraints, just say so and remove all repetitive descriptions ]}
Compared with the duopoly model without demand constraints, the aggregated cross-side benefit ($\alpha^+$) has lower impacts on the participation and prices, while the relative cross-side benefit ($\alpha^-$) has higher impacts (Figure \ref{fig:q_ALPHAalpha_dcon}, \ref{fig:p_ALPHAalpha_con}). Remember that in the duopoly model without constraints, the larger cross-side benefits have dominant effects on participation. The participation increases on the CSP with higher cross-side benefits when $\alpha_W$ and $\alpha_N$ increase by the same amount. The demand constraints weakens the dominant role of the larger cross-side effects on participation, which leads to ``friendlier'' competitions between the two CSPs. The cases when neither of the CSPs makes decent profit are also reduced. Both CSPs make reasonable profit under most parameter settings, which means the CSPs can better co-exist in the market.

%\textbf{[Demand constraints lead to "friendlier" competition is interesting and should be highlighted here; however it needs a nice connection with what has been discussed above]}Actually, it is more reasonable for the relative cross-side benefit ($\alpha^-$) to have higher impact on the equilibrium results. The relative cross-side benefit ($\alpha^-$) reflects the level of competition between the two CSPs, since the only variables in the experiment are $\alpha^+$ and $\alpha^-$. The patterns of profit also prove that the demand constraints improve the duopoly model (Figure \ref{fig:R_ALPHAalpha_con}).

\gap

The above analysis results may help us draw some initial insights on how to \textbf{build CSPs} in practice. First, the envisioned CSP can help enhance the collaboration among CSPs, employers, and commuters, which is beneficial to all players in the market: (i) the CSP can manage to obtain profits from both sides; (ii) the employers can out-source the commuting subsidization to their employees to a third party (in our case, the CSPs) conveniently which helps recruit/retain needed talents; and (iii) the commuters can have more affordable/accessible commuting choices. The profit of a CSP is affected by cross-side positive network effects as well as same-side negative network effects. In practice, cross-side benefits exist because employers and employees value each other's participation. More importantly, employers usually play an important role in the commuting decisions of their employees. For example, the work schedule is often set by employers which determine the departure time of employees, and employers' commuting related programs (e.g., those related to transit passes and parking) also impact their employers' commuting decisions. Therefore, understanding and leveraging the strong interactions of employees and employers to adjust price strategies according to the cross-side effects is very critical for a CSP to attract participation. For example, if worksites value commuters more, a CSP can set lower price to the commuters and set higher price to the worksites to attract more commuters, thus encouraging more worksites to participate. High same-side effects may exist when the participation on a CSP exceeds the maximum number of services the CSP can provide, or the CSP becomes less efficient when the amount of customers increases. Imagine when a CSP hires big vans to pick up employees at their homes and then send them to a worksite. If the number of commuters taking the van increases, the pick up time will increase. Knowing this, some commuters may choose more time efficient ways of commuting and choose not to join the CSP. Thus, when a CSP tries to attract more customers, it is important to develop strategies that can serve the increasing demand efficiently.

\gap

Some existing MaaS, such as ridesourcing companies, can extend their existing platforms to add CSP as a new category of services that are specialized for commuting. For example, a ridesourcing company may have contracts with business owners and assign vehicles to transport commuters to their worksites (each vehicle picks up multiple commuters from the same or different employers) and vice versa. Such ridesourcing vehicles can either send commuters from their homes to the worksites directly, or send them from homes to hubs of public transit. In fact, industry pioneers such as Scoop is doing this by providing carpool services to co-workers of the same company or people living in the same neighborhood. To be more effective, CSP needs to integrate employers and enable (real time) communications between commuters and their employers (e.g., managers) in order to resolve commuting related (and work schedule related) issues promptly, which is currently lacking and should be the key focus of building practical CSPs. This will hopefully help prompt more efficient ridesourcing modes (such as ridesplitting, carpooling, vanpooling) which are currently under utilized \citep{li2019characterization}. The integration of employers into current MaaS platforms will also lead to a win-win-win situation: commuters can choose convenient and less expensive commuting services, MaaS companies have access to larger demands (by working with employers directly) that may result in larger profit, and business owners can also benefit from CSPs because their employees have more convenient ways to get to work and the total commuting trips of their companies are causing less congestion in the adjacent areas (and thus better meet regulations on commuting trip reductions).

%The two-sided market strategies of CSP can be a way to improve the existing MaaS. For example, ride-sourcing companies can have specialized commuting services based on CSP. For example, Uber already developed different types of services based on the value of time and income of the customers, Uber Pool is more appealing to passengers that is flexible in travel time and want to save money, while Uber Express attract more passengers who prefer shortest travel time and are willing to pay more. For companies like Uber, our CSP concept can be a prototype of a new type of service for the morning/afternoon peak hours, either covering the whole trip, or covering the first/last mile trip.

\gap

Our analysis may also help develop the next-generation, CSP-based and \emph{employer-centered} \textbf{TDM strategies} to better leverage the emerging MaaS technologies and to actively engage employers. First, employer-centered TDM strategies can be developed for CSPs to enable and facilitate the communications between employers and employees regarding commuting decisions. For example, a CSP can negotiate with different worksites on the flexible working hours for commuters who live close to each other but work for different worksites. Assume Commuter A and Commuter B live close to each other but work in different worksites (which are also close to each other): A leaves for work at 7:30am and B leaves for work at 8:00am. If both worksites require fixed working hours, A and B need to go to work separately, probably by driving alone (i.e., two cars). If the CSP can negotiate with the worksites and adjust the working hours, A and B may carpool to work together at say 7:45am, reducing the single occupancy vehicles in the road network. Currently, a carpool happens in most cases only if two commuters have the same working hours and close-by worksites. With CSP, there will be more chances for carpool or ridesharing to occur by actively involving employers in the commuting related decisions. As a result, single occupant vehicles can be further reduced due to CSPs.
Second, by understanding/analyzing the interactions/participation of employers and employees to different CSP services (via the two-sided market analysis method discussed above), TDM strategies can be developed to help increase the usage of certain types of commuting services of the CSP that are more beneficial to the urban transportation network. For example, agencies can provide incentives to encourage employers to subscribe for the WF CSP since the platform encourages peak spreading (and thus helps reduce peak hour congestion). This will motivate more commuters (employees) to choose the WF platform. As a result, the actual use of the WF CSP will increase, which could help ease the peak hour congestion. Instead of putting forward regulations that directly guide individual commuters/companies to adopting TDM strategies (such as ridesharing and flexible working hours) as traditionally done, transportation agencies can implement and increase the impact of TDM strategies by working with CSPs who may then have more (positive) influences on the commuting related decisions of employers and commuters.

\gap

Last but not least, our analyses have also shown the value of considering employers and commuters as the two sides of a CSP, and applying the two-sided analysis method to the CSP.
\iffalse
[\textbf{JB: next paragraph may be removed}]
Two-sided markets have unique features compared with one-sided markets: 1) the volume of transactions on the CSP depends on the allocation of price between the two sides but not only on the aggregated price level; 2) the decision of each group of agents affects the outcomes of the other group of agents, typically through an externality, i.e. cross-side positive network effects and same-side negative network effects.

Such an analysis framework can help gain useful insights about the interactions of employers and employees on CSPs, and develop practical CSPs and new TDM strategies, as discussed above.
\fi
\section{Concluding Remarks}
There are now two emerging trends in urban mobility. First, MaaS connects directly demand and supply via mobile platforms, which has been transforming urban mobility in almost all aspects. Second, businesses (employers) are paying more attention to the commuting of their employees due to various reasons (e.g., tighter commuting reduction regulations, the need to recruit/retain talents, etc.). Combing these two trends has motivated innovations in mobility services (such as Scoop / Via) that provide carpool or transit services to co-workers by working closely with businesses. This paper proposed the concept of \textit{commuting service platforms} (CSP) and applied the two-sided market theory to study the demand-price relation and network effects in a market where employers and employees are directly connected by the envisioned CSP. A benchmark model was proposed to clarify the definition of the two-sidedness and the threshold of subsidization. Models for both the monopoly platform and the duopoly platforms were constructed. The duopoly model was further improved with demand constraints, which ensures that the participation rates of worksites and employees are almost the same.

\gap

The analyses presented in this paper allows us to obtain a basic understanding of CSP at the planning level and the interactions of its major players (employers, employees, and the platform), as well as how such interactions may impact the participation of the platform and its prices and profit, as detailed in Section \ref{sec:discussions}. Such analyses and findings can help gain useful insights on how to build CSPs and how to develop associated TDM strategies in practice .
%In the monopoly case, more agents from the two sides will join the CSP when either of their prices decreases. Under some parameter settings, the CSP can use commuters as a loss leader and attract participation from both sides. The increase of the same-side ``congestion'' effects from either side will reduce CSP's profit. However, the CSP is able to maintain its profit for a range of the same-side ``congestion'' effects. In the duopoly model, the participation from the two sides is affected by the relative price of the two CSPs ($p_N^B-p_W^B$ and $p_N^C-p_W^C$). Relatively lower prices on a CSP increase its participation. Consistent with the monopoly model, the cross-side positive effects encourage participation on a CSP. However, high overall cross-side effects intensify the competition between the two CSPs and no CSP can achieve high profits. The proposed demand constraints are effective in controlling the number of employees on each worksite. Under the demand constraints, both CSPs make good profits under most parameter settings ($\alpha^-$ and $\alpha^+$), which indicates that both CSPs will sustain in this two-sided market.

\gap

The proposed CSP and the two-sided market based analysis methods and results are just an initial step toward the emerging MaaS and ESTP that are rapidly evolving with innovations emerging quickly. The proximate commute scenario studied here is also a very simplified version of a general CSP. However, we believe that the envisioned CSP and the two-sided market based analysis method are the first step and a building block to understand and analyze such new trends in transportation. In future research, we will extend the above analysis in several important ways. First, we will investigate how to properly capture and model key components of the two-sided market, such as the key parameters (costs, benefits, etc.), utility functions, and demand functions of employees and worksites/employers in their commuting decisions. For this, understanding their behaviors in terms of commuting decisions is crucial and should be investigated. Second, we will relax Assumption (d) to add demand constraint (i.e., the number of employees) for each worksite in the Proximate Commute problem. Third, we need to extend proximate commute to more general commuting scenarios. For this, we will model how employees are matched with employers, with commuting services as one option provided by employers. We will assume that CSP potentially changes the ease of commuting and thereby the geographic radius through which employees search for jobs and thereby the matching function between them and available job vacancies in the context of a search and matching model. Forth, we will extend the analysis method to study CSP with more service options, e.g., different travel modes (ridesourcing, transit, or a hybrid), which can be modeled as a two-sided market with competitive services. For this, the CSP proposed here is largely an abstract concept. To be more practical, we will need to consider the myriad existing and future mobility services and think about innovative ways to combine them into an integrated CSP to serve practice commuting needs. To do so, we need to study the \emph{operational level} challenges of CSP. For example, designing the optimal number of commuters that each vehicle picks up, the matching of commuters to one serving vehicle, the optimal pick up locations, the optimal routes for the CSP vehicles, how to model the transfers between different modes, how employers' decisions may impact commuting service operations, etc. Such CSP operational issues are also challenging and merits further investigations. At the same time, understanding the behaviors and interactions of the major players (employers, employees, and agencies) with respect to commuting options (e.g., WF and NWF) and the operations of CSPs is crucial, which we can leverage to develop the next-generation, effective TDM strategies for commuting that take advantage the proposed CSP and emerging mobility options. We will pursue these topics in future research and results on these investigations may be reported in subsequent papers.

%To build practical CSPs, we should do

\section{Acknowledgments}
The authors thank Dr. Jacques Lawarree from the Department of Economics of the University of Washington for helping discussions on applying two-sided market theory to transportation applications.

\newpage
%   \citep[chap. 2]{key}    ==>> (Jones et al., 1990, chap. 2)
%   \citep[e.g.,][]{key}    ==>> (e.g., Jones et al., 1990)
%   \citep[see][pg. 34]{key}==>> (see Jones et al., 1990, pg. 34)


%\bibliographystyle{elsarticle-num}

\bibliography{harvard}

\begin{thebibliography}{}

\bibitem[\protect\citename{Andrea~Broaddus, }2009]{TDMcategories2009}
Andrea~Broaddus, Todd~Litman, Gopinath~Menon. 2009.
\newblock {\em Transportation Demand Management Training Document}.
\newblock P. O. Box 5180, 65726 ESCHBORN/GERMANY: German Technical
  Cooperaqtion.
\newblock Chap.~2, pages  19--23.

\bibitem[\protect\citename{Apple~Inc. {\em et~al.},
  }2012]{Employer-Sponsored2012}
Apple~Inc., Best Buy~Co., Inc., {\em et~al.} 2012.
\newblock {\em Success Stories of Employer-Sponsored Transportation Programs}.
\newblock Tech. rept.

\bibitem[\protect\citename{Armstrong, }2006]{armstrong2006competition}
Armstrong, Mark. 2006.
\newblock Competition in two-sided markets.
\newblock {\em The RAND Journal of Economics}, {\bf 37}(3), 668--691.

\bibitem[\protect\citename{Armstrong \& Wright, }2007]{armstrong2007two}
Armstrong, Mark, \& Wright, Julian. 2007.
\newblock Two-sided markets, competitive bottlenecks and exclusive contracts.
\newblock {\em Economic Theory}, {\bf 32}(2), 353--380.

\bibitem[\protect\citename{Caillaud \& Jullien, }2003]{caillaud2003chicken}
Caillaud, Bernard, \& Jullien, Bruno. 2003.
\newblock Chicken \& egg: Competition among intermediation service providers.
\newblock {\em RAND journal of Economics},  309--328.

\bibitem[\protect\citename{{Commute Seattle}, }2016]{commuteSeattle2016}
{Commute Seattle}. 2016.
\newblock {\em Seattle Employer Transportation Benefits Survey Results Report -
  March 2016}.
\newblock Tech. rept.

\bibitem[\protect\citename{{Commute Seattle}, }2017]{commuteSeattle2017}
{Commute Seattle}. 2017.
\newblock {\em ORCA 101 for downtown Seattle Employers}.
\newblock Tech. rept.

\bibitem[\protect\citename{Djavadian \& Chow, }2017]{djavadian2017agent}
Djavadian, Shadi, \& Chow, Joseph~YJ. 2017.
\newblock An agent-based day-to-day adjustment process for modeling 'Mobility
  as a Service' with a two-sided flexible transport market.
\newblock {\em Transportation Research Part B: Methodological}, {\bf 104},
  36--57.

\bibitem[\protect\citename{Dormehl, }2015]{Apple2015Dormehl}
Dormehl, Luke. 2015.
\newblock {\em Apple's Buses Are As Secretive And Efficient As Apple Itself}.
\newblock
  \url{https://www.cultofmac.com/269509/apples-buses-secretive-efficient-apple/
  }.

\bibitem[\protect\citename{Economides \& T{\aa}g, }2012]{economides2012network}
Economides, Nicholas, \& T{\aa}g, Joacim. 2012.
\newblock Network neutrality on the Internet: A two-sided market analysis.
\newblock {\em Information Economics and Policy}, {\bf 24}(2), 91--104.

\bibitem[\protect\citename{ERTICO, }2016]{MaaS2018}
ERTICO. 2016.
\newblock {\em {MaaS Aliance} Kernel Description}.
\newblock \url{https://maas-alliance.eu/}.
\newblock Accessed: 2018-07-02.

\bibitem[\protect\citename{Ferguson, }1990]{ferguson1990transportation}
Ferguson, Erik. 1990.
\newblock Transportation demand management planning, development, and
  implementation.
\newblock {\em Journal of the American Planning Association}, {\bf 56}(4),
  442--456.

\bibitem[\protect\citename{FHWA, }2012]{TDM2012}
FHWA. 2012.
\newblock {\em Transportation Demand Management Strategies}.
\newblock Federal High Way Administration.
\newblock Chap.~5, pages  31--103.

\bibitem[\protect\citename{Fudenberg \& Tirole, }1991]{fudenberg1991game}
Fudenberg, Drew, \& Tirole, Jean. 1991.
\newblock {\em Game theory}.

\bibitem[\protect\citename{{Georgia Institute of Technology}, }1994]{FHAW1994}
{Georgia Institute of Technology}. 1994.
\newblock {\em Overview of Travel Demand Management Measures}.
\newblock Tech. rept.

\bibitem[\protect\citename{Harrington, }2019]{harrington2019commute}
Harrington, Kate. 2019.
\newblock {\em Austin helps companies attract talent by making it easier to
  commute without a car}.
\newblock
  \url{https://mobilitylab.org/2019/01/29/austin-helps-companies-attract-talent-by-making-it-easier-to-commute-without-a-car/}.
\newblock Accessed: 2019-08-19.

\bibitem[\protect\citename{Helft, }2007]{Google2007}
Helft, Miguel. 2007.
\newblock {\em Google's Buses Help Its Workers Beat the Rush}.
\newblock \url{https://www.nytimes.com/2007/03/10/technology/10google.html }.

\bibitem[\protect\citename{Holgu{\'\i}n-Veras {\em et~al.},
  }2011]{holguin2011impacts}
Holgu{\'\i}n-Veras, Jos{\'e}, Wang, Qian, Xu, Ning, \& Ozbay, Kaan. 2011.
\newblock The impacts of time of day pricing on car user behavior: findings
  from the Port Authority of New York and New Jersey’s initiative.
\newblock {\em Transportation}, {\bf 38}(3), 427--443.

\bibitem[\protect\citename{Hotelling, }1929]{hotelling1990stability}
Hotelling, Harold. 1929.
\newblock Stability in competition.
\newblock {\em Economic Journal},  39: 41--57.

\bibitem[\protect\citename{Inrix, }2015]{Inrix2015}
Inrix. 2015.
\newblock {\em ``Traffic Scorecard"}.
\newblock Tech. rept.

\bibitem[\protect\citename{Jeon \& Rochet, }2010]{jeon2010pricing}
Jeon, Doh-Shin, \& Rochet, Jean-Charles. 2010.
\newblock The pricing of academic journals: A two-sided market perspective.
\newblock {\em American Economic Journal: Microeconomics}, {\bf 2}(2), 222--55.

\bibitem[\protect\citename{Kadesh \& Roach, }1997]{kadesh1997commute}
Kadesh, Eileen, \& Roach, William~T. 1997.
\newblock Commute trip reduction—a collaborative approach.
\newblock {\em Energy policy}, {\bf 25}(14-15), 1217--1225.

\bibitem[\protect\citename{Kaiser \& Wright, }2006]{kaiser2006price}
Kaiser, Ulrich, \& Wright, Julian. 2006.
\newblock Price structure in two-sided markets: Evidence from the magazine
  industry.
\newblock {\em International Journal of Industrial Organization}, {\bf 24}(1),
  1--28.

\bibitem[\protect\citename{Lerner, }1934]{lerner1934}
Lerner, Abba. 1934.
\newblock The Concept of Monopoly and the Measurement of Monopoly.
\newblock {\em Review of Economic Studiesl},  157--175.

\bibitem[\protect\citename{Levy, }2016]{Amazon2016levy}
Levy, Nat. 2016.
\newblock {\em Amazon launches stealthy employee shuttle system with
  nondescript buses and little fanfare}.
\newblock
  \url{https://www.geekwire.com/2016/amazon-quietly-debuts-commuter-shuttle-program/
  }.

\bibitem[\protect\citename{Li {\em et~al.}, }2019]{li2019characterization}
Li, Wenxiang, Pu, Ziyuan, Li, Ye, \& Ban, Xuegang~Jeff. 2019.
\newblock Characterization of ridesplitting based on observed data: A case
  study of Chengdu, China.
\newblock {\em Transportation Research Part C: Emerging Technologies}, {\bf
  100}, 330--353.

\bibitem[\protect\citename{{Luum}, }2019]{Luum2019}
{Luum}. 2019.
\newblock {\em {One system for all thins commute}}.
\newblock \url{https://luumbenefits.com/}.
\newblock Accessed: 2019-08-02.

\bibitem[\protect\citename{Mobility-Lab, }2013]{MobilityLab2009}
Mobility-Lab. 2013.
\newblock {\em What is TDM?}
\newblock \url{https://mobilitylab.org/about-us/what-is-tdm/}.

\bibitem[\protect\citename{Mullins, }1995]{Mullins1995}
Mullins, G. \&~Mullins, C. 1995.
\newblock {\em Proximate Commuting: A Demonstration Project of a Strategic
  Commute Reduction Program}.
\newblock Washington State Department of Transportation.

\bibitem[\protect\citename{Mullins, }1999]{ProximateCommute2018}
Mullins, Gene. 1999.
\newblock {\em Proximate Commute: innovative web-based transportation and
  work/family benefits program}.
\newblock \url{http://www.proximatecommute.com/}.
\newblock Accessed: 2018-07-02.

\bibitem[\protect\citename{Rochet \& Tirole, }2003]{rochet2003platform}
Rochet, Jean-Charles, \& Tirole, Jean. 2003.
\newblock Platform competition in two-sided markets.
\newblock {\em Journal of the european economic association}, {\bf 1}(4),
  990--1029.

\bibitem[\protect\citename{Rochet \& Tirole, }2006]{rochet2006two}
Rochet, Jean-Charles, \& Tirole, Jean. 2006.
\newblock Two-sided markets: a progress report.
\newblock {\em The RAND journal of economics}, {\bf 37}(3), 645--667.

\bibitem[\protect\citename{Rochet \& Tirole, }2008]{rochet2008tying}
Rochet, Jean~Charles, \& Tirole, Jean. 2008.
\newblock Tying in two-sided markets and the honor all cards rule.
\newblock {\em International journal of industrial organization}, {\bf 26}(6),
  1333--1347.

\bibitem[\protect\citename{Rysman, }2009]{rysman2009economics}
Rysman, Marc. 2009.
\newblock The economics of two-sided markets.
\newblock {\em Journal of economic perspectives}, {\bf 23}(3), 125--43.

\bibitem[\protect\citename{{Scoop}, }2019]{Scoop2019}
{Scoop}. 2019.
\newblock {\em {Dramatically improve your commute}}.
\newblock \url{https://www.takescoop.com/}.
\newblock Accessed: 2019-08-02.

\bibitem[\protect\citename{Shaheen \& Ismail, }2016]{SharedMobility2016}
Shaheen, S., Adam~C., \& Ismail, Z. 2016.
\newblock {\em ``Shared Mobility: Current Practices and Guiding Principles"}.
\newblock Tech. rept.

\bibitem[\protect\citename{Wang {\em et~al.}, }2016]{wang2016pricing}
Wang, Xiaolei, He, Fang, Yang, Hai, \& Gao, H~Oliver. 2016.
\newblock Pricing strategies for a taxi-hailing platform.
\newblock {\em Transportation Research Part E: Logistics and Transportation
  Review}, {\bf 93}, 212--231.

\bibitem[\protect\citename{{Wikipedia}, }Accessed: 2018]{wiki:commuting}
{Wikipedia}. Accessed: 2018.
\newblock {\em Commuting}.

\bibitem[\protect\citename{Wright, }2004]{wright2004one}
Wright, Julian. 2004.
\newblock One-sided logic in two-sided markets.
\newblock {\em Review of Network Economics}, {\bf 3}(1).

\bibitem[\protect\citename{WSDOT, }2009]{WSDOTctr2009}
WSDOT. 2009.
\newblock {\em Commute Trip Reduction Program}.
\newblock Tech. rept.

\bibitem[\protect\citename{Young, }1992]{Ernst1992}
Young, Ernst~\&. 1992.
\newblock {\em Regulation XV Cost Survey}.
\newblock Tech. rept.

\bibitem[\protect\citename{Yushimito {\em et~al.}, }2014]{yushimito2014two}
Yushimito, Wilfredo~F, Ban, Xuegang, \& Holgu{\'\i}n-Veras, Jos{\'e}. 2014.
\newblock A two-stage optimization model for staggered work hours.
\newblock {\em Journal of Intelligent Transportation Systems}, {\bf 18}(4),
  410--425.

\bibitem[\protect\citename{Yushimito {\em et~al.},
  }2015]{yushimito2015correcting}
Yushimito, Wilfredo~F, Ban, Xuegang, \& Holgu{\'\i}n-Veras, Jos{\'e}. 2015.
\newblock Correcting the market failure in work trips with work rescheduling:
  an analysis using bi-level models for the firm-workers interplay.
\newblock {\em Networks and Spatial Economics}, {\bf 15}(3), 883--915.

\bibitem[\protect\citename{Zha {\em et~al.}, }2016]{zha2016economic}
Zha, Liteng, Yin, Yafeng, \& Yang, Hai. 2016.
\newblock Economic analysis of ride-sourcing markets.
\newblock {\em Transportation Research Part C: Emerging Technologies}, {\bf
  71}, 249--266.

\end{thebibliography}
\bibliographystyle{authordate1}

\begin{appendices}
\section{Duopoly model when workistes multi-home} \label{Append_oneside_multihome_platform}
	Worksites tend to view the CSPs as homogenous. In contrast, commuters tend to view the CSPs as heterogeneous because they often evaluate the level of service of a CSP based on various factors: (i) how convenient it is to choose a CSP based on a commuter's schedule; (ii) total vehicle travel time on a CSP, which is likely to be different on the WF CSP and the NWF CSP; (iii) the comfort level of the CSP services, etc. In section \ref{sec:BC_single-home}, we assume that the same-side ``congestion'' effects are high enough (condition \text{(B2)}) so that no agents will multi-home. Here we relax this constraint and allow worksites to multi-home. Here are the conditions for the duopoly model where the worksites multi-home and the commuters single-home:
	
	\gap
	
	\textbf{(C1)} $U_0^B = 0$ and $U_0^C$ is high enough so that commuters wish to join at least one of the CSPs\\
	\textbf{(C2)} (i) $t^B =0$, the cost of joining a CSP for worksites is low, so that it is possible for worksites to choose both CSPs; (ii) $t^C > \beta_N q_N^B+ \beta_W q_W^B$, the same-side ``congestion'' effects of commuters are high, ensuring that commuters single-home\\
	\textbf{(C3)} $f_W^B < \min \{ \frac{\alpha_W}{3} \}$, $f_i^C < \min \{ \frac{3\alpha_W}{4} \}$: ensures that both CSPs are willing to serve worksites.
	
	\gap
	
	To find the equilibrium in this setting, the consistent demand configurations need to be characterized. We first list all the possible configurations of worksites in a duopoly model, and then show that worksites will always choose Configuration 1 under conditions (C1) $\sim$ (C3). Commuters single-home under condition \textbf{(C2)} since \textbf{Lemma \ref{lem:single-home}} still applies.
	
	\gap

	\textit{Configuration 1: worksites multi-home}
	
	\gap
	
	Given that $Q^B = 1, q_W^B = q_N^B = 0$, the fraction of commuters joining the WF CSP is determined by the Hotelling Model,
	\begin{align}
	q_W^C = \frac{1}{2} + \frac{p_N^C - p_W^C + \beta_W - \beta_N}{2t^C} \label{eq:commuter_hotelling}
	\end{align}
	The number of commuters joining the NWF CSP is $1-q_W^C$. We assume that when a worksite is indifferent between join and not join a CSP, it will join the platform. It is optimal for worksites to mutlti-home when $U_{NW}^B \ge \max\{ U_N^B, U_W^B, 0 \}$, i.e., $-(p_W^B + p_N^B) - t^B + \alpha_W q_W^C + \alpha_N q_N^C \ge \max \{ -p_W^B - t^B x^B + \alpha_W q_W^C, -p_N^B -t^B(1-x^B) + \alpha_N q_N^C, 0  \}$, which implies that multi-homing is preferred over single-homing or not joining any of the CSPs. The first two inequalities can be written as,
	\begin{align}
	p_W^B \le (\frac{1}{2} + \frac{p_N^C - p_W^C + \beta_W - \beta_N}{2t^C}) \alpha_W  \label{ineq:conf1_W} \\
	p_N^B \le (\frac{1}{2} + \frac{p_W^C - p_N^C + \beta_N - \beta_W}{2t^C}) \alpha_N  \label{ineq:conf1_N}
	\end{align}
	Profits of the CSPs are,
	\begin{align}
	R_W =& p_W^B - f_W^B + (p_W^C - f_W^C)(\frac{1}{2} + \frac{p_N^C - p_W^C + \beta_W - \beta_N}{2t^C}) \\
	R_N =& p_N^B - f_N^B + (p_N^C - f_N^C)(\frac{1}{2} + \frac{p_W^C - p_N^C + \beta_N - \beta_W}{2t^C})
	\end{align}
		
	\gap
	
	\noindent\textit{Configuration 2: worksites single-home on the WF CSP}
		
	\gap
	
	The fraction of commuters joining the WF CSP is,
	\begin{align}
	q_W^C = \frac{1}{2} + \frac{p_N^C - p_W^C + \beta_W }{2t^C}
	\end{align}
	Worksites choose to single-home on the WF CSP when $U_W^B \ge \max\{ U_N^B, U_{NW}^B, 0 \}$, i.e., $-p_W^B - t^B x^B + \alpha_W q_W^C \ge \max \{   -p_N^B -t^B(1-x^B) + \alpha_N q_N^C, -(p_W^B + p_N^B) - t^B + \alpha_W q_W^C + \alpha_N q_N^C, 0  \}$. These inequalities can be written as,
	\begin{align}
	p_W^B \le (\frac{1}{2} + \frac{p_N^C - p_W^C + \beta_W }{2t^C}) \alpha_W \label{ineq:single_home_W}\\
	p_N^B \ge (\frac{1}{2} + \frac{p_W^C - p_N^C  - \beta_W}{2t^C}) \alpha_N
	\end{align}
	Profits of the CSPs are,
	\begin{align}
	R_W =& p_W^B - f_W^B + (p_W^C - f_W^C)(\frac{1}{2} + \frac{p_N^C - p_W^C + \beta_W}{2t^C}) \\
	R_N =& (p_N^C - f_N^C)(\frac{1}{2} + \frac{p_W^C - p_N^C  - \beta_W}{2t^C})
	\end{align}
	\textit{Configuration 3: worksites single-home on the NWF CSP}
			
	\gap
	
	The fraction of commuters on the WF CSP is,	
	\begin{align}
	q_W^C = \frac{1}{2} + \frac{p_N^C - p_W^C - \beta_N }{2t^C}
	\end{align}
	Worksites choose to single-home on the WF CSP when $U_N^B \ge \max\{ U_W^B, U_{NW}^B, 0 \}$, i.e., $-p_N^B -t^B(1-x^B) + \alpha_N q_N^C \ge \max \{ -p_W^B - t^B x^B + \alpha_W q_W^C  , -(p_W^B + p_N^B) - t^B + \alpha_W q_W^C + \alpha_N q_N^C, 0  \}$. These inequalities can be written as,
	\begin{align}
	p_W^B \ge (\frac{1}{2} + \frac{p_N^C - p_W^C - \beta_N}{2t^C}) \alpha_W \\
	p_N^B \le (\frac{1}{2} + \frac{p_W^C - p_N^C + \beta_N}{2t^C}) \alpha_N \label{ineq:single_home_N}
	\end{align}
	Profits of the CSPs are,
	\begin{align}
	R_W =&  (p_W^C - f_W^C)(\frac{1}{2} + \frac{p_N^C - p_W^C - \beta_N}{2t^C}) \\
	R_N =& p_N^B - f_N^B + (p_N^C - f_N^C)(\frac{1}{2} + \frac{p_W^C - p_N^C  + \beta_N}{2t^C})
	\end{align}
		
	\gap
	
	\textit{Configuration 4: worksites join neither of the CSPs}
	
	\gap
	
	The proportion of commuters joining the WF CSP is the same as that in Configuration 1. Worksites do not want to join any CSP when $0 \ge \max \{ -p_W^B - t^B x^B + \alpha_W q_W^C  , -p_N^B -t^B(1-x^B) + \alpha_N q_N^C,-(p_W^B + p_N^B) - t^B + \alpha_W q_W^C + \alpha_N q_N^C, 0  \}$. This requires each inequality \eqref{ineq:conf1_W} and \eqref{ineq:conf1_N} be reversed. The CSPs only make profits from the commuter side.	There exists price range when some of the configurations overlap. To make the explanation more concise, we assume that the two CSPs set the same prices, i.e., $p_W^B=p_N^B=p^B, p_W^C=p_N^C=p^C$. Configuration 1, 2 and 3 are all consistent when $\max \{ (\frac{1}{2} - \frac{\beta_N}{2 t^C})\alpha_N , (\frac{1}{2} - \frac{\beta_N}{2 t^C})\alpha_W  \} \le p^B \le \min \{ (\frac{1}{2} + \frac{\beta_W - \beta_N}{2 t^C}) \alpha_W , (\frac{1}{2} + \frac{\beta_N - \beta_W}{2 t^C}) \alpha_N \} $. Configuration 4 is the reverse of configuration 1, so it is impossible for them to overlap. Configuration 2, 3 and 4 are consistent at the same time when $\max \{ (\frac{1}{2} + \frac{\beta_W - \beta_N}{2 t^C}) \alpha_W , (\frac{1}{2} + \frac{\beta_N - \beta_W}{2 t^C}) \alpha_N \} \le p^B \le \min \{ (\frac{1}{2} + \frac{\beta_W }{2 t^C}) \alpha_W , (\frac{1}{2} + \frac{\beta_N}{2 t^C}) \alpha_N \} $.
	
	\subsubsection{One-sided cross-side network effects}
	The analysis is straightforward when the cross-side benefits are one-sided. In this section we will discuss about such cases. The simplified network effects will still unveil important insights from the duopoly model. If we do not consider the cross-side benefits of commuters, then $\beta_W = \beta_N = 0$. In this case, the equilibrium is described in the following proposition,
	\begin{prop} \label{Prop_onesided_competitive}
		Let condition \textbf{(C1)-(C3)} hold and assume $\beta_W = \beta_N = 0$. Then the equilibrium is unique, CSPs will serve both sides of the market, with worksites multi-home and commuters single-home. The optimal prices for worksites are $p_W^B = (\frac{1}{2}+\frac{f_N^C-f_W^C - \alpha^-}{6t^C}) \alpha_W, p_N^B = (\frac{1}{2}+\frac{f_W^C-f_N^C + \alpha^-}{6t^C}) \alpha_N $. The equilibrium prices of commuters are depend on the parameter settings of the model. If $\frac{f_N^C}{3} + \frac{2(f_W^C)}{3} + t^C \ge \frac{ \alpha_N}{3} + \frac{2 \alpha_W}{3}$ and $\frac{2(f_N^C)}{3} + \frac{f_W^C}{3} + t^C \ge \frac{ 2\alpha_N}{3} + \frac{ \alpha_W}{3}$, the optimal prices for commuters are,
		\begin{align}
		p_W^C =& \frac{f_N^C- \alpha_N}{3} + \frac{2(f_W^C- \alpha_W)}{3} + t^C  \\
		p_N^C =& \frac{2(f_N^C- \alpha_N)}{3} + \frac{f_W^C- \alpha_W}{3} + t^C
		\end{align}
		CSPs make profits,
		\begin{align}
		R_W =&  - f_W^B + (\frac{ f_N^C - f_W^C -\alpha^-}{6t^C} + \frac{1}{2})( \frac{ f_N^C-f_W^C -\alpha^-}{3}+t^C) \label{eq:RW_oneside_1}\\
		R_N =&  - f_N^B + (\frac{f_W^C - f_N^C + \alpha^-}{6t^C} + \frac{1}{2})( \frac{ f_W^C-f_N^C + \alpha^- }{3}+t^C) \label{eq:RN_oneside_1}
		\end{align}
		If $\frac{f_N^C}{3} + \frac{2(f_W^C)}{3} + t^C < \frac{ \alpha_N}{3} + \frac{2 \alpha_W}{3}$ and $\frac{2(f_N^C)}{3} + \frac{f_W^C}{3} + t^C < \frac{ 2\alpha_N}{3} + \frac{ \alpha_W}{3}$ , the equilibrium prices for commuters are $p_W^C=0, p_N^C=0$. The profits of the CSPs are,
		\begin{align}
		R_W =&  - f_W^B + (\frac{ f_N^C - f_W^C -\alpha^-}{6t^C} + \frac{1}{2})(\alpha_W - f_W^C) \label{eq:RW_oneside_2}\\
		R_N =&  - f_N^B + (\frac{f_W^C - f_N^C + \alpha^-}{6t^C} + \frac{1}{2})(\alpha_N - f_N^C) \label{eq:RN_oneside_2}
		\end{align}
	\end{prop}
	
	\begin{myproof}
		It takes 2 steps to proof \textbf{Proposition \ref{Prop_onesided_competitive}}. First we show how we derive the equilibrium prices when both CSPs are willing to serve worksites. In the second step, we will explain why each CSP is better off by serving worksites.
				
		\gap
		
		Step (i): Suppose both CSPs are willing to serve the worksites, then the equilibrium prices follow the expressions presented in \textbf{Proposition \ref{Prop_onesided_competitive}}.
		
		\gap
		
		Since the decisions of commuters are not affected by worksites ($\beta_W=0, \beta_N=0$), the participation of commuters takes the form of equation \eqref{eq:commuter_hotelling}. Worksites, knowing the decisions of commuters, choose the WF CSP if $p_W^B \le q_W^C \alpha_W$, or choose the NWF CSP if $p_N^B \le q_N^C \alpha_N$, as characterized in inequalities \eqref{ineq:single_home_W} and \eqref{ineq:single_home_N}. Worksites' decisions of joining one CSP is independent of their decisions of joining the other CSP. Thus, CSPs will fully extract the surplus from worksites. In other words, each CSP will set the prices to worksites as high as possible. Set $p_W^B = q_W^C \alpha_W$, yields the profit function of the WF CSP. Similarly, we obtain the profit function of the NWF CSP when setting  $p_N^B = q_N^C \alpha_N$.
		\begin{align}
		R_W =&  - f_W^B + (p_W^C + \alpha_W  - f_W^C)(\frac{1}{2} + \frac{p_N^C - p_W^C }{2t^C}) \\
		R_N =&  - f_N^B + (p_N^C + \alpha_N  - f_N^C)(\frac{1}{2} + \frac{p_W^C - p_N^C }{2t^C})
		\end{align}
		From the perspective of CSPs, the revenue from worksites can be regarded as a reduction to the marginal cost of commuters, from $f_i^C$ to $f_i^C - \alpha_i$. Given that both CSPs serve worksites, the equilibrium is unique. Under condition \textbf{(C3)}, the profits of CSPs are non-negative. The profit maximization problems of CSPs are,
		\begin{align*}
		\frac{\partial R_i}{p_i^C} = 0 \quad \forall i \in \{ W,N\}
		\end{align*}
		which yield the price structure in \textbf{Proposition \ref{Prop_onesided_competitive}}.
		
		\gap
		
		\noindent Step (ii): Each CSP is better off when serving worksites.
		
		\gap
		
		In this part, we are going to prove that the WF CSP is better off when serving worksites. Similar proof applies to the NWF CSP. First, we study the cases when the equilibrium prices of commuters are positive, i.e., $\frac{f_N^C}{3} + \frac{2(f_W^C)}{3} + t^C \ge \frac{ \alpha_N}{3} + \frac{2 \alpha_W}{3}$ . Profit functions are given in equation \eqref{eq:RW_oneside_1} and \eqref{eq:RN_oneside_1}. Suppose on the contrary, the WF CSP stops serving worksites, its profit is,
		\begin{align}
		R_W =&  (p_W^C - f_W^C)(\frac{1}{2} + \frac{\frac{2(f_N^C- \alpha_N)}{3} + \frac{f_W^C- \alpha_W}{3} + t^C - p_W^C }{2t^C})
		\end{align}
		when worksite set price $p_W^C$ to commuters. The profit function is maximized at $p_W^C = \frac{f_N^C}{3} + \frac{f_W^C}{3} -\frac{\alpha_W}{6} - \frac{\alpha_N}{3} + t^C$. The WF CSP makes positive profits only if the price is larger than cost, i.e., $t^C > \frac{f_W^C - f_N^C}{3} + \frac{\alpha_W}{6} + \frac{\alpha_N}{3} $, in which case it obtains profit,
		\begin{align*}
		\widetilde R_W = \frac{1}{72t^C}(2\alpha_N + \alpha_W -2f_N^C + 2f_W^C -6t^C)^2
		\end{align*}
		Under condition \textbf{(C3)}, $\widetilde R_W$ is less than the profit we obtained in \textbf{Proposition \ref{Prop_onesided_competitive}} equation \eqref{eq:RW_oneside_1}. The proof is as follows.

\gap

	From condition \textbf{(C3)}, we know that $f_i^C < \frac{3}{4}\alpha_W$, thus, $-\alpha_W^2 + \frac{4}{3}\alpha_W (f_W^C - f_N^C) <0$, so that,
	\begin{align*}
	& \widetilde R_W < R_W \\
	\Longleftrightarrow \quad & \frac{1}{72t^C}(2\alpha_N + \alpha_W -2f_N^C + 2f_W^C -6t^C)^2 < -f_W^B + (\frac{f_N^C - f_W^C - \alpha^-}{6t^C} + \frac{1}{2})(\frac{f_N^C - f_W^C - \alpha^-}{3} + t^C) \\
	\Longleftrightarrow \quad  & 4\alpha_N \alpha_W + \alpha_W^2 - 4\alpha_W f_W^C - 12\alpha_W t^C < -8\alpha_N \alpha_W + 4\alpha_W^2 + 8 \alpha_W f_N^C - 8 \alpha_W f_W^C + 24 \alpha_W t^C - 72 f_W^B t^C \\
	\Longleftrightarrow \quad &  \quad\quad\quad\quad 12 \alpha_W (\alpha_N - 3 t^C) < 3 \alpha_W - 72 f_W^B t^C + 4 \alpha_W (f_W^C - f_N^C) \\
	\Longleftrightarrow \quad & \quad\quad\quad\quad  4 \alpha_W (3t^C - \alpha_N) > -\alpha_W^2 + 24 f_W^B t^C + \frac{4}{3}\alpha_W (f_W^C - f_N^C) \\
	\Longleftrightarrow \quad  & \quad\quad\quad\quad  8 \alpha_W t^C > -\alpha_W^2 + 24 f_W^B t^C + \frac{4}{3}\alpha_W (f_W^C - f_N^C)  \quad (-\alpha_W^2 + \frac{4}{3}\alpha_W (f_W^C - f_N^C) <0)\\
	\Longleftrightarrow \quad  &  \quad\quad\quad\quad  \alpha_W t^C >  3 f_W^B t^C \\
	\Longleftrightarrow \quad  &  \quad\quad\quad\quad  \frac{1}{3} \alpha_W  > f_W^B
	\end{align*}
	Condition \textbf{(C3)} states that $\frac{1}{3} \alpha_W  > f_W^B$. So the inequality $\frac{1}{72t^C}(2\alpha_N + \alpha_W -2f_N^C + 2f_W^C -6t^C)^2 < -f_W^B + (\frac{f_N^C - f_W^C - \alpha^-}{6t^C} + \frac{1}{2})(\frac{f_N^C - f_W^C - \alpha^-}{3} + t^C)$ holds, which proves that $\widetilde R_W < -f_W^B + (\frac{f_N^C - f_W^C - \alpha^-}{6t^C} + \frac{1}{2})(\frac{f_N^C - f_W^C - \alpha^-}{3} + t^C)$.
	In the same way, we can show that when $\frac{f_N^C}{3} + \frac{2(f_W^C)}{3} + t^C < \frac{ \alpha_N}{3} + \frac{2 \alpha_W}{3}$, the WF CSP will lose profit if not serving worksites.
	\end{myproof}
	
	\textbf{Proposition \ref{Prop_onesided_competitive}} finds the conditions under which an equilibrium exists where worksites multi-home and have their surplus fully extracted.

\end{appendices}





% \end{thebibliography}

\end{document}